\documentclass[twocolumn,aps,pra,superscriptaddress,amsfonts,longbibliography]{revtex4-1}
\usepackage[tight]{subfigure}
\usepackage{physics,tensor,siunitx}
\usepackage{amsmath,dsfont}
\usepackage{mathtools}
\usepackage{amssymb}
\usepackage{booktabs}
\usepackage{mathrsfs}
\usepackage{varwidth}
\usepackage{graphicx,color,caption}
\usepackage{bm}
\usepackage{cases}
\usepackage{array}
\usepackage{dcolumn}
\usepackage{chngcntr}
\usepackage{hepparticles}
\usepackage{heppennames}
\usepackage{hepnicenames}
\usepackage{epstopdf}
\usepackage{newfloat}
\usepackage{flushend}
\usepackage{nicefrac}
\usepackage{comment}
\usepackage{subfigure}
\usepackage{svg}
\usepackage{etex}
\usepackage[version=4]{mhchem}
\usepackage{adjustbox}
\DeclareFloatingEnvironment[name={Extended Data Figure}]{suppfigure}
\DeclareUnicodeCharacter{2212}{-}
\usepackage{hyperref}
\hypersetup{
    colorlinks=true,
    linkcolor=blue,     % color of internal links
    citecolor=blue,    % color of citation links
    urlcolor={}         % color of external links
}
\usepackage[capitalise]{cleveref}
\usepackage{tikz}
\usetikzlibrary{positioning,calc}
\usetikzlibrary{decorations.pathmorphing} % noisy shapes
\usetikzlibrary{fit}					% fitting shapes to coordinates
\usetikzlibrary{backgrounds}	% drawing the background after the foreground
\usetikzlibrary{shapes.geometric} %
\usetikzlibrary{positioning}
\usetikzlibrary{arrows.meta}
\usetikzlibrary{quantikz2}
\usepackage{yquant}

\newcommand{\todo}[1]{%
  \fcolorbox{red}{red!15}{%
    \parbox{0.9\linewidth}{\textcolor{black}{#1}}%
  }%
}

\begin{document}

\title{Tensor-Network-Based Distributed Quantum Dynamics on Independent Quantum Computers}
%\title{An asynchronous, distributed quantum algorithm for Wavepacket Dynamics and Vibrational spectroscopy using tensor networks: Implementation on ion-trap quantum computers for the study of small protonated water clusters}

\author{Anurag Dwivedi}
\affiliation{Department of Chemistry, Indiana University, Bloomington, Indiana 47405, USA}
\affiliation{Indiana University Quantum Science and Engineering Center, Bloomington, Indiana 47405, USA}
\author{Melissa C. Revelle}
\affiliation{Sandia National Laboratories, Albuquerque, New Mexico 87123, USA}
\author{Daniel S. Lobser}
\affiliation{Sandia National Laboratories, Albuquerque, New Mexico 87123, USA}
\author{Brian K. McFarland}
\affiliation{Sandia National Laboratories, Albuquerque, New Mexico 87123, USA}
\author{Edward C. Tortorici}
\affiliation{Sandia National Laboratories, Albuquerque, New Mexico 87123, USA}
\author{Christopher G. Yale}
\affiliation{Sandia National Laboratories, Albuquerque, New Mexico 87123, USA}
\author{Susan M. Clark}
\affiliation{Sandia National Laboratories, Albuquerque, New Mexico 87123, USA}
\author{Philip Richerme}
\affiliation{Department of Physics, Indiana University, Bloomington, Indiana 47405, USA}
\affiliation{Indiana University Quantum Science and Engineering Center, Bloomington, Indiana 47405, USA}
\author{Srinivasan S. Iyengar}
\email{iyengar@iu.edu}
\affiliation{Indiana University Quantum Science and Engineering Center, Bloomington, Indiana 47405, USA}
\affiliation{Department of Chemistry, Indiana University, Bloomington, Indiana 47405, USA}
\date{\today}

\begin{abstract}
%\textbf{\abstractname}
We present an approach based on tensor networks for distributed quantum computing simulation of chemical wavepacket dynamics in a continuous variable representation. The central idea is that the tensor-network representation of the multidimensional time-evolution operator naturally induces an elevated Hilbert space in which the global dynamics decomposes into a collection of independent lower-dimensional propagations. This transformation converts an entangled quantum evolution into a set of parallel computational tasks that can be executed asynchronously across heterogeneous quantum and classical computing architectures. 
%One key insight that arises from our work is that the entanglement entropy directly determines the size of the elevated Hilbert space and the number of parallel computational streams in this formalism. 
The resulting formalism establishes a direct connection between tensor-network decompositions, uniformly controlled quantum circuits, and asynchronous distributed quantum computing.
The approach is developed with a goal towards hybrid quantum/classical implementation, and %in fact 
is appropriate for a general heterogeneous mixture of quantum hardware systems; it is demonstrated here on %a stream of 
ion-trap quantum computers. The experimental realization of the asynchronously distributed quantum processes that arise from the tensor-network decomposition are carried out on the Sandia National Laboratories' trapped-ion quantum computer, where the circuits are compiled using native partial-entangling $XX(\theta)$ gates, reducing the expected two-qubit gate infidelity by more than 30\% relative to conventional fully entangling decompositions. We demonstrate the methodology by quantum computing the vibrational spectral properties of a small protonated water cluster that shows critical quantum nuclear behavior. Such water cluster systems have been found to be challenging for classical computation and for experimental action spectroscopy and here, for the first time, we provide results for vibrational spectroscopy that are in agreement with the respective classical results to within 4cm$^{-1}$, thus allowing for the potential for spectroscopic accuracy from quantum computations. To compute molecular vibrations efficiently, we also introduce here a modified version of the phase estimation algorithm to directly obtain energy differences (instead of absolute energies) that are most relevant as spectroscopic observables. More broadly, we note that while the methods developed here have been demonstrated for chemical dynamics and vibrational spectroscopy, the general prescription of introducing an elevated space where the dynamical components are decoupled and treated as a parallel tasks, with individual components being represented as continuous variables, is also appropriate for other continuous systems such as entangled optical beams and correlated fluid elements. These extensions provide a pathway toward scalable distributed quantum simulation on future heterogeneous quantum-computing architectures.
\end{abstract}
\maketitle

\section{Introduction}
\label{Sec:Introduction}
Simulating the quantum dynamics of interacting many-body systems remains a grand challenge\cite{Feynman1982,Feynman-Comp,hibbs} in multiple areas of the physical sciences. This difficulty is particularly pronounced in molecular systems, with coupled nuclear degrees of freedom, such as in hydrogen-bond networks in water, as well as in hydrogen, hydride, and proton
%and in other hydrogen/hydride/proton 
transfer reaction problems in biological and atmospheric systems, where an accurate treatment of high-dimensional wavepacket dynamics on correlated potential energy surfaces is needed. Despite significant advances in classical computing algorithms, such as the multiconfigurational time-dependent Hartree (MCTDH) approximation\cite{MCTDH-Meyer1,Manthe1992-kh,mctdh-meyer,Meyer2009-em,MLMCTDHWang,Burghardt-MLMCTDH,Otto_POTFIT} %and tensor network methods\cite{Orus2014-gg,Chan-While-MPS-MPO,Nicole-TN}, 
the exponential scaling of both computation and storage, %Hilbert space 
with system size\cite{Otto_POTFIT,Nicole-TN,Batista-TTsoft-2017}, continues to limit predictive simulations of such complex systems.

At the same time, quantum computing appears to offer a fundamentally different route to  address such challenges by encoding the wavefunction directly on a quantum register and evolving these under unitary dynamics\cite{Nielsen-Chuang-QuantComp,preskill,Preskill2021-qo}. In principle, algorithms based on quantum phase estimation (QPE)\cite{Nielsen-Chuang-QuantComp} enable access to spectral features through time evolution and Fourier transforms. However, practical implementations of such approaches on near-term quantum devices are severely restricted by circuit depth\cite{frag-QC-Harry,frag-QC-2}, entangling gate counts\cite{abughanem2024two,yan2025limitations}, 
%{\marginpar {\footnotesize {\color{red} {Need more references here to do with circuit depth specifically and how this affects QC.}}}}
and limited qubit connectivity\cite{yuan2023does}. These limitations are particularly acute for quantum dynamical simulations, where repeated application of time-evolution operators may be needed.

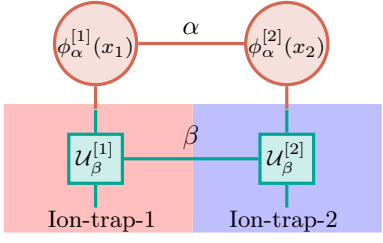
\begin{figure}
    \centering
    % Color palette
\definecolor{bblue}{rgb}{0.19, 0.55, 0.91}
\definecolor{green}{rgb}{0.0, 0.65, 0.58}
\definecolor{purple}{rgb}{0.6, 0.4, 0.8}
\definecolor{orange}{rgb}{0.83, 0.4, 0.32}

    \tikzstyle{psi} = [rectangle, very thick, minimum width=2cm, minimum height = 0.6cm, inner sep=0.0cm,draw=purple, fill=purple!20]
    
    \tikzstyle{phi} = [circle, very thick, minimum size=0.6cm, inner sep=0.0mm, draw=orange, fill=orange!20]
    
    \tikzstyle{phinew} = [circle, very thick, minimum size=0.6cm, inner sep=0.0mm, draw=bblue, fill=bblue!20]
    
    \tikzstyle{phiempty} = [circle, very thick, minimum size=0.cm, inner sep=0.0mm, draw=bblue, fill=bblue!20]

    % Time-Evolution operator tensor cores
    \tikzstyle{U} = [rectangle, very thick, minimum size=0.6cm, inner sep=0.1cm,draw=green, fill=green!20]
    
    % Define arrow style
    \tikzset{arw/.style={-Stealth, thick}}

    \begin{tikzpicture}[>=latex, node distance=0.7cm,
        every edge quote/.style={font=\fontsize{7}{1}}]
%        \node (psi1) [psi,font=\fontsize{9}{1}]{$\psi(x_1,x_2)$};
      
%        \node (psix01) [below=0mm of psi1, xshift = -7.5mm, minimum size=0mm, inner sep=0mm]{};
%        \node (psix02) [below=0mm of psi1, xshift = +7.5mm, minimum size=0mm, inner sep=0mm]{};

%        \node (psix00) [below=2mm of psi1, minimum size=0mm, inner sep=0mm]{};
        
%        \node (psix1) [below=2mm of psi1, xshift = -7.5mm, minimum size=0mm, inner sep=0mm]{};
%        \node (psix2) [below=2mm of psi1, xshift = +7.5mm, minimum size=0mm, inner sep=0mm]{};
        
        \fill[fill=blue!100!white!25](0,-2.3)--(0,-0.8)--(2.5,-0.8)--(2.5,-2.5)--(0,-2.5);
        \fill[fill=red!100!white!25](0,-2.3)--(0,-0.8)--(-2.5,-0.8)--(-2.5,-2.5)--(0,-2.5);

        \node (pphiempty) [phiempty, font=\fontsize{7}{1}]{};

        \node (pphi1) [phi, left=7mm of pphiempty, font=\fontsize{8}{1}]{$\phi^{[1]}_\alpha(x_1)$};
        \node (pphi2) [phi, right=14mm of pphi1, font=\fontsize{8}{1}]{$\phi^{[2]}_\alpha(x_2)$};
        
        \node (ppx1) [below=3mm of pphi1, minimum size=0mm, inner sep=0mm]{};
        \node (ppx2) [below=3mm of pphi2, minimum size=0mm, inner sep=0mm]{};
        
        \node (V1) [U, below=3mm of ppx1, font=\fontsize{8}{1}]{$\mathcal{U}^{[1]}_\beta$};
        \node (V2) [U, below=3mm of ppx2, font=\fontsize{8}{1}]{$\mathcal{U}^{[2]}_\beta$};         

    %    \node (u1) [U, below=2mm of K1, font=\fontsize{8}{1}]{$\mathcal{V}_{1}$};
     %   \node (u2) [U, below=2mm of K2, font=\fontsize{8}{1}]{$\mathcal{V}_{2}$};
             
        \node (ux1) [below=3mm of V1, minimum size=0mm, inner sep=0mm]{};
        \node (ux2) [below=3mm of V2, minimum size=0mm, inner sep=0mm]{};

%        \draw[-To, very thick, black, to path={-| (\tikztotarget)}] (psix00) -- (ppx01);
        
        \path[-]
            
%           \draw[arw,thick,draw=black] (eql0sgn.south) -| (eqlsgn.north);
           
%            (eql0sgn) edge[very thick, purple] (eqlsgn)
%            (psix01) edge[very thick, purple] node[left,black, yshift=-1mm, font=\fontsize{10}{1}] {} (psix1)
%            (psix02) edge[very thick, purple] 
%            node[left,black, yshift=-1mm,font=\fontsize{10}{1}] {} (psix2)

%            (psix00) edge[thick, black] 
%            node[left,black, yshift=-1mm,font=\fontsize{10}{1}] {} (ppx01)
%            (psix00) edge[thick, black] 
%           node[left,black, yshift=-1mm,font=\fontsize{10}{1}] {} (ppx02)
            
            (pphi1) edge[very thick, orange] node[above,black,font=\fontsize{10}{1}] {$\alpha$} (pphi2)
           
            (pphi1) edge[very thick, orange] node[left,black,yshift=-1mm,font=\fontsize{7}{1}] {} (ppx1)
            (pphi2) edge[very thick, orange] node[left,black,yshift=-1mm,font=\fontsize{7}{1}] {} (ppx2)
            
            (V1) edge[very thick, green] node[above,black,font=\fontsize{10}{1}] {$\beta$} (V2)
            
   %         (u1) edge[very thick, green] node[below,black,font=\fontsize{10}{1}] {} (u2)

            (V1) edge[very thick, green] (ppx1)
            (V1) edge[very thick, green] node[left,black,yshift=-0.5mm,font=\fontsize{7}{1}] {}(ux1)
            (V2) edge[very thick, green] (ppx2)
            (V2) edge[very thick, green] node[left,black,yshift=-0.5mm,font=\fontsize{7}{1}] {}(ux2)
            
            %(u1) edge[very thick, green] (K1)
            %(u2) edge[very thick, green] (K2)
           
            ;
        \filldraw[black] (1.2,-2.1) circle (0pt)node[below]{Ion-trap-2};
        \filldraw[black] (-1.2,-2.1) circle (0pt)node[below]{Ion-trap-1};
%        \filldraw[black] (0.0,-3.2) circle (0pt)node[below]{In general, $\alpha\times\beta$ times};
%        \filldraw[black] (0.0,-3.6) circle (0pt)node[below]{number of dimensions};
%        \filldraw[black] (0.0,-4) circle (0pt)node[below]{distributed ion traps,};        
%        \filldraw[black] (0.0,-4.4) circle (0pt)node[below]{working in parallel.};        
%        \filldraw[black] (2.55,-1.4) circle (0pt)node[above]{Ion-trap-2};
%        \filldraw[black] (2.55,-1.8) circle %(0pt)node[above]{Quantum};
%        \filldraw[black] (2.55,-2.2) circle (0pt)node[above]{Computer-2};
%        \filldraw[black] (-0.90,-1.4) circle (0pt)node[above]{Ion-trap-1};
  %      \filldraw[black] (-0.90,-1.8) circle (0pt)node[above]{Quantum};
   %     \filldraw[black] (-0.90,-2.2) circle (0pt)node[above]{Computer-1};
           
    \end{tikzpicture}

\begin{comment}
\begin{tikzpicture}[scale=0.75]
\fill[fill=blue!100!white!35](1.25,-1)--(1.25,1)--(3.75,1)--(3.75,-1)--(1.25,-1);
\fill[fill=red!100!white!35](1.25,1)--(1.25,3)--(3.75,3)--(3.75,1)--(1.25,1);

\filldraw[black] (0,0) circle (3pt)node[left]{$\ket{\phi^\alpha_2}$};
\filldraw[black] (0,2) circle (3pt)node[left]{$\ket{\phi^\alpha_1}$};
\draw[black,thick](0,0)--(2.5,0);
\draw[black,thick](0,2)--(2.5,2);
\filldraw[black] (2.5,0) circle (3pt)node[below]{Ion-trap-2};
\filldraw[black] (2.5,2) circle (3pt)node[above]{Ion-trap-1};
\draw[black,thick](2.5,0)--(4.5,0);
\draw[black,thick](2.5,2)--(4.5,2);
\draw[black,thick](2.5,0)--(2.5,2)node[midway,right]{$\beta$};
\end{tikzpicture}
\end{comment}
    \caption{The essence of the distributed algorithm presented here which appears from a tensor network formalism. There is one set of parallel streams of computational tasks for each value of the pair of variables $(\alpha,\beta)$. Thus, there are $[\alpha\times\beta\times {\text {number of nuclear dimensions}}]$ parallel computing tasks, each executed here as a separate ion-trap computation, possible due to elevation of the entanglement variables. This yields an enormous computational advantage as the number of dimensions increase. Future developments will involve a heterogeneous set of classical and hardware systems all runing these parallel set of tasks.}
    \label{fig:TN-parallel}
\end{figure}

In this publication, we demonstrate a tensor-network-based framework for quantum wavepacket dynamics and show that this formalism yields an asynchronous and parallelizable quantum computing algorithm, which we implement on Sandia National Labs' QSCOUT ion-trap quantum computing system\cite{IonQ-Anurag,TN-Miguel-Anurag}. This approach enables a decomposition of the time-evolution operator into a structured set of lower-dimensional operations suitable for distributed quantum execution. By expressing both the wavefunction and propagator in a tensor network form, we show that the global unitary evolution can be recast as a block-diagonal operator in an elevated, higher-dimensional space, whose block-components correspond to conditionally applied product-state time-evolution operators. This representation establishes a direct mapping between tensor network contractions,   uniformly controlled quantum circuits\cite{bergholm2005uniformcontrolgate,2025uniformcontrolgate}, and distributed/hybrid quantum computing architectures where entanglement indices are used to create an {\em elevated} space and are also used as control quantum registers. %Furthermore, the entanglement entropy determines the number of distributed quantum computing channels, and  becomes a resource for computation. 
%{\marginpar {\footnotesize {\color{red} {I think figure 1 doesn't represent the idea talked here as the two parallelism are kind of different, one is where the unitary is written in terms of block diagonal product of uniraties for each sets of $\beta$ and the other is where every dimension is parallel, so one is parallel across entanglement bond dimension and the other is across dimensions. Figure 1 talked about the paralellism of the later.}}}} 
As a result, the overall dynamics is exactly decomposed into a family of independent, reduced-dimensional quantum tasks that may be executed in parallel\cite{Preskill2018-NISQ,Nai-Hui-Hybrid-QC,IonQ-Anurag,frag-QC-Harry,frag-QC-2} across multiple quantum processors with potential for use as a hybrid quantum/classical algorithm. Figure \ref{fig:TN-parallel} captures the essence of this idea. 

In addition to the above, we have also utilized here a {\em quantum resource-optimized variant of the phase estimation algorithm} tailored for quantum dynamical simulations on NISQ environments. Here, the time-evolution stage is implemented on quantum hardware, while the Fourier transform is performed classically. This hybrid approach significantly reduces circuit complexity while preserving access to spectroscopic observables, including energy differences and vibrational features derived from time-correlation functions. Together with the tensor-network decomposition, this enables a practical pathway for performing quantum dynamical simulations on near-term devices with limited coherence and connectivity.

We demonstrate this framework through the simulation of vibrational dynamics in a protonated water-wire system, implemented on a distributed set of trapped-ion quantum processors. The water-cluster systems have been considered to be a major challenge to both experiment\cite{johnson-jordan-21mer,Johnson-21mer-2016} and theory\cite{HDMeyer-Zundel-1,HDMeyer-Zundel-2,HDMeyer-Zundel-3,Scott-proj,admp-21mer,admp-21mer-2}. The study of protonated water clusters have deep fundamental as well as applied implications due to prevalence in biological ion-channels\cite{allen2003gramicidin}, inside enzyme active sites\cite{Nagle:78,baciou1995interruption,Guo}, in polymer electrolyte fuel cells\cite{ye2012water} and in the earth's atmosphere\cite{Solomon-Rowland-O3,Finlayson-Pitts-Review,Turco-HCl,SolomonO3,Pilling-Review,Troe,Logan-1981-HOX-ozone,McEwanPhillips:Review,Castleman-ChemRev,McEwanchemistryoftheatmosphere,Waynechemistryoftheatmosphere,Francisco-ACR,Hoffman-N2-H2-redelim,Thorneley-Lowe-PCET-N2,Schrock-N2-NCET}. The study of such systems is complicated by, multi-dimensional quantum nuclear effects arising from hydrogen bonded networks as evidenced by the Grothhuss mechanism of proton transfer\cite{grotthuss}. 
By decomposing the multidimensional Hamiltonian for such systems into a collection of effective one-dimensional components as formally allowed by the aforementioned tensor network formalism, we realize parallel quantum simulations whose combined results reproduce the key spectral features of the system. Indeed, the agreement between quantum hardware results and exact classical simulations is of the order of $\approx 4cm^{-1}$. This is at the {\em spectroscopic accuracy level of agreement between the quantum simulation and classical simulation} and validates the approach and suggests its potential for scaling to more complex systems in future.

This paper is organized as follows: In \cref{Sec:TN,Sec:PEA-Vib}, we present the theoretical framework underlying the tensor-network-based parallel quantum dynamics approach and the phase estimation algorithm developed for resource-optimized quantum vibrational spectroscopy. In particular, \cref{Sec:TN} introduces the tensor-network decomposition of the multidimensional propagator, the resulting block-diagonal structure in the elevated representation, and the distributed quantum execution strategy for parallel implementation on quantum hardware. \cref{Sec:PEA-Vib} describes the modified phase estimation protocol for extracting vibrational spectra with reduced circuit depth requirements. The experimental implementation of the trapped-ion quantum simulations is described in \cref{Sec:Experimental}. In \cref{Sec:Results}, we demonstrate the framework through simulations of proton-transfer vibrational dynamics in a protonated water-wire system implemented on a trapped-ion quantum processor at Sandia National Laboratories.
% Modifed by MCR
In particular, we present the reduced-dimensional potential energy surfaces, the quantum wavepacket dynamics obtained from trapped-ion quantum hardware, vibrational spectra extracted from the quantum dynamics, and a detailed comparison between the quantum-computed spectra and exact multidimensional calculations, including a quantitative error analysis. Finally, conclusions and future directions are discussed in \cref{Sec:Outlook}.

\section{Tensor network based parallel quantum dynamics simulations on quantum computers}
\label{Sec:TN}
The central challenge in quantum dynamical simulations lies in the efficient representation and implementation of the time-evolution operator acting on a high-dimensional Hilbert space. For systems with multiple coupled degrees of freedom, such as  wavepackets for molecular systems, the dimensionality of the Hilbert space grows exponentially with system size\cite{Otto_POTFIT,Feynman1982,Feynman-Comp}, rendering direct implementations of the time-evolution operator to be prohibitively expensive on classical hardware architectures.

To address this challenge, we begin with a tensor-network-based decomposition of quantum dynamics that exposes a structured representation of the propagator amenable to distributed quantum execution. The key idea is to exploit the entanglement structure of both the wavefunction and the propagator to create a global evolution strategy that decomposes exactly into a collection of lower-dimensional operations and may then be executed in parallel on multiple hardware systems.

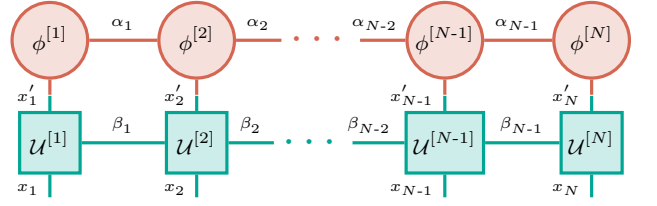
\begin{figure}[tbp]
\centering
% Color palette
\definecolor{bblue}{rgb}{0.19, 0.55, 0.91}
\definecolor{green}{rgb}{0.0, 0.65, 0.58}
\definecolor{purple}{rgb}{0.6, 0.4, 0.8}
\definecolor{orange}{rgb}{0.83, 0.4, 0.32}

    \tikzstyle{phi} = [circle, very thick, minimum size=1cm, inner sep=0.2mm, draw=orange, fill=orange!20]
    
    \tikzstyle{phinew} = [circle, very thick, minimum size=1cm, inner sep=0.2mm, draw=bblue, fill=bblue!20]
    
    % Time-Evolution operator tensor cores
    \tikzstyle{U} = [rectangle, very thick, minimum size=0.8cm, inner sep=0.1cm,draw=green, fill=green!20]
    
    % Define arrow style
    \tikzset{arw/.style={-Stealth, thick}}

    \begin{tikzpicture}[>=latex,node distance=3cm,
        every edge quote/.style={font=\fontsize{7}{1}},
        ]
		\node (pphi1) [phi,font=\fontsize{8}{1}]{$\phi^{[1]}$};
        \node (pphi2) [phi, right=9mm of pphi1,font=\fontsize{8}{1}]{$\phi^{[2]}$};
        \node (pphidots) [right=6mm of pphi2, minimum size=8mm,font=\fontsize{18}{1},orange]{$\cdots$};
        \node (pphiN_1) [phi, right=6mm of pphidots,font=\fontsize{8}{1}]{$\phi^{[N\text{-}1]}$};
        \node (pphiN) [phi, right=9mm of pphiN_1,font=\fontsize{8}{1}]{$\phi^{[N]}$};        
        \node (ppx1) [below=2mm of pphi1, minimum size=0mm, inner sep=0mm]{};
        \node (ppx2) [below=2mm of pphi2, minimum size=0mm, inner sep=0mm]{};
        \node (ppxN_1) [below=2mm of pphiN_1, minimum size=0mm, inner sep=0mm]{};
        \node (ppxN) [below=2mm of pphiN, minimum size=0mm, inner sep=0mm]{};
        
        \node (u1) [U, below=2mm of ppx1, font=\fontsize{8}{1}]{$\mathcal{U}^{[1]}$};
        \node (u2) [U, below=2mm of ppx2, font=\fontsize{8}{1}]{$\mathcal{U}^{[2]}$};
        \node (udots) [right=6mm of u2, minimum size=8mm,font=\fontsize{18}{1},green]{$\cdots$};
        \node (uN_1) [U, below=2mm of ppxN_1, font=\fontsize{8}{1}]{$\mathcal{U}^{[N\text{-}1]}$};
        \node (uN) [U, below=2mm of ppxN, font=\fontsize{8}{1}]{$\mathcal{U}^{[N]}$};        
        \node (ux1) [below=3mm of u1, minimum size=0mm, inner sep=0mm]{};
        \node (ux2) [below=3mm of u2, minimum size=0mm, inner sep=0mm]{};
        \node (uxN_1) [below=3mm of uN_1, minimum size=0mm, inner sep=0mm]{};
        \node (uxN) [below=3mm of uN, minimum size=0mm, inner sep=0mm]{};
                
        \path[-]
            
%           \draw[arw,thick,draw=black] (eql0sgn.south) -| (eqlsgn.north);
           
%            (eql0sgn) edge[very thick, purple] (eqlsgn)
            
            (pphi1) edge[very thick, orange] node[above,black] {$_{\alpha_1}$} (pphi2)
            (pphi2) edge[very thick, orange] node[above,black] {$_{\alpha_2}$} (pphidots)
            (pphidots) edge[very thick, orange] node[above,black,xshift=-1mm] {$_{\alpha_{N\text{-}2}}$}  (pphiN_1)
            (pphiN_1) edge[very thick, orange] node[above,black] {$_{\alpha_{N\text{-}1}}$}  (pphiN)
            (pphi1) edge[very thick, orange] node[left,black,yshift=-1mm] {$_{x'_1}$} (ppx1)
            (pphi2) edge[very thick, orange] node[left,black,yshift=-1mm] {$_{x'_2}$} (ppx2)
            (pphiN_1) edge[very thick, orange] node[left,black,yshift=-1mm] {$_{x'_{N\text{-}1}}$} (ppxN_1)
            (pphiN) edge[very thick, orange] node[left,black,yshift=-1mm] {$_{x'_N}$} (ppxN)
            
%            (u1) edge[double,very thick, green] node[above,black] {$_{\beta_1}$}  node[below,black] {$_{\gamma_1}$} (u2)
            (u1) edge[very thick, green] node[above,black] {$_{\beta_1}$}  (u2)
%            (u2) edge[double,very thick, green] node[above,black] {$_{\beta_2}$}  node[below,black] {$_{\gamma_2}$} (udots)
            (u2) edge[very thick, green] node[above,black] {$_{\beta_2}$}  (udots)
%            (udots) edge[double,very thick, green] node[above,black,xshift=-1.2mm] {$_{\beta_{N\text{-}2}}$} node[below,black,xshift=-1.2mm] {$_{\gamma_{N\text{-}2}}$} (uN_1)
            (udots) edge[very thick, green] node[above,black,xshift=-1.2mm] {$_{\beta_{N\text{-}2}}$} (uN_1)
%            (uN_1) edge[double,very thick, green] node[above,black] {$_{\beta_{N\text{-}1}}$}  node[below,black] {$_{\gamma_{N\text{-}1}}$} (uN)
            (uN_1) edge[very thick, green] node[above,black] {$_{\beta_{N\text{-}1}}$} (uN)
            (u1) edge[very thick, green] (ppx1)
            (u1) edge[very thick, green] node[left,black,yshift=-0.5mm] {$_{x_1}$}(ux1)
            (u2) edge[very thick, green] (ppx2)
            (u2) edge[very thick, green] node[left,black,yshift=-0.5mm] {$_{x_2}$}(ux2)
            (uN_1) edge[very thick, green] (ppxN_1)
            (uN_1) edge[very thick, green] node[left,black,yshift=-0.5mm] {$_{x_{N\text{-}1}}$}(uxN_1)
            (uN) edge[very thick, green] (ppxN)
            (uN) edge[very thick, green] node[left,black,yshift=-0.5mm] {$_{x_N}$}(uxN)
            
            ;

    \end{tikzpicture}
\caption{Time evolution of a wavepacket in a tensor-network representation as given by Eq. (\ref{Eq:MPO_MPS}). The initial state is parameterized by entanglement variables $\bar{\alpha}$, while the time-evolution propagator is characterized by $\bar{\beta}$. }
\label{Fig:MPS_tEvo_ND}
\end{figure}

We begin by representing the initial wavepacket, $\chi_0$, as a matrix product state (MPS)~\cite{tensor_network}. (Although we use MPS states here, this is not a restriction and other graph based tensor-network schemes\cite{frag-TN-Anup} can also be used within our formalism.) In this representation, a wavepacket with $N$ nuclear degrees of freedom is expressed as a multi-configurational expansion~\cite{mctdh-meyer,MLMCTDHWang,Burghardt-MLMCTDH,Otto_POTFIT} that may be graphically represented as in the top row (orange) in Figure \ref{Fig:MPS_tEvo_ND}:
\begin{align}
\chi_0(\vb{\bar{x}}) = \sum_{\bar{\boldsymbol\alpha}}
\qty[\prod_{j=1}^{N} \tensor*{\phi}{^{[j]}_{}^{x_j}_{\alpha_{j-1},\alpha_j}}] = \sum_{\bar{\boldsymbol\alpha}} \tensor*{\phi}{^{\bar{x}}_{\bar{\alpha}}} ,
\label{Eq:MPS_WF}
\end{align}
where, the dependence on the discretized version of the continuous position basis, $\vb{\bar{x}} \equiv (x_1, x_2, \ldots, x_N)$, is encoded in tensor elements parameterized by $x_j$ in the superscript of, $\tensor*{\phi}{^{[j]}_{}^{x_j}_{\alpha_{j-1},\alpha_j}}$. These tensors $\phi^{[j]}$, %which are second or third order tensors, 
can thus be interpreted as reduced-dimensional functions associated with each coordinate $x_j$. The superscripts in square brackets $[\cdot]$ label the tensor cores and the terms $\tensor*{\phi}{^{\bar{x}}_{\bar{\alpha}}}$ refer to the associated product states. The summation indices $\bar{\boldsymbol\alpha} \equiv (\alpha_0, \alpha_1, \ldots, \alpha_N)$  are commonly referred to as the \emph{bond dimensions} (or equivalently, entanglement dimensions or Schmidt ranks).
%, and these determine the number of retained singular values at each bipartition of the system in the Schmidt decomposition~\cite{Nicole-TN}.

\begin{comment}
\begin{figure}[tbp]
    \subfigure[]{\input{Figures/Figures_MPS_prop}
        \label{Fig:MPS_tEvo_trotter-a}}
    \subfigure[]{\input{Figures/Figure-TN-first-trotter}\label{Fig:MPS_tEvo_trotter-b}}
   \caption{Figure \ref{Fig:MPS_tEvo_trotter-a} illustrates the tensor-network representation of the time evolution of a wavepacket. Figure \ref{Fig:MPS_tEvo_trotter-b} shows the corresponding evolution in its Trotterized formulation. In this representation, the initial state is parameterized by entanglement variables $\alpha$, while the Trotterized propagator is characterized by entanglement variables $\beta$. Together, they define the composite index $\mu = \alpha \times \beta$ that labels the evolved state.}
    \label{Fig:MPS_tEvo_trotter}
\end{figure}
\end{comment}

The time-evolution operator $\hat{U}$ is expressed in the matrix product operator (MPO) form~\cite{MPO,MPOR} as
\begin{align}
\mel{\bar{\vb{x}}'}{\hat{U}}{\bar{\vb{x}}} 
&= \sum_{\bar{\boldsymbol\beta}}^{\bar{\boldsymbol\eta}} %U_{\bar{\beta}}^{\bar{x},\bar{x}'}
%^{N_{\bar{\boldsymbol\beta}}} 
\left[ \prod_{j=1}^{N} 
\tensor*{\mathcal{U}}{^{[j]}_{}^{x_j x'_j}_{\beta_{j-1},}_{\beta_j}} 
\right] = \sum_{\bar{\boldsymbol\beta}}^{\bar{\boldsymbol\eta}} U_{\bar{\beta}}^{\bar{x},\bar{x}'},
\label{Eq:propagator-TN}
\end{align}
where the indices $\bar{\boldsymbol\beta} \equiv (\beta_0, \beta_1, \ldots, \beta_N)$, bounded by $\bar{\boldsymbol\eta} \equiv (\eta_0 = 1, \eta_1, \ldots, \eta_{N-1}, \eta_N = 1)$, encode the entanglement structure of the operator across different degrees of freedom. Each tensor $\mathcal{U}^{[j]}$ represents the reduced-dimensional effective propagator that evolves the corresponding lower-dimensional function $\phi^{[j]}$ in \cref{Eq:MPS_WF} and $U_{\bar{\beta}}^{\bar{x},\bar{x}'}$ represent the respective product state operators. These operators propagate the MPS state as per
%Acting on an initial state $\ket{\chi_0}$, written in matrix product state (MPS) form (\cref{Eq:MPS_WF}), the time-evolved wavefunction becomes
\begin{align}
\chi_{t}(\bar{\vb{x}}) 
&= \sum_{\bar{\boldsymbol\alpha},\bar{\boldsymbol\beta}} 
\left[
\prod_{j=1}^{N} 
\int dx'_j \;
\tensor*{\mathcal{U}}{^{[j]}_{}^{x_j x'_j}_{\beta_{j-1},}_{\beta_j}}
\;\tensor*{\phi}{^{[j]}_{}^{x'_j}_{\alpha_{j-1},\alpha_j}}
\right] \nonumber \\ &= \sum_{\bar{\boldsymbol\alpha},\bar{\boldsymbol\beta}} \int d\bar{x}' U_{\bar{\beta}}^{\bar{x},\bar{x}'} \tensor*{\phi}{^{\bar{x}'}_{\bar{\alpha}}}. 
\label{Eq:MPO_MPS}
\end{align}
This expression corresponds to a contraction of the MPO with the MPS, resulting in a new state whose bond structure is governed jointly by $\bar{\boldsymbol\alpha}$ and $\bar{\boldsymbol\beta}$. These equations are illustrated in Figure \ref{Fig:MPS_tEvo_ND}. 

\subsection{Block-diagonal structure of the tensor-network propagator in an {\em elevated} representation}
The propagator in Eq. (\ref{Eq:propagator-TN}) can be recast as a block-diagonal matrix acting on an enlarged Hilbert space. 
%,
%\begin{align}
%\hat{U} = \bigoplus_{\boldsymbol{\bar{\beta}}} U_{\boldsymbol{\bar{\beta}}},
%\end{align}
%where each block acts on a product state labeled by the entanglement indices and given by
%\begin{align}
%\hat{U} = %\bigoplus_{\bar{\beta}}
%\sum_{\bar{\beta}}
%\left[ \mathcal{U}^{[1]}_{1,\beta_1}
%\;\mathcal{U}^{[2]}_{\beta_1,\beta_2}
%\cdots
%\;\mathcal{U}^{[N]}_{\beta_{N-1},1} \right].
%\end{align}
\begin{figure}[tbp]
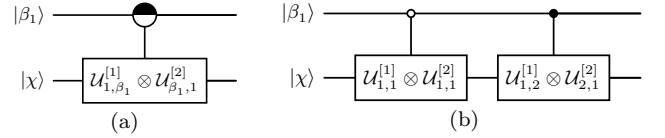

\centering
\subfigure[]{%
    \resizebox{0.38\linewidth}{!}{\input{Figures/dim2_beta2}}}
\hfill
\subfigure[]{%
   \resizebox{0.58\linewidth}{!}{\input{Figures/Figure_dim2_beta2_expanded}}}
\caption{(a) Uniformly controlled gate-like structure with single control qubit encoding the bond-dimension index $\beta_1$. (b) An illustration of Figure (a) with maximum bond dimension $\eta_1=2$ (i.e., upper bound to the quantity $\beta_1$). Also see Eq. (\ref{Block-2D-1Bond}). Critical to note that the individual blocks $\mathcal{U}_{1,1}^{[1]} \otimes \;\mathcal{U}_{1,1}^{[2]}$ and $\mathcal{U}_{1,2}^{[1]} \otimes \;\mathcal{U}_{1,2}^{[2]}$ are direct product operators that act on product states of dimensions ``$[1]$'' and ``$[2]$''.}
\label{fig:2D_2beta_ucg}
\end{figure}
%\begin{figure*}[tbp]
\begin{figure}[tbp]
    \centering
    \resizebox{\columnwidth}{!}{%
        \input{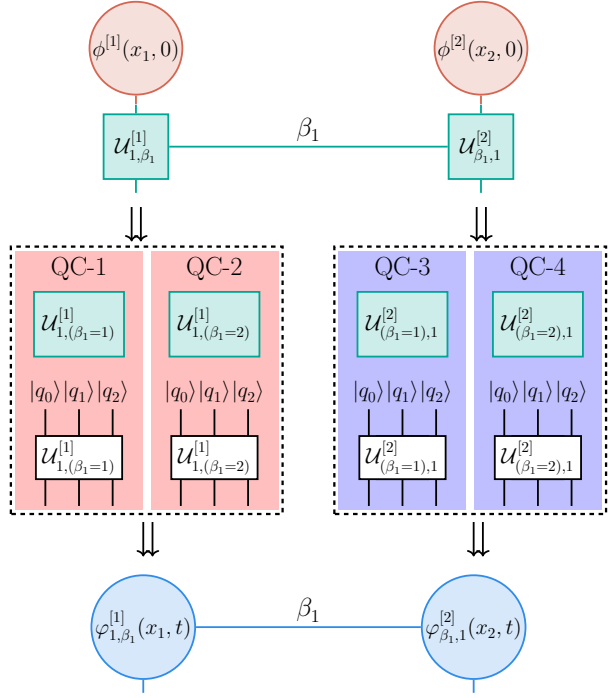}
    }
    \caption{The circuit in \cref{fig:2D_2beta_ucg} can be distributed, in principle, across four quantum computers, where each constituent sub-circuit can be executed simultaneously. In this paper, this is all done sequentially on one quantum computer using three qubits. The aggregation (the double vertical arrow at the bottom of the figure) is also done classically here through measurement, and the final state inherits the entanglement from the unitary above. As described in Section \ref{Logham}, such entanglement arises from the potential energy surface.}
    \label{Fig:MPS_tEvo_trotter-e}
\end{figure}
%\end{figure*}
We illustrate this idea here for the case $N = 2$ and $\beta_1= 2$, where the time-evolution operator,
\begin{align}
%U_{1} \oplus U_{2} \nonumber\\ 
\mathcal{U}^{[1]}_{1,1}\,
\;\mathcal{U}^{[2]}_{1,1} + %\oplus 
\mathcal{U}^{[1]}_{1,2}\,
\;\mathcal{U}^{[2]}_{2,1}
\end{align}
may be rewritten in an enlarged Hilbert space as, 
\begin{align}
\hat{U} =
\begin{bmatrix}
\mathcal{U}_{1,1}^{[1]} \otimes\;\mathcal{U}_{1,1}^{[2]} & 0 \\
0 & \mathcal{U}_{1,2}^{[1]} \otimes\;\mathcal{U}_{2,1}^{[2]}
\end{bmatrix}
\label{Block-2D-1Bond}
\end{align}
Here, %and it is clear from here as to how 
the size of the resultant vector space has been increase by a factor corresponding to the extent of entanglement, defined here by the upper bound to the quantity $\beta_1$, say $\eta_1$. That is, the size of the unitary in Eq. (\ref{Block-2D-1Bond}) is now $D^2*\eta_1$ for $D$ basis discretizations per dimension. 
Thus, when Eq. (\ref{Block-2D-1Bond}) acts on a product initial state $\left[ \tensor*{\phi}{^{[1]}_{}^{x_1}} \tensor*{\phi}{^{[2]}_{}^{x_2}} \right]$, we obtain 
\begin{align}
\begin{bmatrix}
\tensor*{\phi}{^{[1]}_{}^{x_1}_{\beta_{1}=1}} \tensor*{\phi}{^{[2]}_{}^{x_2}_{\beta_{1}=1}} \\[1.75ex]
\tensor*{\phi}{^{[1]}_{}^{x_1}_{\beta_{1}=2}} \tensor*{\phi}{^{[2]}_{}^{x_2}_{\beta_{1}=2}}
\end{bmatrix} &=
\begin{bmatrix}
\mathcal{U}_{1,1}^{[1]} \otimes\;\mathcal{U}_{1,1}^{[2]} & 0 \\[1.75ex]
0 & \mathcal{U}_{1,2}^{[1]} \otimes\;\mathcal{U}_{2,1}^{[2]}
\end{bmatrix}
\begin{bmatrix}
\tensor*{\phi}{^{[1]}_{}^{x_1}} \tensor*{\phi}{^{[2]}_{}^{x_2}} \\[1.75ex]
\tensor*{\phi}{^{[1]}_{}^{x_1}} \tensor*{\phi}{^{[2]}_{}^{x_2}}
\end{bmatrix} 
\nonumber \\ &=
\begin{bmatrix}
\mathcal{U}_{1,1}^{[1]} \tensor*{\phi}{^{[1]}_{}^{x_1}} \otimes \mathcal{U}_{1,1}^{[2]} \tensor*{\phi}{^{[2]}_{}^{x_2}}\\[1.75ex]
\mathcal{U}_{1,1}^{[2]} \tensor*{\phi}{^{[1]}_{}^{x_1}} \otimes \mathcal{U}_{1,2}^{[1]}  \tensor*{\phi}{^{[2]}_{}^{x_2}}
\end{bmatrix} 
\label{Block-2D-1Bond-prop}
\end{align}
%Consequently, the associated maximum entanglement entropy supported by this MPS structure is $\ln \eta_1$. %See \cref{fig:2D_2beta_ucg}. (More precisely, the entanglement entropy of the bipartite system  is given by the von Neumann entropy of the reduced density matrix, equivalently the Shannon entropy of the squared Schmidt coefficients\cite{tensor_network}, and is bounded by $\ln \eta_1$.)
where the elevated vector space is clearly marked based on the additional index $\beta_1$. 
%now enlarged by an amount corresponding to the extent of entanglement.

%The associated bipartite entanglement entropy, ``S", is given by the von Neumann entropy of the reduced density matrix, equivalently the Shannon entropy of the squared Schmidt coefficients \cite{tensor_network}, and is bounded by $S \le \ln \eta_1$.

%But, the gain here is that the resultant operations can be treated as a parallel stream of operations acting on product states as can be seen from the tensor product notation used for each diagonal block. This aspect is also clear from  Figure \ref{Fig:MPS_tEvo_trotter-e}. Additionally, this structure can also be implemented using a uniformly controlled gate with a single set of control qubits, as outlined in \cref{fig:2D_2beta_ucg}, but this will require multiple control gates. The real advantage is that multiple hardware systems can be simultaneous used for parallel streams depicted in Figure \ref{Fig:MPS_tEvo_trotter-e} and hence this method is really appropriate for a hybrid quantum-HPC platform. (General versions of the algorithm are presented below.)

The block-diagonal operator in Eq.(\ref{Block-2D-1Bond}) can be implemented directly as a quantum circuit using  uniformly controlled gates (UCG)\cite{bergholm2005uniformcontrolgate} with a single set of control qubits encoding the bond index $(\beta_1)$, as shown in \cref{fig:2D_2beta_ucg}. However, an important feature of Eq.(\ref{Block-2D-1Bond}) is that each diagonal block is itself a tensor product of lower-rank unitary operators. See Eq. (\ref{Block-2D-1Bond-prop}). %Consequently, the action of each block decomposes into independent operations acting on product states, rather than requiring the simulation of a single monolithic operator.
This tensor-product structure exposes an additional level of parallelism beyond that provided by the UCG representation. Specifically, the constituent operators within each diagonal block can be executed as independent computational streams, as illustrated in \cref{Fig:MPS_tEvo_trotter-e}. While the UCG formulation provides a compact circuit-level representation, it still requires the implementation of the corresponding controlled operations. In contrast, the block-diagonal decomposition naturally lends itself to a distributed execution environment where the constituent operators within each diagonal block may be evolved independently and in a completely asynchronous manner %simultaneously 
on separate hardware resources. Bipartite systems with greater entanglement are discussed in Appendix \ref{Bipartite-beta2}.

The principal advantage of this formulation is therefore its compatibility with large-scale parallel execution. Multiple quantum processors, or quantum processors coupled to classical high-performance computing (HPC) resources, can be employed %concurrently 
to evaluate the independent streams associated with the different tensor-product sectors. This makes the approach particularly well suited for hybrid quantum--HPC architectures. %(Generalizations of this construction are presented below.)

\begin{figure}[tbp]
\centering
    \resizebox{0.45\textwidth}{!}{\input{Figures/general_ucg_matrix}}
\caption{Block diagonal form of the overall unitary in arbitrary dimensions. The dimensionality of the problem is expanded by a factor of $\left[ \prod \bar{\eta} \right]$ (see Section \ref{scaling}) which results in a block diagonal representation with each block acting on a product state. As noted earlier, the individual blocks, $U_{\bar{\beta}}^{\bar{x},\bar{x}'} \equiv \left[ \prod_{j=1}^{N} 
\tensor*{\mathcal{U}}{^{[j]}_{}^{x_j x'_j}_{\beta_{j-1},}_{\beta_j}} 
\right]$ are product operators where each operator acts on lower dimensional function as noted in Eqs (\ref{Eq:MPO_MPS-D1}) and (\ref{Eq:MPO_MPS}).}
\label{fig:general_ucg_matrix}
\end{figure}
\begin{figure*}[tbp]
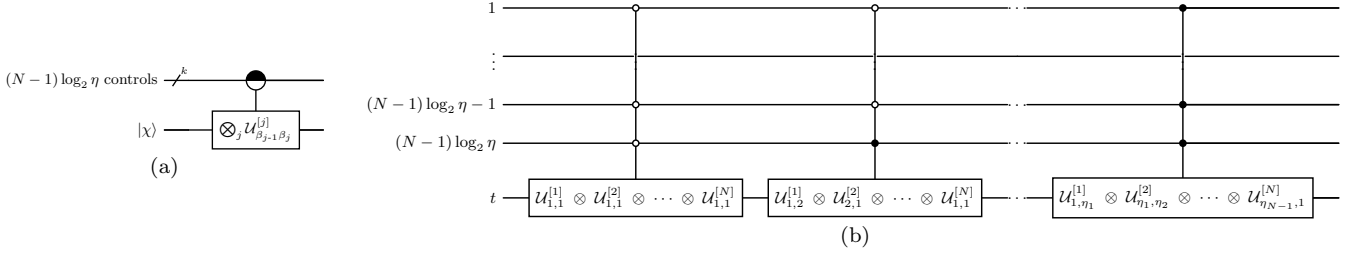

\centering
\subfigure[]{%
    \resizebox{0.25\textwidth}{!}{\input{Figures/general_circuit1}}}
\hfill
\subfigure[]{%
    \resizebox{0.73 \textwidth}{!}{\input{Figures/general_ucg_circuit}}}
\caption{(a) Uniformly controlled gate (UCG) representation of the block-diagonal operator shown in \cref{fig:general_ucg_matrix}.}
\label{fig:general_ucg_circuit}
\end{figure*}
\begin{comment}
\begin{figure*}[tbp]
    \centering
    \resizebox{0.90\textwidth}{!}{%
        \input{Figures/Figure-TN-ND-2-2SV}
    }
    %\input{Figures/Figure-TN-2SV-U2}
    \caption{The circuit in \cref{fig:general_ucg_circuit} can, in principle, be distributed across several quantum computers, and this figure shows the full-scale parallel generalization of our algorithm. The ion-traps are indexed using the dimension (first index), and the two entanglement indices following the dimension.}
    \label{Fig:MPS_tEvo_trotter-g}
\end{figure*}
\end{comment}
\begin{figure*}[tbp]
    \centering
    \resizebox{0.90\textwidth}{!}{%
        \input{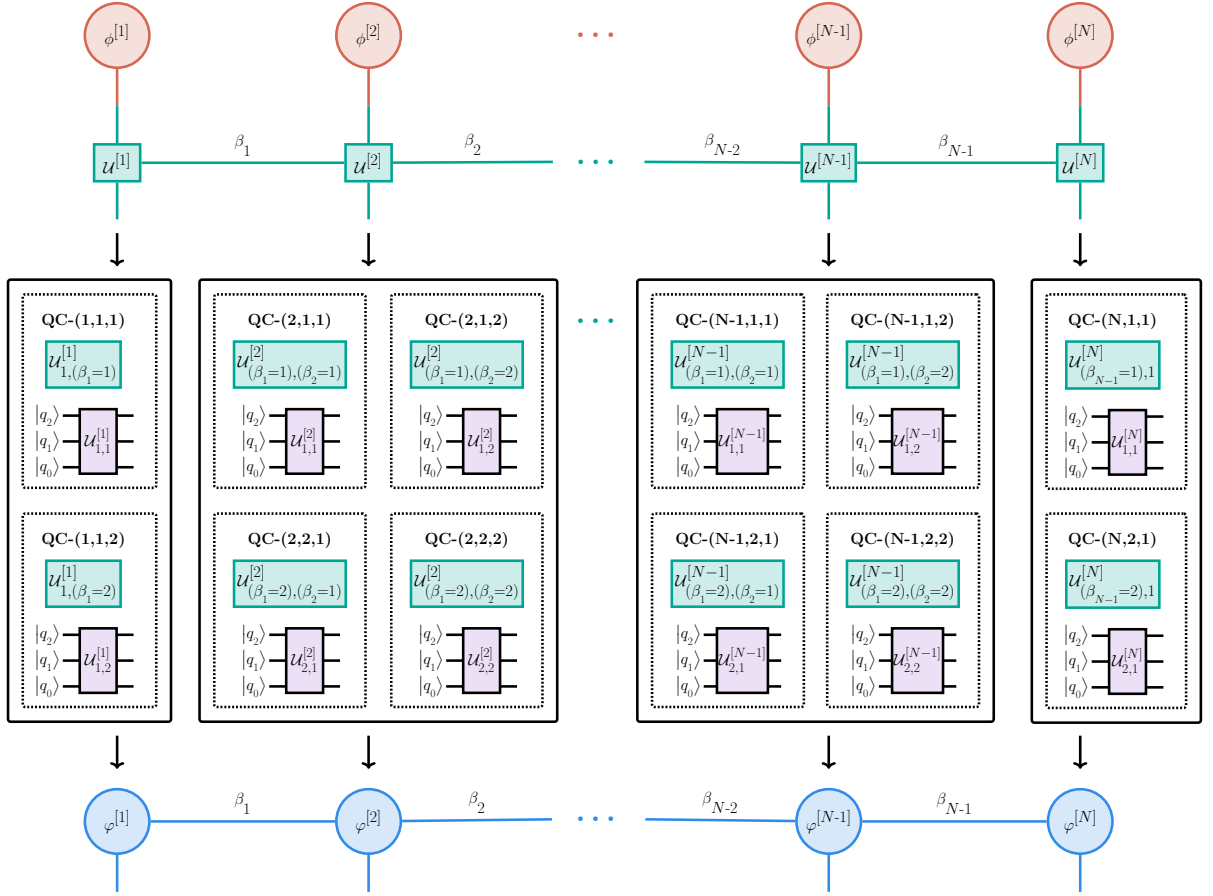}
    }
    \caption{The circuit in \cref{fig:general_ucg_circuit} can, in principle, be distributed across several quantum computers, and this figure shows the full-scale parallel generalization of our algorithm. The ion-traps are indexed here using the dimension (first index), and the two entanglement indices following the dimension. So for example, ``\textbf{QC-(2,1,1)}'' implies the specific quantum hardware system (trapped ions in our case) that computes the second dimension for entanglement variables $\beta_1=\beta_2=1$. As is clear, the number of parallel streams (dotted boxes) %is approximately equal to four times the number of dimensions, which 
    is equal to $\sum_i \eta_i^2$. By contrast the number of blocks in \cref{fig:general_ucg_matrix} is $\left[ \prod \bar{\eta} \right]$ as is the number of constraint operations in Figure \ref{fig:general_ucg_circuit}.} 
    %The generic ``U'' inside each ciruit box refers to the unitary mentioed right above in green. }
    \label{Fig:MPS_tEvo_trotter-g}
\end{figure*}

%\subsection{Distributed and asynchronous quantum execution from tensor-network decomposition: Formal Scaling}
\subsection{Entanglement Entropy as a computational resource: Formal Scaling of Distributed Quantum Execution from Tensor-Network Decompositions}
\label{scaling}
\begin{comment}
Within this representation, the full operator can be expressed as
\begin{align}
\hat{U} =
\sum_{\boldsymbol{\beta}}
\ket{\boldsymbol{\beta}}\bra{\boldsymbol{\beta}}
\otimes U_{\boldsymbol{\beta}},
\end{align}
which corresponds precisely to the structure of a uniformly controlled unitary acting on quantum hardware, as depicted in \cref{fig:general_ucg_circuit}. 
\end{comment}
%As noted, 
The main features of the algorithm presented above are depicted in Figures \ref{fig:general_ucg_matrix} \ref{fig:general_ucg_circuit} and \ref{Fig:MPS_tEvo_trotter-g} for arbitrary dimensions. As before, these figures are obtained from the block-diagonal nature of the propagator in an elevated Hilbert space that contains both the system degrees of freedom and the entanglement degrees of freedom. In this section, we present a formal analysis of the reduction in complexity and the extent of parallelism afforded by this algorithm. In general, for a system with $N$ dimensions and $D$ basis discretizations per dimension, the formal computational scaling of quantum propagation involves matrix-vector operations that are potentially ${\cal O}(D^N)$. The algorithm presented here provides an efficient quantum computing strategy for this problem that automatically reduces to an asynchronous and distributed set of quantum processes that can be executed on a hybrid set of quantum and classical hardware platforms. 

For an $N$-dimensional system, the block-diagonal structure for the evolution operator is illustrated in \cref{fig:general_ucg_matrix}. As before, this structure maps onto a uniformly controlled gate (UCG) configuration where the entanglement indices $\bar{\beta}$ are encoded in control registers. The associated quantum circuit is provided in Figure \ref{fig:general_ucg_circuit}. The number of required control qubits may scale as $\left\{ \ln \left[ \prod \bar{\eta} \right] \right\}$, where $\left[ \prod \bar{\eta} \equiv \eta_1 \eta_2 \cdots \right]$, that is product of the individual maximum bond dimensions, or the maximumm entanglement volume within this representation.% for
%or the hypervolume (and hence total entanglement) 
% the bond dimensions $\bar{\beta}$. 
The quantity $\bar{\eta}$ is the upper bound in Eq. (\ref{Eq:propagator-TN}). For the case where all bonds are cut, the quantity $\left\{ \ln \left[ \prod \bar{\eta} \right] \right\}$ represents the maximum possible entanglement entropy of the associated tensor network state. 
Thus, the number of control qubits may scale as the maximum possible entanglement entropy of the system. 
The individual operations in each block may be viewed as a collection of $\left[ \prod_i \eta_i \right]$ independent unitary blocks, where each block corresponds to a single operation ($U_{\bar{\beta}}^{\bar{x},\bar{x}'}\tensor*{\phi}{^{\bar{x}'}_{\bar{\alpha}}}$) acting on a product-state sector of dimension $D^N$. The key observation, however, is that each block is itself composed of a direct product of $N$ lower-rank unitary operators, one associated with each physical dimension. As shown in \cref{Eq:MPO_MPS,Block-2D-1Bond-prop} (see also \cref{Eq:MPO_MPS-product-chi0} below), these operators act independently on product states and therefore admit a highly decomposed computational structure.

However, when the individual one-dimensional streams are propagated in parallel, one can parse these our separately and the number of such individual streams is bounded by 
\begin{align}
    2\eta_1+2\eta_N + \sum_{i=2}^{N-1} \eta_i \eta_{i+1} \approx \sum_i \eta_i^2
\end{align}
for large $N$, 
as shown in \cref{Fig:MPS_tEvo_trotter-g}. 
%{\em Thus,  the number of parallel streams available for computation scale as $\sum_i \eta_i^{2}$.}
%\begin{align}
%m = (N-1)\log_2(\eta_{\max}),
%\end{align}
%with $\eta_{\max} = \max_j \eta_j$ denoting the maximum bond dimension across all sites as specified in Eq. (\ref{Eq:MPS_WF}).
%It must be emphasized this is a completely rigorous step with no approximations except for numerical truncations as needed that curtail the number of critical values in the set $\left\{ {\beta}_1 {\alpha}_1, {\beta}_12 {\alpha}_2, \cdots \right\}$. The advantage here is {\em extreme parallelism} where each subtask being reduced dimensional and hence controllable circuit depth. This aspect is seen from Figures \ref{fig:2D_2beta_ucg}, \ref{fig:2D_4beta_ucg}. \ref{fig:general_ucg_matrix} and \ref{fig:general_ucg_circuit}. 
%The matrix in \cref{fig:general_ucg_matrix} has dimension $\left[\prod_i \eta_i\right]D^N$. Equivalently, it 
This is because while the block-diagonal form of \cref{fig:general_ucg_matrix} appears to contain $N\prod_i \eta_i$ independent unitary operations that could be executed in parallel, many of these operations are repeated across different blocks and therefore need not be recomputed when treated in an asynchronous manner. For example, the operator $\mathcal{U}^{[1]}_{1,1}$ appears in every block for which the remaining bond indices vary while $\beta_1=1$. Consequently, this operator need only be evaluated once and reused wherever it appears. Accounting for such redundancies, the number of distinct computational tasks is reduced to a maximum of $\sum_i \eta_i^2$ operations,
each corresponding to an ${\cal O}(D^1)$ operation of the form
\begin{align}
\int dx'_j \;
\tensor*{\mathcal{U}}{^{[j]}_{}^{x_j x'_j}_{\beta_{j-1},}_{\beta_j}}
\;\tensor*{\phi}{^{[j]}_{}^{x'_j}_{\alpha_{j-1},\alpha_j}}
\label{Eq:MPO_MPS-D1}
\end{align} 
which also follows directly from Eq. (\ref{Eq:MPO_MPS}). A similar reduction may also be possible in the UCG implementation, but will require additional control gates which in current implementations may be impractical. 

%This decomposition highlights the intrinsically parallel nature of the tensor-network formulation and motivates the hybrid execution strategy illustrated in Fig.\ref{Fig:MPS_tEvo_trotter-g}. Rather than treating the propagator as a single large-scale operation, the formalism generates a family of $N\sum_i \eta_i^2$ independent (${\cal O}(D)$) tasks associated with the local operators
\begin{comment}
$\left\{
\tensor*{\mathcal{U}}{^{[j]}_{}^{x_j x'_j}_{\beta_{j-1},}_{\beta_j}}
\right\}.$
\end{comment}
%These tasks can, in principle, be executed asynchronously on separate quantum processors and subsequently combined to reconstruct the full evolution. In the present implementation all such operations are performed sequentially on a single ion-trap quantum computer; however, the formalism itself places no such restriction. The individual tasks may be distributed across multiple quantum processors, potentially employing different hardware architectures. This flexibility is reflected in the labeling scheme used in Fig.~\ref{Fig:MPS_tEvo_trotter-g}. For example, the label \textbf{QC-(2,2,1)} denotes the quantum processor responsible for evaluating the circuit associated with the second physical dimension for the bond-index configuration ($\beta_1=2$) and ($\beta_2=1$). The resulting framework therefore provides a natural pathway toward large-scale hybrid quantum--HPC implementations of tensor-network-based quantum dynamics.

This parallel and hybrid aspect is emphasized in Figure \ref{Fig:MPS_tEvo_trotter-g}. 
%As can be seen, our tensor network formalism provides a family of $\left\{ \sum_i \eta_i^2 \right\}$ ${\cal O}(D^1)$ operations, where the ${\cal O}(D^1)$ operators are $\left\{ \tensor*{\mathcal{U}}{^{[j]}_{}^{x_j x'_j}_{\beta_{j-1},}_{\beta_j}} \right\}$. This 
The family of operators may be used to spawn a set of {\em asynchronous} tasks that can be performed on separate ion-trap quantum computers. In the current implementation these ${\cal O}(D^1)$ operations are performed on a single ion-trap, sequentially, but this is not the only possible implementation. In principle, these separate tasks can be performed, as noted, in a completely asynchronous way, on multiple quantum computers, some of which may involve ion-traps and some may involve other quantum architectures. The labeling scheme of quantum hardware in Figure \ref{Fig:MPS_tEvo_trotter-g} suggests this generality. For example, the label ``\textbf{QC-(2,2,1)}'' implies the specific quantum hardware system that computes the action of the circuit corresponding to the second dimension for entanglement variables $\beta_1=2$ and $\beta_2=1$. %The operator used in each circuit is represented within the green box immediately above each circuit representation in  Figure \ref{Fig:MPS_tEvo_trotter-g}.

In summary, a uniformly controlled gate implementation of Figure \ref{fig:general_ucg_circuit} may require the number of ancilla to be equal to up to a maximum of $\ln \left[ \prod \bar{\eta} \right]$, which is the maximum possible entanglement entropy that can be supported by the tensor network. By contrast, an asynchronous and distributed implementation needs a maximum of $\sum_i \eta_i^2$ number of parallel streams, which may in general scale as the area of the entanglement hyperspace defined by edges $\left\{ \eta_i \right\}$ to capture the quantum dynamics, as can be seen in Figure \ref{Fig:MPS_tEvo_trotter-g}. As hardware systems evolve, future versions of this algorithm can include both aspects by constructing parallel tasks that may contain a small set of uniformly controlled gate implementations.

We wish to add one caveat to this argument. Although the general trend is clear, one may ask how the method scales for systems that truly contain exponential entanglement. In such cases $\left\{ \eta_i \right\}$ may grow exponentially with system size, and while the current approach still partitions the full problem into manageable chunks, the number of such independent processes could grow exponentially. For such problems, a hybrid approach, such as the one mentioned at the end of the previous paragraph, could be useful to consider. However, for most practical problems\cite{Hastings_2007-arealaw} we expect the proposed methodology to produce efficient algorithms. 

Thus, the decomposition introduced here provides a pathway for mitigating two of the primary limitations of near-term quantum devices: (a) circuit depth and (b) limited qubit connectivity. By transforming a global, potentially highly entangled unitary into a family of %conditionally applied 
product-state operations ($U_{\bar{\beta}}^{\bar{x},\bar{x}'}\tensor*{\phi}{^{\bar{x}'}_{\bar{\alpha}}}$) that may be thought to be conditioned by the entanglement index $\bar{\beta}$, the approach reduces both the entangling gate count and the effective circuit depth of each sub-task.
The resulting structure is inherently parallel, making it well-suited for emerging distributed quantum computing architectures and hybrid Quantum-HPC in which multiple quantum (and classical) processors are coordinated to perform large computations. Figure \ref{fig:general_ucg_circuit} shows the extent parallelism possible within this formalism. % on the higher (real-space times entanglement) dimensional space. 

%This formulation establishes a direct correspondence between tensor-network contractions, the associated block-diagonal operator structure, and their realization as quantum circuits. In particular, the entanglement indices of the tensor network are {\em elevated} to control qubits, while each contraction path through the network defines a conditionally applied unitary evolution on the target system. 

\subsection{Obtaining $\left\{ U_{\bar{\beta}} \right\}$ from electronic structure}
\label{Logham}
As specified, the formalism here pertains to chemical dynamics, where the overall Hamiltonian may be written as a sum of the nuclear kinetic energy and potential energy operators represented here on a grid representation,

\begin{align}
\hat{H}(\vb{\bar{x}},\vb{\bar{x}}^\prime)
= \sum_i K(x_i,x_i^\prime) 
+ \delta(\vb{\bar{x}}-\vb{\bar{x}}^\prime)\,\hat{V}(\vb{\bar{x}}),
\label{Eq:Hamiltonian}
\end{align}
where $\hat{V}(\vb{\bar{x}})$ is the (local) potential energy operator, diagonal in the coordinate representation and obtained from electronic structure calculations. The varianble $\bar{x}$ now represents a nuclear geometry where this electronic structure is computed. In contrast to the kinetic term, the potential energy is naturally not separable across dimensions, which introduces additional complexity.

%A further challenge in quantum dynamics lies in approximating the action of the time-evolution operator on a state $\ket{\Psi}$, namely $e^{-i\hat{H}\Delta t/\hbar}\ket{\Psi}$. In this work, we adopt the Trotter--Suzuki decomposition~\cite{Campbell1897-tj,Baker1901-cj,Trotter,Suzuki1976-us,Nelson-Trotter,qwaimd}, which provides a systematic approximation whose accuracy is controlled by the time step $\Delta t$.

The time evolution operator $\hat{U} = e^{-i\hat{H}\Delta t/\hbar}$ is approximated using a first-order Trotter decomposition~\cite{Trotter,Nelson-Trotter}: 
\begin{align}
e^{-i\hat{H}\Delta t/\hbar}
&= e^{-i\hat{V}(\vb{\bar{x}})\Delta t/\hbar}
\left\{ \prod_i e^{-i\hat{K}(x_i,x_i^\prime)\Delta t/\hbar} \right\}
+ \mathcal{O}(\Delta t^2) %\nonumber \\
%&= e^{-i\hat{V}(\vb{\bar{x}})\Delta t/\hbar}
%\left\{ \prod_i \tensor*{\mathcal{K}}{^{[i]}}(x_i,x_i^\prime) \right\}
%+ \mathcal{O}(\Delta t^2),
\label{Eq:first-order-Trotter}
\end{align}
%where $\tensor*{\mathcal{K}}{^{[i]}}(x_i,x_i^\prime)$ denotes the kinetic propagator associated with the $i$th coordinate.  
%constructed using the DAF representation introduced in \cref{DAFderivative}.
and a
key complication arises from the non-separable nature of the potential energy operator in the coordinate representation. To address this, we express the potential propagator also as an MPS, 
%a tensor network. In two dimensions, this takes the form
\begin{align}
e^{-i\hat{V}(\vb{\bar{x}})\Delta t/\hbar}
= \sum_{\bar{\boldsymbol\beta}}^{\bar{\boldsymbol\eta}} 
\qty[\prod_{j=1}^{N} \tensor*{\mathcal{V}}{^{[j]}_{}^{x_j}_{\beta_{j-1},\beta_j}}],
\label{Eq:UV-TN-1}
\end{align}
and using this expression, Eq. (\ref{Eq:propagator-TN}) may now be rewritten for a product state initial wavepacket as
\begin{align}
\chi_{t}(\bar{\vb{x}}) 
&= \sum_{\bar{\boldsymbol\beta}} 
\left[
\prod_{j=1}^{N} 
\int dx'_j \;
e^{-i\hat{K}(x_i,x_i^\prime)\Delta t/\hbar} 
\tensor*{\mathcal{V}}{^{[j]}_{}^{x'_j}_{\beta_{j-1},\beta_j}}
\tensor*{\phi}{^{[j]}_{}^{x'_j}}
\right] \nonumber \\ &= \sum_{\bar{\boldsymbol\beta}} \int d\bar{x}' U_{\bar{\beta}}^{\bar{x},\bar{x}'} \tensor*{\phi}{^{\bar{x}'}}. 
\label{Eq:MPO_MPS-product-chi0}
\end{align}
It is thus clear that all the entanglement in the propagated state arises from the potential. Again, as noted above, this paper deals with an MPS representation of wavepackets and potential propagators, but also allows more general graph-based tensor-network treatments as discussed in Ref. \onlinecite{frag-TN-Anup}. (Ref. \onlinecite{frag-TN-Anup} provides general expressions for tensor network treatment of potential propagators when graph-based molecular fragmentation techniques are used to obtain the potential energy surface in high dimensions\cite{Xiao-LLM} with post-Hartree-Fock accuracy.)

However, the factorization of the potential propagator above brings about an extremely subtle aspect that needs to be further discussed. The individual $\left\{ \tensor*{\mathcal{V}}{^{[j]}_{}^{x'_j}_{\beta_{j-1},\beta_j}} \right\}$ propagators may be rewritten as
\begin{align}
\nonumber
\tensor*{\mathcal{V}}{^{[j]}_{}^{x'_j}_{\beta_{j-1},\beta_j}}
&=
\exp{-\imath \tensor*{{V}}{^{[j]}_{}^{x'_j}_{\beta_{j-1},\beta_j}}  t/\hbar},
\end{align}
where the effective reduced dimensional entangled potentials $\left\{ \tensor*{{V}}{^{[j]}_{}^{x'_j}_{\beta_{j-1},\beta_j}} \right\}$ may now be complex thus making the individual operators, $U_{\bar{\beta}}^{\bar{x},\bar{x}'}$ potentially non-unitary. However, this non-unitarity is critical in that it only arises due to the entanglement enforced by the potential. For example, if the $\left\{ U_{\bar{\beta}}^{\bar{x},\bar{x}'} \right\}$ turn out to be completely unitary, there is no effective population redistribution between the various modes after each propagation step, thus rendering all dimensions to be orthogonal. Thus non-unitarity of the individual propagated branches of the tensor network is a critical feature in this algorithm. 

\subsection{Key experimental advances that make the action of $\left\{ U_{\bar{\beta}} \right\}$ efficient on ion-traps}
In addition to the theoretical aspects presented here, it is critical to mention an experimental advance that makes the associated calculations possible. Experimental details are presented in Section \ref{Sec:Experimental}. As might be clear, now the key elements of this algorithm reduces to a family of parallel operations given by $\left\{ \tensor*{\mathcal{U}}{^{[j]}_{}^{x_j x'_j}_{\beta_{j-1},}_{\beta_j}} \tensor*{\phi}{^{[j]}_{}^{x_j}_{\alpha_{j-1},\alpha_j}} \right\}$. In Ref. \cite{saha2025qsd} we present a general Quantum Shannon Decomposition (QSD) algorithm to perform these individual  operations. In this QSD circuit though, arbitrary decompositions of three-qubit unitary propagators obtained during the implementaion of operations $\tensor*{\mathcal{U}}{^{[j]}_{}^{x_j x'_j}_{\beta_{j-1},}_{\beta_j}} \tensor*{\phi}{^{[j]}_{}^{x_j}_{\alpha_{j-1},\alpha_j}}$ in this paper, require 24 fully-entangling CNOT gates or $XX(\pi/2)$ gates. In Section \ref{Sec:Experimental} and in Appendix \ref{Sec:QSD}, we show that we extend the QSD decomposition to allow $XX$ gates with arbitrary angle $\theta$ and reconstruct the desired unitary at each timestep. The result is a circuit for each dimension with 6 fully-entangling $XX(\pi/2)$ gates, and 18 partial-entangling $XX(\theta)$ gates, with $0 < \theta < \pi/2$. Partial-angle gates are implemented by keeping the gate time constant (to satisfy phase-space closure constraints) and scaling the overall laser amplitude to generate the desired entanglement\cite{yale2025realization}. This gate count is lower than that obtained from the standard QSD algorithm\cite{saha2025qsd}.
%in Appendix \ref{Sec:QSD}. 

%To get a proper picture of the single qubits and entangling gates reduction completely,c, \cref{tab:gates_comparison} comparison between 

%Additionally, each unitary $\left\{ U_{\boldsymbol{\beta}} \right\} $ is a product state unitary and leads to further simplifications in computation.

%Below, we present one algorithm to handle the reduced dimensional operations using Quantum Shannon Decomposition. 

In the following section, we combine this tensor-network decomposition with a resource-optimized variant of quantum phase estimation to construct a practical algorithm for extracting dynamical and spectroscopic information from quantum simulations.

%Each quantum operation is represented here using QSD given in Appendix A.

\begin{comment}
For a two-dimensional system, there is a single entanglement index $\beta$, and for a product initial state the expression simplifies to
\begin{align}
\chi_{\Delta t}(\bar{\vb{x}}) 
&= \sum_{\beta} 
\left[
\prod_{j=1}^{2} 
\int dx'_j \;
\tensor*{\mathcal{U}}{^{[j]}{}^{x_j x'_j}_{\beta}}
\tensor*{\phi}{^{[j]}{}^{x'_j}}
\right], 
\label{Eq:MPS_WF-2D-prod-init}
\end{align}
which can be expressed as the action of a family of $\{U\chi\}$ operations. In principle, these operations can be constructed and executed in parallel on a Quantum-HPC architecture. This parallel structure is illustrated in \cref{Fig:MPS_tEvo_ND,Fig:MPS_tEvo_trotter-d}. 
\end{comment}

\begin{figure}[htbp]
  \centering
  % Define colors
  \definecolor{purple}{rgb}{0.6, 0.4, 0.8}
  \definecolor{orange}{rgb}{0.83, 0.4, 0.22}
  \definecolor{ggray}{rgb}{0.55, 0.55, 0.55}

  % Wrap the circuit in a node and overlay lines safely
  \begin{tikzpicture}
    % Put quantikz inside a node (no conflict with TikZ overlay)
    \node[inner sep=0pt] (circ) {
      \begin{quantikz}[row sep=0.7cm, column sep=0.4cm]
        \lstick{$\ket{a_1}$} & \gate{H} & \ctrl{3} & \qw & \qw & \gate[wires=3][1cm]{QFT} & \meter{} \\
        \lstick{$\ket{a_2}$} & \gate{H} & \qw & \ctrl{2} & \qw & \qw & \meter{} \\
        \lstick{$\ket{a_3}$} & \gate{H}  & \qw & \qw & \ctrl{1} & \qw & \meter{} \\
        \lstick{$\ket{q}\equiv\ket{\chi_0}$} & \qw & \gate{U} & \gate{U^2} & \gate{U^4} & \qw & \meter{} \\
      \end{quantikz}
    };

    % Overlay red dotted lines (positions tuned manually)
    \draw[red, densely dotted, line width=1.8pt] ($(circ.south west)+(2.6,0.6)$) -- ($(circ.north west)+(2.6,0.2)$)
      node[above, black, yshift=0.2cm]{\textbf{Init}};
    \draw[red, densely dotted, line width=1.8pt] ($(circ.south west)+(5.7,0.6)$) -- ($(circ.north west)+(5.7,0.2)$)
      node[above, black, yshift=0.2cm]{\textbf{Prop}};
    \draw[red, densely dotted, line width=1.8pt] ($(circ.south west)+(7.1,0.6)$) -- ($(circ.north west)+(7.1,0.2)$)
      node[above, black, yshift=0.2cm]{\textbf{Final}};
  \end{tikzpicture}

  \caption{\label{Fig:PEA-ckt1} Illustration of the phase estimation algorithm for quantum wavepacket dynamics. \( U = \exp\{-\imath H \Delta t / \hbar\} \). $\ket{q}$ refers to a family of qubits operated upon by powers of $U$.}
\end{figure}
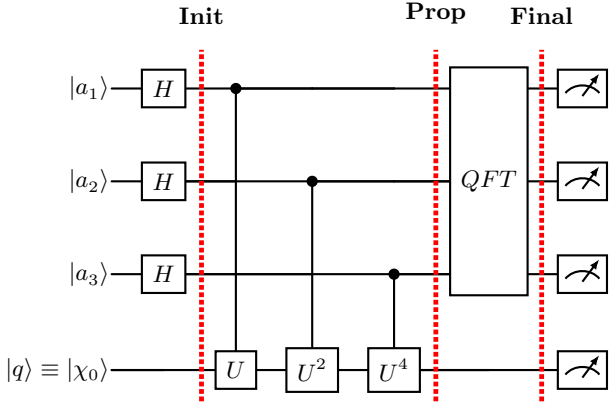
\section{Modified phase estimation algorithm for resource optimized quantum vibrational spectroscopy}
\label{Sec:PEA-Vib}
In quantum phase estimation, illustrated in \cref{Fig:PEA-ckt1}, a unitary time-evolution operator is used to construct the final quantum state as
\begin{align}
\ket{\mathrm{Final}}
&=
\frac{1}{\sqrt{2^a2^q}}
\sum_{e=0}^{2^a-1}
\sum_{k=0}^{2^q-1}
\ket{e}_a
\ket{x_k}
\nonumber\\
&\quad\times
\left[
\sum_{m=0}^{2^a-1}
\frac{1}{\sqrt{2^a}}
e^{\imath (e\Delta E)(m\Delta t)/\hbar}
\chi(x_k;m\Delta t)
\right]
\nonumber\\
&=
\frac{1}{\sqrt{2^a2^q}}
\sum_{e=0}^{2^a-1}
\sum_{k=0}^{2^q-1}
\ket{e}_a
\ket{x_k}
\xi(x_k,E_e),
\label{PEA-Final-main}
\end{align}
where
$\xi(x_k,E_e)$
denotes the time-to-energy Fourier transform of
$\chi(x_k,t_m)$.
Here, the multidimensional quantum wavepacket is represented on a discrete coordinate grid mapped onto the computational basis of the $q$-qubit register according to
\begin{align}
x_k
\longrightarrow
\ket{k}_q,
\end{align}
with
\begin{align}
\left\{
\ket{k}_q
\in
\left\{
\ket{00\cdots00}_q,
\ket{00\cdots01}_q,
\cdots,
\ket{2^q-1}_q
\right\}
\right\},
\label{discrete-Compbasis-xk}
\end{align}
where $x_k$ labels the discrete coordinate grid points of the wavepacket representation. Appendix \ref{PEA-SDO} provides a detailed dervation of Eq. (\ref{PEA-Final-main}). More details on the general map between the continuous representation $\ket{x}$ and discrete qubit representation for the nuclear dynamics problem treated here in discussed in Section \ref{mapping}. Here, the index $j$ is reserved exclusively for labeling the physical dimensions of the multidimensional system. Similarly, the ancilla register, after Fourier transform, encodes a discrete energy grid,
\begin{align}
\left\{
\ket{e}_a
\in
\left\{
\ket{00\cdots00}_a,
\ket{00\cdots01}_a,
\cdots,
\ket{2^a-1}_a
\right\}
\right\},
\end{align}
where $\ket{e}_a$ represents the computational basis states associated with the discretized energy values $E_e$.

Measurement of
$\ket{\mathrm{Final}}$
therefore yields the states
$\left\{
\ket{k}_q;
\ket{e}_a
\right\}$
with probability
$\left|
\xi(x_k,E_e)
\right|^2$,
corresponding to a specific coordinate grid point $x_k$ and energy value $E_e$. The resulting probability distribution forms the Fourier power spectrum associated with the vibrational eigenstates of the system.

This standard phase-estimation procedure requires a large number of controlled unitary operations, which remain challenging to implement efficiently on current quantum hardware. In the following, we therefore introduce an alternative approach that substantially reduces the number of controlled operations while naturally integrating with the tensor-network formalism developed above.

In this publication, we provide two effective ways to help implement the above algorithm, using tensor networks on current quantum computers. (a) We first split the algorithm depicted in Figure \ref{Fig:PEA-ckt1} into two parts. The first part does the quantum propagation to obtain multiple time samples of the quantum state and arrives at ``Prop'' (for propagated) in Figure \ref{Fig:PEA-ckt1}. The second part is the Fourier transform. In our current implementation, we perform the Fourier transform on a classical computer whereas the first part, quantum propagation, is done on a quantum computer. (b) We introduce an alternative algorithm that yields eigen-energy differences, as opposed to absolute energies as these are experimentally observable. 
%This is done purely for implementational simplicity, and this restriction can be lifted in future.   

This second aspect is a key distinction between the phase estimation algorithm and the approach used here. As noted, the phase estimation algorithm provides an approach to compute the eigenstates of a Hamiltonian from time-propagation followed by Fourier transforms. But, in most chemical applications, including vibrational spectroscopy, energy differences are experimentally observed and have chemical meaning, but absolute energies do not have any physical meaning. For example, in molecular spectroscopy, it is energy differences between eigenvalues that are measured and not absolute eigenenergies. Compare this aspect with Eq. (\ref{PEA-Final-main}) which results in spectral intensities at absolute energy values given by $E_e$.  Hence we begin our approach here towards energy differences to arrive at a reduced quantum resource phase estimation using the Fourier transform of the density-density autocorrelation function and arrive at the following final expression: 
\begin{align}  
    {\cal P} (\omega) & = \int dx {\left\vert \int_{-\infty}^{+\infty} dt\,  e^{{\imath \omega t}} \; \rho(x,x; t) \right\vert}^2 \\
    & = \int dx {\left\vert \sum_{i,j} \delta(\omega-(E_{i}-E_{j})) c_{i}(0)c_{j}^{*}(0) {\phi_{i}(x)}
    {\phi_{j}(x)} \right\vert}^2 ,
    \label{Eq:Density-timecorrelation-FT-Final} 
\end{align}
As can be seen from above, this expression yields spectral energy differences directly. 
Furthermore, using tensor networks, we use an approximate form of the expression above given by, 
\begin{align}  
    {\cal P} (\omega) \approx& \sum_{\bar{\boldsymbol\alpha}}^{\bar{\boldsymbol\eta}} \prod_{j=1}^{N} \int dx_j  {\left\vert \int_{-\infty}^{+\infty} dt\,  e^{{\imath \omega t}} \; {\left\vert \tensor*{\phi}{^{[j]}_{}^{x_j}_{\alpha_{j-1},\alpha_j}}\right\vert}^2 \right\vert}^2
    \label{Eq:Intensity_FT}
\end{align}
Let's define, 
\begin{align}
\tensor*{P}{^{[j]}_{}^{}_{\alpha_{j-1},\alpha_j}}(\omega)
=
\int \dd{x_j}\,
{\left\vert \int_{-\infty}^{+\infty} dt\,  e^{{\imath \omega t}} \;
\left\vert \tensor*{\phi}{^{[j]}_{}^{x_j}_{\alpha_{j-1},\alpha_j}}\right\vert^2\right\vert}^2
\label{Eq:Intensity_FT1}
\end{align}

Then, \cref{Eq:Intensity_FT} can be written as:

\begin{align}  
    {\cal P} (\omega) \approx& \sum_{\bar{\boldsymbol\alpha}}^{\bar{\boldsymbol\eta}} \prod_{j=1}^{N} \tensor*{P}{^{[j]}_{}^{}_{\alpha_{j-1},\alpha_j}}(\omega)
    \label{Eq:Intensity_FT2}
\end{align}
As we will see, this expression provides a $\approx 4cm^{-1}$-level agreement between the quantum computing results and classical computing results. 
%It is however critical to rewrite the above expression using the discrete computational basis so as to make a direct connection to the experimental advances in the next section and the QSD algorithm in Appendix \ref{Sec:QSD}. 

%As stated in Eq. (\ref{PEA-Final-main}),  a quantum wavepacket can also be represented on a multi-dimensional discrete coordinate representation that is mapped onto the computational basis of the q-qubit space, $x_k \rightarrow \ket{k}_q$, and this expression in \cref{Eq:Intensity_FT} becomes
%\begin{align}
%     {\cal P} (\omega) \approx& 
%     \frac{1}{\sqrt{2^q}} \sum_{\bar{\boldsymbol\alpha}}^{\bar{\boldsymbol\eta}} \sum_{e=0}^{2^a-1} \sum_{k=0}^{2^q-1} \ket{e}_a \ket{x_k} \nonumber \\ & 
%     \left[ \sum_{j'=0}^{2^a-1} \frac{1}{\sqrt{2^a}} e^{\imath (e\Delta E)(j'\Delta t) / \hbar} {\left\vert \tensor*{\phi}{^{[j]}_{}^{x_j;j'\Delta t)}_{\alpha_{j-1},\alpha_j}}\right\vert}^2  \right]  
     \label{Iw-Final-main}
%\end{align}
%where the summation over ``e" is performed classically. 
%\todo{May need more in this section above to emphasize parallelism}

\section{Experimental implementation}
\label{Sec:Experimental}
Experiments are performed using $^{171}$Yb$^+$ ions confined in a surface electrode trap (Sandia Peregrine \cite{Revelle2020Peregrine}). Qubit states are encoded using the $\ket{F=0, m_F=0}\equiv\ket{0}$ and $\ket{F=1, m_F=0}\equiv\ket{1}$ hyperfine levels of the $^2$S$_{1/2}$ manifold \cite{olmschenk2007manipulation}. Doppler cooling and state initialization into $\ket{0}$ are performed using near-resonant laser beams at 369 nm, and resolved sideband cooling prepares the ions near their ground motional states. State-dependent fluorescence from the ions at 369 nm is captured using a high numerical aperture objective (NA=0.6) and imaged onto a multimode fiber array to provide site-resolved readout of the ion qubit state \cite{clark2021engineering}. State preparation and measurement (SPAM) errors are estimated to be $0.7\%$.

Quantum gate operations are engineered using pairs of Raman beams at 355 nm, with a propagation direction aligned to one of the radial modes of the ion crystal. The Raman beams have two path options, one of which is shaped elliptically and focused to an $8~\mu$m$\times160~\mu$m spot size at the center of the trap, such that it globally illuminates all ions. The other passes through a multi-channel acousto-optic modulator \cite{Harris2020aom} such that each ion may be individually addressed with independent laser amplitudes, frequencies, and phases. The Raman beams can be operated in either a co-propagating configuration, such that both Raman tones pass along the same path, or a counter-propagating configuration that is sensitive to ion motion.

Single qubit gates are driven by tuning the beatnote frequency between Raman beams onto resonance with the hyperfine transition freq of 12.642819 GHz. Depending on the chosen phase and laser pulse duration, this executes rotations on the Bloch sphere of form $R_x(\theta)$ or $R_y(\theta)$. Rotations around the $z-$axis of the Bloch sphere are implemented virtually. Rotations around the $x-$ and $y-$axes are performed in the co-propagating Raman configuration with 99.5(3)$\%$ typical fidelity for $\pi/2$ rotations. Due to the large number of $\pi/2$ gates in the QSD of our unitary propagator, we dynamically correct these gates for small amplitude and frequency errors in the laser pulse \cite{Piliouras2026,Piliouras2026SCSQ,Amer2025,Barnes2022} to improve the overall circuit coherence.

Two qubit gates are implemented between pairs of ions by individually addressing them with counter-propagating Raman beams to drive M\o{}lmer-S\o{}rensen interactions \cite{molmer1999multiparticle}. These gates take the Ising-type form $XX(\theta)$, which are equivalent to CNOT gates (up to single-qubit rotations) when $\theta$ is set equal to $\pi/2$ \cite{maslov2017basic}. The amplitude of the laser addressing beam is modulated with a Gaussian envelope to minimize phase-space non-closure errors and to reduce high-frequency drive components which may off-resonantly couple to nearby transitions \cite{ruzic2024leveraging}. In addition, wrapper gates are added around the M\o{}lmer-S\o{}rensen gates to maintain phase correlation with the co-propagating single qubit rotations \cite{yale2025realization}. Typical gate times for full entanglement are 200~$\mu$s with two-qubit gate fidelities of 98.0(3)$\%$. 

A key aspect of our experimental implementation is the use of partial-angle two-qubit entangling gates, which can be performed at higher fidelity than fully-entangling gates \cite{yale2025realization}. In the standard QSD circuit, arbitrary decompositions of three-qubit unitary propagators require 24 fully-entangling CNOT gates or $XX(\pi/2)$ gates. Here, we extend the decomposition to allow $XX$ gates with arbitrary angle $\theta$ and reconstruct the desired unitary at each timestep. The result is a circuit with 6 fully-entangling $XX(\pi/2)$ gates, and 18 partial-entangling $XX(\theta)$ gates, with $0 < \theta < \pi/2$. Partial-angle gates are implemented by keeping the gate time constant (to satisfy phase-space closure constraints) and scaling the overall laser amplitude generate the desired entanglement. Careful calibrations have been performed between $\pi/32 \leq \theta \leq \pi/2$ to minimize effects of gain non-linearity and variable light shifts. To improve the overall fidelity, single- and two-qubit gates with rotation angles $\theta < \pi/32$ and $\theta > 31\pi/32$ are approximated as $\theta=0$ and $\theta=\pi/2$, respectively. We estimate that our use of partial-entangling gates has led to a $>30\%$ infidelity reduction when implementing QSD circuits on our trapped ion hardware. Also see discussion in Appendix \ref{Sec:QSD}.

\section{Quantum wavepacket dynamics of protonated water clusters on ion-trap quantum computers}
\label{Sec:Results}
%{\todo {Anurag: Background on water cluster spectroscopy from Anup’s paper, may be also Xiao’s LLM paper, discussions spectroscopic challenges in water cluster systems and how experiments and theory dont agre creating controversy. It is discussed in water papers from the group. A key challenge of water clusters is that these are multi-dimensional quantum nuclear and electron correlation has a critical role due to polarization of electronic structure due to nuclear motion. Again citations are key here and can help create the basis for PRX or Science.}}

Water clusters provide a uniquely controlled environment for understanding how hydrogen bonding, many-body interactions, and quantum effects give rise to the unusual properties of water \cite{HDMeyer-Zundel-1,protonwire1,atmosph-clusters1,atmosph-clusters2,pomesroux2,protonwire3,protonwire4,teeter,bio-clusters2,lipscombpeek,turowlett,Scott-proj,hesselmann2018correlation,znamenskiy2007quantum,stillinger1980water}. Additionally, protonated water wires and water clusters are an important class of molecular systems found in many constrained environments such as biological membranes and enzyme active sites \cite{Nagle:78,baciou1995interruption,Guo}, ions channels \cite{allen2003gramicidin,pullman1983gramicidin}, carbon nanotubes \cite{WW3,ww11,hummer_rasaiah_noworyta_2001}, and fuel cells \cite{ye2012water}. Water wires are also present in the photosynthetic reaction center of \textit{Rhodobactersphaeroides} where they are responsible for proton transfer to a secondary quinone group \cite{baciou1995interruption}. 
Furthermore, the lightweight hydrogen nucleus makes quantum nuclear effects important in such cases
%for hydrogen transfer reactions
\cite{Cptuckerman3,H+OH-solv, schmittVothProtonTransport1999, schmitt1998multistate}; additionally the multidimensional quantum nuclear effects in such systems is also known to be critical\cite{johnson-jordan-21mer,Johnson-Jordan-Zundel-Science}.  

Consequently, they have been studied extensively using high-resolution infrared (IR), terahertz vibration–rotation–tunneling (VRT), and rotational spectroscopy across cluster sizes ranging from the water dimer to larger aggregates\cite{johnson-jordan-21mer,Johnson-21mer-2016,Johnson-Jordan-Zundel-Science,saykally2001water,saykally1996waterscience}. These experiments reveal highly fluxional structures, dense manifolds of tunneling states, and vibrational spectra that are strongly influenced by cooperative hydrogen bonding\cite{johnson-jordan-21mer,Johnson-21mer-2016,Johnson-Jordan-Zundel-Science,saykally2001water,saykally1996waterscience}. Even small water clusters exhibit signficantly complex quantum effects \cite{stillinger1980water,tokmachev2024networks} and vibrational mode couplings\cite{admp-21mer,SSI-Review1-QC-ES-QN,fragIJQC-review}, and their spectra are extremely sensitive to both the underlying potential energy surface and the treatment of multi-dimensional quantum nuclear effects\cite{saykally1996waterscience,fragIJQC-review,frag-AIMD-multitop-2,frag-ML-Xiao,Johnson-21mer-2016,Johnson-Jordan-H+-water-cluster-sizes,Johnson-Jordan-Zundel-Science,Johnson-Zundel-OH-H2O-quantum,HDMeyer-Zundel-1,HDMeyer-Zundel-2,HDMeyer-Zundel-3,Asmis-Zundel,Fournier-Asmis-21mer}. 

Despite decades of experimental and theoretical effort, significant discrepancies between measured spectra and theoretical predictions have persisted for many protonated as well as hydroxide-rich water clusters\cite{Johnson-21mer-2016,AsthagiriOH,H2O6OH-Xiaohu}, leading to long-standing controversies regarding structural assignments and spectral interpretation\cite{Voth2021resolving,saykally2001water}. A notable example relates to the spectroscopy of the hydrated excess proton, where competing models historically favored either the Eigen-like $\mathrm{H_9O_4^+}$ structure or the Zundel-like $\mathrm{H_5O_2^+}$ motif \cite{zundel2012hydration,Cptuckerman3, H+OH-solv, schmittVothProtonTransport1999,schmitt1998multistate,Voth2021resolving}. Subsequent studies revealed that the hydrated proton actually samples a highly fluxional environment in which these structures represent limiting configurations connected by large-amplitude proton transfer and hydrogen-bond rearrangements\cite{marx2006proton,Marx:00,TuckermanOH,TuckermanNature:2002,MarxNature:1999,admp-21mer}. More broadly, the difficulty arises because water clusters are intrinsically multidimensional quantum nuclear systems in which collective intermolecular vibrations strongly influence the spectra\cite{saykally2001water,saykally1996waterscience}. Accurately describing these effects requires quantum nuclear dynamics on high-dimensional potential energy surfaces that also capture electron correlation, many-body polarization, and dispersion\cite{hesselmann2018correlation,polarizability1,eisenberg2005structure}. The simultaneous need for high-level electronic structure and multidimensional quantum dynamics has made quantitative agreement between theory and experiment extremely challenging, and water cluster spectroscopy therefore remains an important benchmark for developing predictive methods quantum nuclear dynamics\cite{saykally2001water,marx2006proton,Voth2021resolving}.

\begin{figure}[tbp]
    
    \begin{minipage}[t]{0.33\columnwidth}
        \centering
        \raisebox{0.3cm}{\includegraphics[width=0.99\columnwidth]{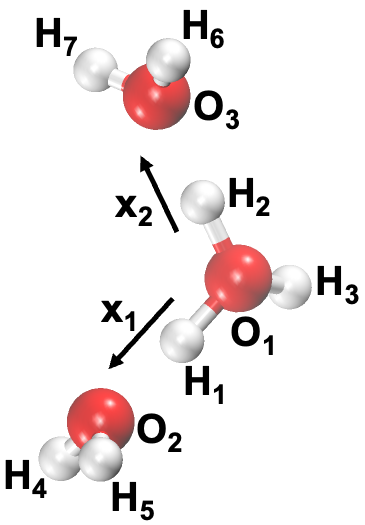}}
    \end{minipage}
    \hfill
    \begin{minipage}[t]{0.63\columnwidth}
        \centering
        \includegraphics[width=0.999\columnwidth]{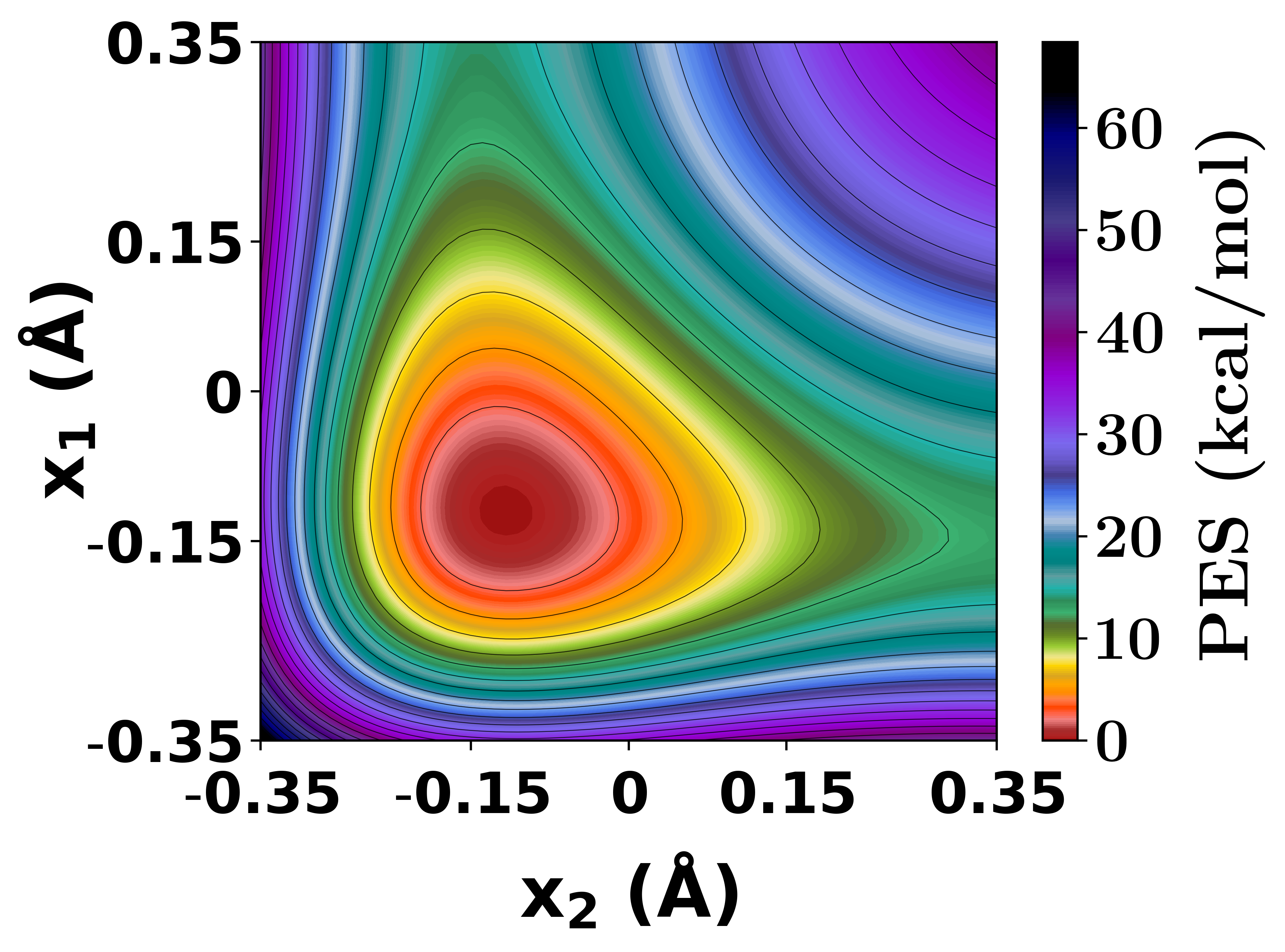}
    \end{minipage}
    \caption{Molecular structure of the \ce{H7O3+} molecule (left) and the corresponding two-dimensional potential energy surface (right) plotted along the shared hydrogen coordinates $x_1$ and $x_2$. Each coordinate represents the proton shared between the water molecules.
    }
    \label{Fig:2D_molecule_and_pes}
\end{figure}

The system considered in this work is a protonated water wire, \ce{H7O3+} \cite{Johnson-Jordan-H+-water-cluster-sizes}. This system acts a bridge between the key Zundel and Eigen cations and we show here that these kinds of systems can now be studied on current quantum computers, thus taking a critical step towards future predictive vibrational spectrocopy calculations on quantum computers, beyond the harmonic approximation.
%, a minimal model for proton transport along hydrogen-bonded networks\cite{protonwire1,pomesroux2,protonwire3,protonwire4}. 

%Such motifs arise ubiquitously in condensed-phase and biological environments, including membrane systems\cite{Nagle:78,baciou1995interruption,Guo}, ion channels\cite{domene2003potassium}, and catalytic sites\cite{baciou1995interruption}, where they facilitate efficient charge migration. 
%Proton transport in these networks proceeds via the Grotthuss mechanism, in which the excess proton is relayed through a hydrogen-bonded chain via concerted rearrangements of the molecular framework, rather than through physical diffusion of a single ionic species\cite{protonwire1}.

%Owing to the low mass of the proton, nuclear quantum effects\cite{Cptuckerman3,H+OH-solv,schmittVothProtonTransport1999,schmitt1998multistate} play a central role in governing the dynamics of these systems. In particular, the interplay of anharmonicity and multidimensional coupling leads to strongly non-classical behavior that cannot be captured within a purely classical description\cite{admp-21mer,SSI-Review1-QC-ES-QN,fragIJQC-review}. For this reason, the \ce{H7O3+} water wire provides a stringent and physically relevant platform for benchmarking multidimensional quantum dynamical methods and potential energy surface representations.

\begin{table}[tbp]
\centering
\renewcommand{\arraystretch}{1.25}
\setlength{\tabcolsep}{8pt}
\caption{Geometric parameters for the \ce{H7O3+} system. The atom numbering follows that presented in the left panel of Fig.~\ref{Fig:2D_molecule_and_pes}.}
\label{tab:geom_parameters2D}

\begin{tabular*}{\columnwidth}{@{\extracolsep{\fill}} l c}
\hline\hline

\textbf{Donor-acceptor distances} & \textbf{Value} \\
\hline\\[-3mm]
Hydrogen bond OO distance (O$_1$–O$_2$) & \SI{2.715}{\angstrom} \\
Hydrogen bond OO distance (O$_1$–O$_3$) & \SI{2.715}{\angstrom} \\

\hline
\textbf{Hydrogen bond angles} & \\
\hline\\[-3mm]
Hydrogen bond OHO angle (O$_1$–H$_1$–O$_2$) & \SI{175.27}{\degree} \\
Hydrogen bond OHO angle (O$_1$–H$_2$–O$_3$) & \SI{177.69}{\degree} \\

\hline
\textbf{Water molecule bond angles} & \\
\hline\\[-3mm]
Water bond angle (H$_4$–O$_2$–H$_5$) & \SI{107.11}{\degree} \\
Water bond angle (H$_6$–O$_3$–H$_7$) & \SI{106.89}{\degree} \\

\hline\hline
\end{tabular*}
\end{table}

\begin{table}[tbp]
\centering
\renewcommand{\arraystretch}{1.25}
\setlength{\tabcolsep}{8pt}
\caption{Characteristics of the grid over which the two-dimensional potential surface is created.}
\label{tab:grid}

\begin{tabular*}{\columnwidth}{@{\extracolsep{\fill}} cc}
\hline\hline

\textbf{Parameter} & \textbf{Value} \\
\hline

\begin{tabular}{@{}c@{}} No. of grid points along \\ the $x_1$-dimension\end{tabular} & 8\\
\begin{tabular}{@{}c@{}} No. of grid points along \\ the $x_2$-dimension\end{tabular} & 8\\

\begin{tabular}{@{}c@{}} Grid size along the \\ $x_1$-dimension  \end{tabular} & 0.70 \AA\\
\begin{tabular}{@{}c@{}} Grid size along the \\ $x_2$-dimension \end{tabular} & 0.70 \AA\\

Level of theory & \begin{tabular}{@{}c@{}} B3LYP/6-311++G(d,p) \end{tabular}\\

\hline\hline
\end{tabular*}

\end{table}

\begin{table*}[tbp]
    \centering
    \small
    \setlength{\tabcolsep}{5pt}
    \renewcommand{\arraystretch}{1.2}
    \caption{Initial wavepacket characteristics. The initial state is defined on a discrete grid using the Kronecker delta $\delta_{i,j}$, where $\delta_{i,j}=1$ if $i=j$ and $0$ otherwise.}
    \label{tab:initial_wavepacket_characteristics}
    \begin{tabular*}{\textwidth}{@{\extracolsep{\fill}} l c c}
        \hline\hline
        \textbf{Wavepacket Formula} & \textbf{Bond Dimension} & \textbf{Average Energy} \\
        \hline\\[-3mm]
        $\chi_0(x_1,x_2)=\delta_{x_1,0.25}\,\frac{1}{\sqrt{2}}\left(\delta_{x_2,-0.15}+\delta_{x_2,-0.05}\right)$
        & $1$ & 40.99 kcal/mol \\
        \\[-3mm]
        \hline\hline
    \end{tabular*}
\end{table*}

\begin{table}[tbp]
    \centering
    \small
    \setlength{\tabcolsep}{6pt}
    \renewcommand{\arraystretch}{1.2}
    \caption{Simulation parameters for the water wire system \ce{H7O3+}.}
    \label{tab:simulation_parameters}
    \begin{tabular*}{\columnwidth}{@{\extracolsep{\fill}} l c}
        \hline\hline
        \textbf{Parameter} & \textbf{\ce{H7O3+}} \\
        \hline\\[-3mm]
        Time step $\Delta t$ & 1 fs \\
        Total time $T$       & 150 fs \\
        Qubit counts $D$          & 5 \\
        Number of shots           & 1000 \\
        \\[-3mm]
        \hline\hline
    \end{tabular*}
\end{table}

\subsection{Molecular geometry and reduced dimensional potential energy surfaces}
\label{Sec:PES}
To investigate the structural, dynamical, and vibrational properties of the \ce{H7O3+} water-wire system (\cref{Fig:2D_molecule_and_pes}), we perform quantum wavepacket dynamics simulations on Born–Oppenheimer potential energy surfaces constructed at the density functional theory level using the B3LYP/6-311++G(d,p) method. The optimized geometrical parameters associated with these surfaces are summarized in \cref{tab:geom_parameters2D}, while the details of the multidimensional grid used to represent the wavepacket are provided in \cref{tab:grid}. The characteristics of the initial wavepackets employed in the simulations are listed in \cref{tab:initial_wavepacket_characteristics} and simulation parameters are given in  Table \ref{tab:simulation_parameters}.

For the \ce{H7O3+} water-wire system, the hydrogen-bonded network formed by three water molecules supports two coupled proton-transfer coordinates. The shared proton stretch coordinates are the only degrees of freedom treated quantum mechanically in this study and future studies will work towards scaling up to the full system with 33 nuclear degrees of freedom. The treatment of the proton  stretches here gives rise to an effective two-dimensional PES, as shown in \cref{Fig:2D_molecule_and_pes}. The coordinates $x_1$ and $x_2$ describe proton motion along the respective donor–acceptor oxygen axes associated with each hydrogen bond. The interaction between these coordinates introduces coupling between the hydrogen-bonding degrees of freedom, resulting in a multidimensional energy landscape.

The PES is discretized on a direct-product grid points along each coordinate, symmetrically distributed about the respective grid centers along the donor–acceptor axes. 
%This representation enables the capture of coupled proton-transfer dynamics and associated anharmonic effects within the hydrogen-bonded network. 
The kinetic energy operator 
may be approximated in a number of ways. One approach is to recognize that this operator is diagonal in the momentum representation and hence fast Fourier transforms are commonly employed\cite{feit1,feit2,feit3,dev,dkos1,dkos2}. 
In this paper, we employ an analytic banded Toeplitz distributed approximating functional (DAF)
\cite{DAFprop-PRL,discreteDAF,Debadrita-Mapping-1D-3Qubits,qwaimd-TCAreview} representation for the 
grid representation of 
the kinetic energy operator:  
\begin{align}
K(x,x^{\prime}) =& K(\left\vert x-x^{\prime}\right\vert) =
\frac{-\hbar^2}{4m\sigma^3\sqrt{2\pi}}
\exp \left\{ -\frac{ {\left( x - x^\prime \right)}^2}
{2 {\sigma}^2} \right\} \nonumber \\ 
&\sum_{n=0}^{M_{DAF}/2} {\left( \frac{-1}{4} \right)}^n \frac{1}{n!} 
H_{2n+2} \left( \frac{ x - x^\prime }{ \sqrt{2} \sigma} \right).
\label{DAFfreeprop+derivative}
\end{align}
The analytic banded-Toeplitz representation of the DAF approximation for the kinetic energy operator is one where the matrix elements, $K_{ij} \equiv K{(\vert} i-j {\vert )}$, 
The quantities $\left\{ H_{2n+2}\left( \frac{ x - x^\prime }{ \sqrt{2} \sigma} \right) \right\}$ in Eq. (\ref{DAFfreeprop+derivative}) are the even order Hermite polynomials that only depend on the spread separating the grid basis vectors, $\ket{x}$ and $\ket{x^{\prime}}$, and $M_{DAF}$ and $\sigma$ are parameters that together determine the accuracy and efficiency of the resultant approximate kinetic energy operator. 

The system is subsequently investigated through quantum wavepacket propagation implemented using quantum computing algorithms.
The time evolution of the wavepacket is obtained using a quantum algorithm implemented on the \ce{QSCOUT} platform at Sandia National Laboratories, following the approach described in \cref{Sec:PEA-Vib}. (Mapping the grid representation to the qubit representation is discussed in Section \ref{mapping}.) This framework enables the simulation of multidimensional quantum nuclear dynamics and the extraction of vibrational information directly from the propagated quantum states.

All the quantum computing results here are compared with the exact treatment of time-evolution operator represented in the eigenstates basis of the nuclear Hamiltonian, with a detailed error analysis provided in \cref{Sec:Vibrational}. All calculations are performed using a Python-based classical and quantum algorithm software developed within the Iyengar group. 

%{\todo {Anurag: grid distributions for classical simulation, electronic structure potentials, etc.}}
%The reduced-dimensional Born–Oppenheimer potential energy surfaces (PESs), shown in \cref{Fig:2D_molecule_and_pes}, were constructed at the B3LYP/6-311++G(d,p) level of density functional theory. The corresponding nuclear wavepacket grid parameters are summarized in \cref{tab:grid}.

\subsection{Mapping the continuous grid representation onto the discrete computational  basis from qubits}
\label{mapping}
To enable simulation on a gate-based quantum computer, the discretized nuclear-coordinate basis is mapped onto the computational basis of a qubit register. Let
\begin{align}
\mathcal{G}=\left\{\ket{g_i}\right\}_{i=1}^{D}
\end{align}
denote the set of $D$ grid basis points obtained from the discretization of the continuous one-dimensional nuclear-coordinate space for the evolution of each unitary in $\left\{ \tensor*{\mathcal{U}}{^{[j]}_{}^{x_j x'_j}_{\beta_{j-1},}_{\beta_j}} \right\}$. %Here each basis state $\ket{g_i}$ corresponds to a specific multidimensional grid point $\mathbf{g}_i$. 
Each grid point is assigned a unique integer label ``$i$'', which is subsequently encoded in binary form using $n=\left\lceil \log_2 D \right\rceil$ qubits. The resulting correspondence is
\begin{align}
\ket{g_i}
\;\longleftrightarrow\;
i
\;\longleftrightarrow\;
\ket{i}_q,
\label{correspondence}
\end{align}
where $\ket{i}_q = \ket{b_{n-1} b_{n-2}\cdots b_0}$, is a Hamming space vector, and $b_k\in{0,1}$ are the binary components of the integer (i). The computational basis states belong to the Hamming space %set
\begin{align}
\mathcal{B}_n
=
\left\{
\ket{00\cdots 00},
\ket{00\cdots 01},
\ldots,
\ket{11\cdots 11}
\right\}.\nonumber
\end{align}
and the discussion here establishes a one-to-one correspondence between the discretized nuclear configurations and the computational basis states of the qubit register.

Using this mapping, an arbitrary nuclear wavepacket, say $\tensor*{\phi}{^{[j]}_{}^{x_j}_{\alpha_{j-1},\alpha_j}}$ in Eqs. (\ref{Eq:MPO_MPS}) or (\ref{Eq:MPO_MPS-D1}), that is used in the quantum propagation formalism discussed here, can be represented as
\begin{align}
\ket{\tensor*{\phi}{^{[j]}_{}_{\alpha_{j-1},\alpha_j}}}
=
\sum_{i=1}^{D}
c_i \ket{g_i}
=
\sum_{i=1}^{D}
c_i \ket{i}_q,
\end{align}
and $\left\{ \ket{g_i} \right\}$ are now the discritized grid representation for dimension $x_j$ over which the state  $\bra{x_j}\ket{\tensor*{\phi}{^{[j]}_{}_{\alpha_{j-1},\alpha_j}}} \equiv \tensor*{\phi}{^{[j]}_{}^{x_j}_{\alpha_{j-1},\alpha_j}}$ is represented. This 
allows the initial wavepacket to be prepared directly on the qubit register and subsequently evolved under the encoded unitary dynamics. The molecular nuclear dynamics can therefore be simulated entirely within the computational basis of the quantum processor.
\subsection{Quantum Wavepacket Dynamics on a Trapped-Ion Quantum Computer}
\label{Sec:Wavepacket_Dynamics}

\begin{comment}
\begin{figure}[tbp]
    \centering
    {\input{Figures/Figure-TN-first-trotter}}
    \caption{\label{Fig:MPS_tEvo_trotter-c} Derived from \cref{Fig:MPS_tEvo_ND}, but for a two-dimensional system and depicts the simultaneous study of quantum dynamics for individual one-dimensional wavefunctions on distinct ion-trap quantum computers.}
\end{figure}
\end{comment}

This section presents the implementation of multidimensional quantum wavepacket dynamics on distributed trapped-ion quantum hardware. 

%\begin{comment}
%\marginpar{\tiny{Is Table \ref{tab:eigenenergies_first10} needed}}
\begin{table}[tbp]
    \centering
    \small
    \setlength{\tabcolsep}{5pt}
    \renewcommand{\arraystretch}{1.2}
    \caption{First 10 eigenenergies of the \ce{H7O3+} molecule, reported in kcal/mol.}
    \label{tab:eigenenergies_first10}
    \begin{tabular*}{\columnwidth}{@{\extracolsep{\fill}} c c}
        \hline\hline
        \textbf{Eigenstate} & \textbf{Energy (kcal/mol)}\footnote{1kcal/mol = 349.76 cm$^{-1}$ = 10.49 THz. Energy differences here are generally greater than the resolution imposed by choice of total propagation time of 150fs apart from the difference between eigenstates 2 and 3.} \\
        \hline\\[-3mm]
        1  & 8.3884 \\
        2  & 15.3090 \\
        3  & 15.9444 \\
        4  & 21.0859 \\
        5  & 21.4919 \\
        6  & 23.4887 \\
        7  & 27.4780 \\
        8  & 27.7994 \\
        9  & 28.9209 \\
        10 & 30.2794 \\
        \\[-3mm]
        \hline\hline
    \end{tabular*}
\end{table}
%\end{comment}
Each one-dimensional wavepacket was propagated  on the quantum computer for a total of 150 fs using a timestep of 1 fs, as summarized in \cref{tab:simulation_parameters}. These parameters were selected based on the following practical considerations. Firstly, for hydrogen bond dynamics, time-steps of the order of 1fs are commonly used given that oscillation periods are of the order of $\approx$10fs in such systems. This choice provides good resolution for the higher frequency hydrogen bond oscillations. The second choice of total propagation time of 150fs %arises from computational expense associated with the quantum computer as it arises from number of simulations and number measurements. 
%to accurately resolve the vibrational spectra extracted from the time-domain dynamics. 
%Furthermore, the total propagation time of 150 fs 
corresponds to a frequency resolution of approximately 6.67 THz (222.5 cm$^{-1}$ or 0.64kcal/mol) in the Fourier- spectrum. %From a vibrational spectroscopy perspective, 
This %is a large 
resolution window, as we will see, is sufficient to resolve all energy differnces in the current vibrational problem %the current problem does not have closely spaced energy eigenvalues 
(see Table \ref{tab:eigenenergies_first10}). 
%and hence this large resolution window suffices to capture the spectrum accurately. 
Other problems may require longer propagation times. 
%enabling the separation of closely spaced vibrational features. 
The 1 fs sampling interval yields a Nyquist frequency of 1000 THz, which very much exceeds the largest energy gap among all the eigenenergies of the lower-dimensional subsystems. Consequently, the chosen temporal discretization captures the full range of physically relevant vibrational frequencies without aliasing while providing sufficient spectral resolution to identify individual vibrational modes.
\begin{figure}[tbp]
\centering
\subfigure[]{%
    \includegraphics[width=0.48\columnwidth]{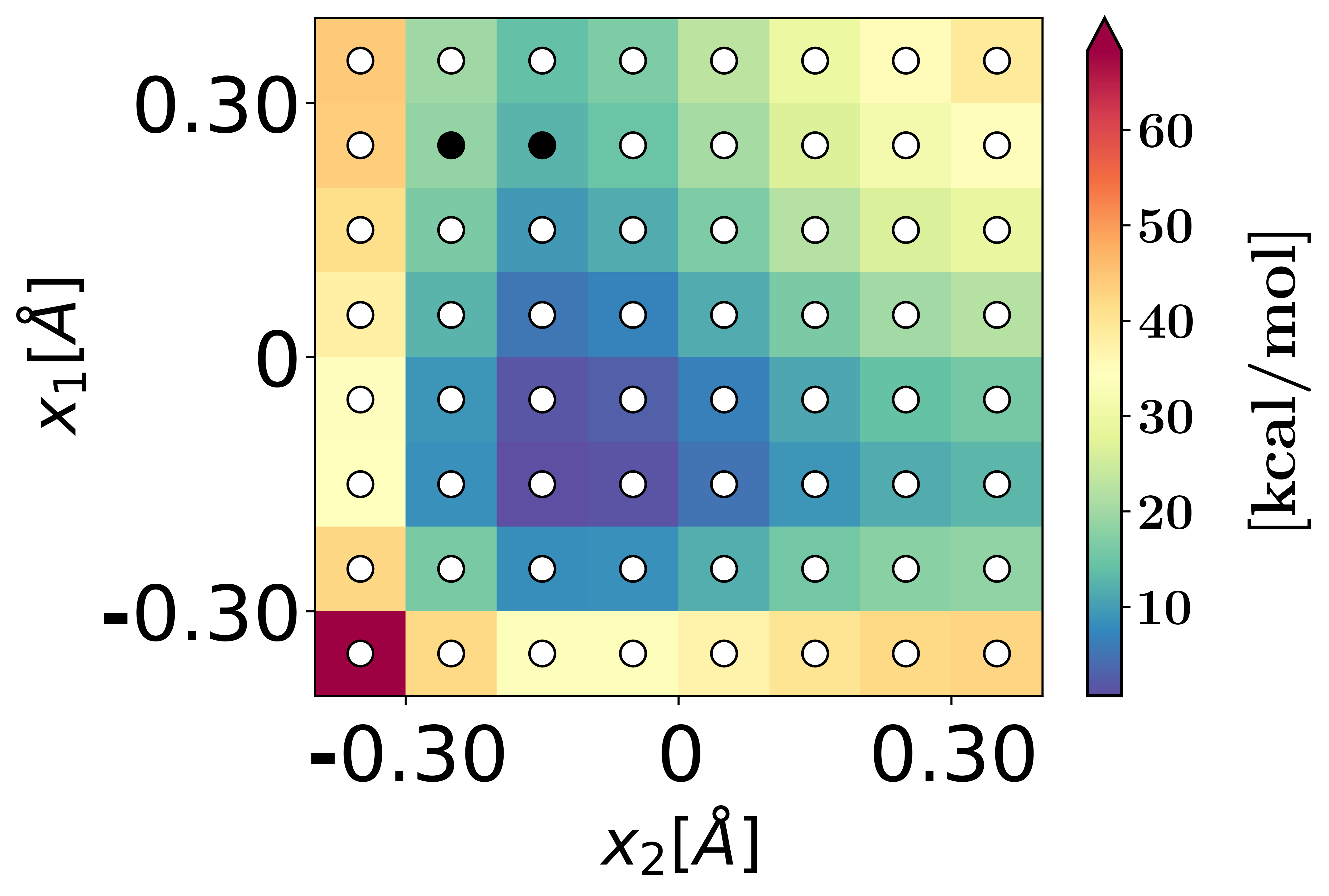}
}
\hfill
\subfigure[]{%
    \includegraphics[width=0.48\columnwidth]{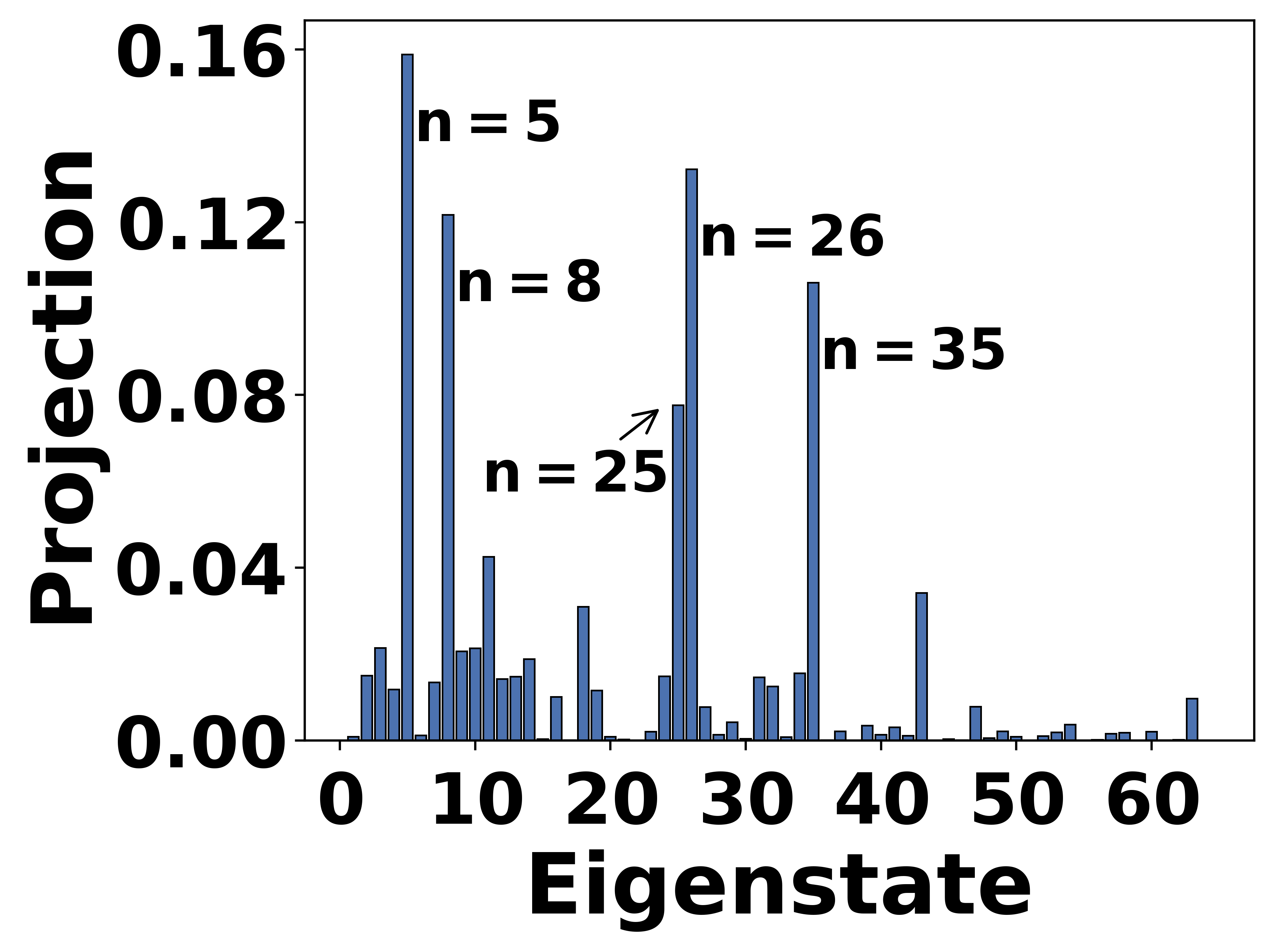}
}
\caption{
(a) Potential energy surface (PES) on $8\times8$ grid, with the initial discrete wavepacket overlaid on a uniform grid. The wavepacket is defined using a Kronecker delta distribution, where filled (black) markers denote grid points with unit amplitude and open (white) markers indicate zero amplitude. The grid spans the interval $[-0.35,\,0.35]$ along both $x_1$ and $x_2$, discretized into 8 points per dimension ($\Delta x = 0.10$). The wavepacket is localized at $x_1 = 0.25$ and forms an equal superposition at $x_2 = -0.15$ and $x_2 = -0.05$, consistent with $\psi(x_1,x_2,0)=\delta_{x_1,0.25}\,\frac{1}{\sqrt{2}}(\delta_{x_2,-0.15}+\delta_{x_2,-0.05})$.
(b) Projection of the initial wavepacket onto the eigenstates of the system, shown as $|\langle \phi_n | \psi(0) \rangle|^2$ versus eigenstate index $n$. The five largest contributions are explicitly labeled.}
\label{fig:pes_wavepacket_combined}
\end{figure}
%These bond dimensions quantify the degree of entanglement between nuclear coordinates and determine the number of effective one-dimensional propagators required to represent the full multidimensional dynamics. Consequently, the tensor-network decomposition naturally generates multiple quantum replicas that can be propagated independently on distributed quantum hardware. The total effective bond dimension, represented by $\alpha\times\beta$ and compactly denoted by $\mu$, is explicitly illustrated in \cref{Fig:MPS_tEvo_ND,Fig:MPS_tEvo_trotter-c}.
\begin{figure*}[t]
 	\centering
    \subfigure[Quantum (blue) and classical (yellow) time-traces for $\ket{x_1}$]{\includegraphics[width=0.45\textwidth]{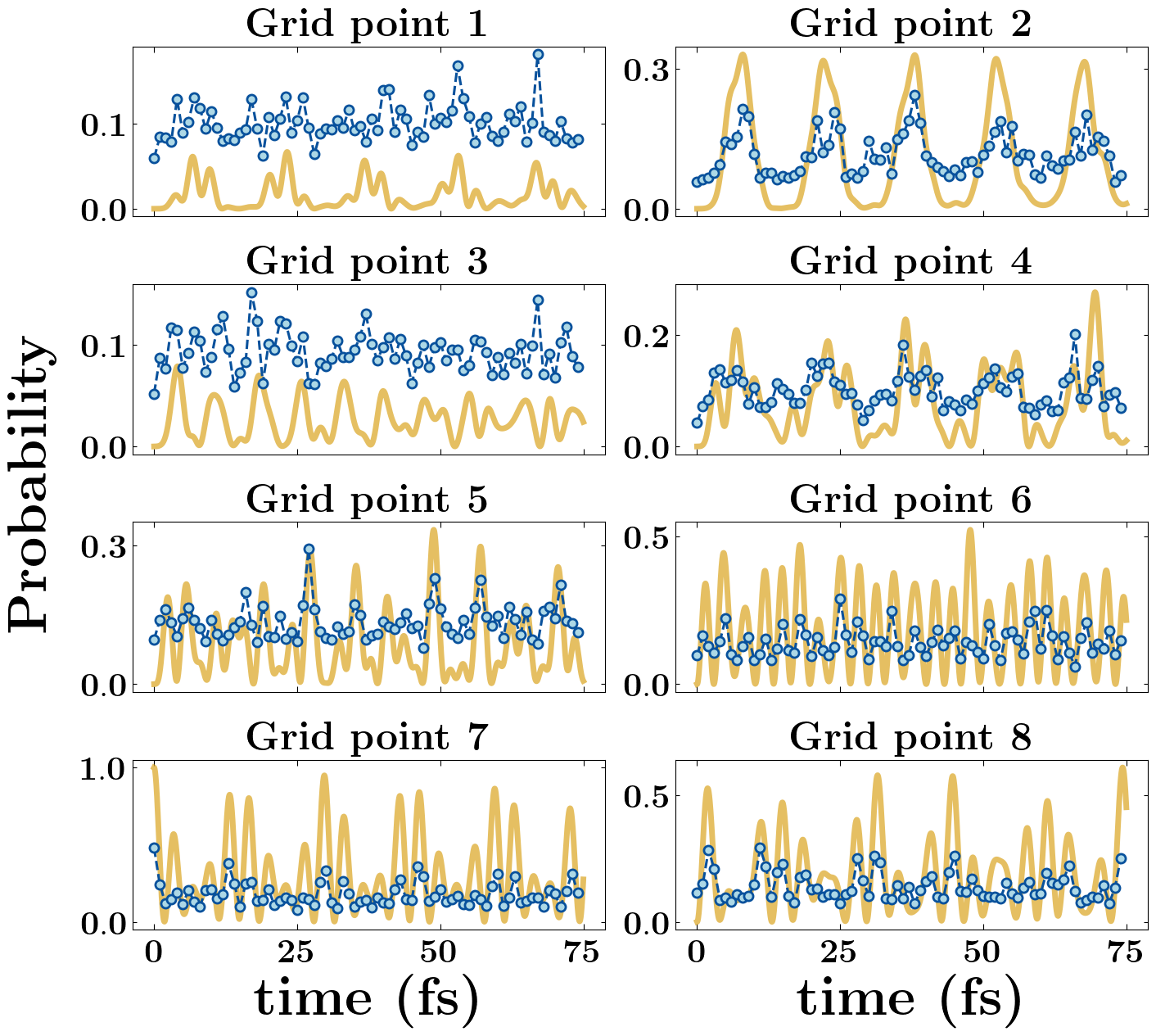}\label{Fig:average_left1_wp}}
    \subfigure[Quantum (blue) and classical (yellow) time-traces of $\ket{x_1}$]{\includegraphics[width=0.45\textwidth]{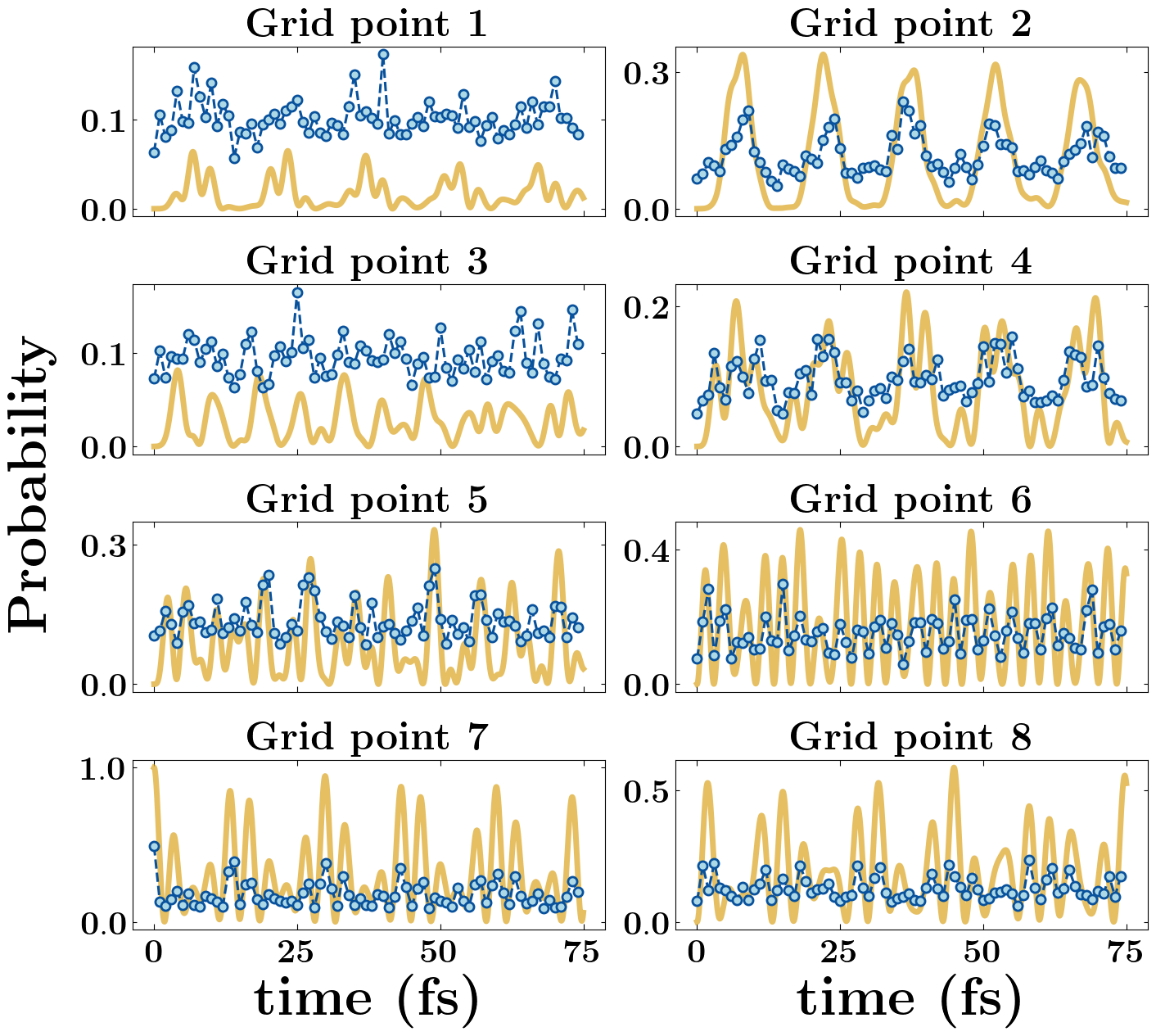}\label{Fig:average_left2_wp}}\\
    \subfigure[Quantum (blue) and classical (yellow) time-traces of $\ket{x_2}$]{\includegraphics[width=0.45\textwidth]{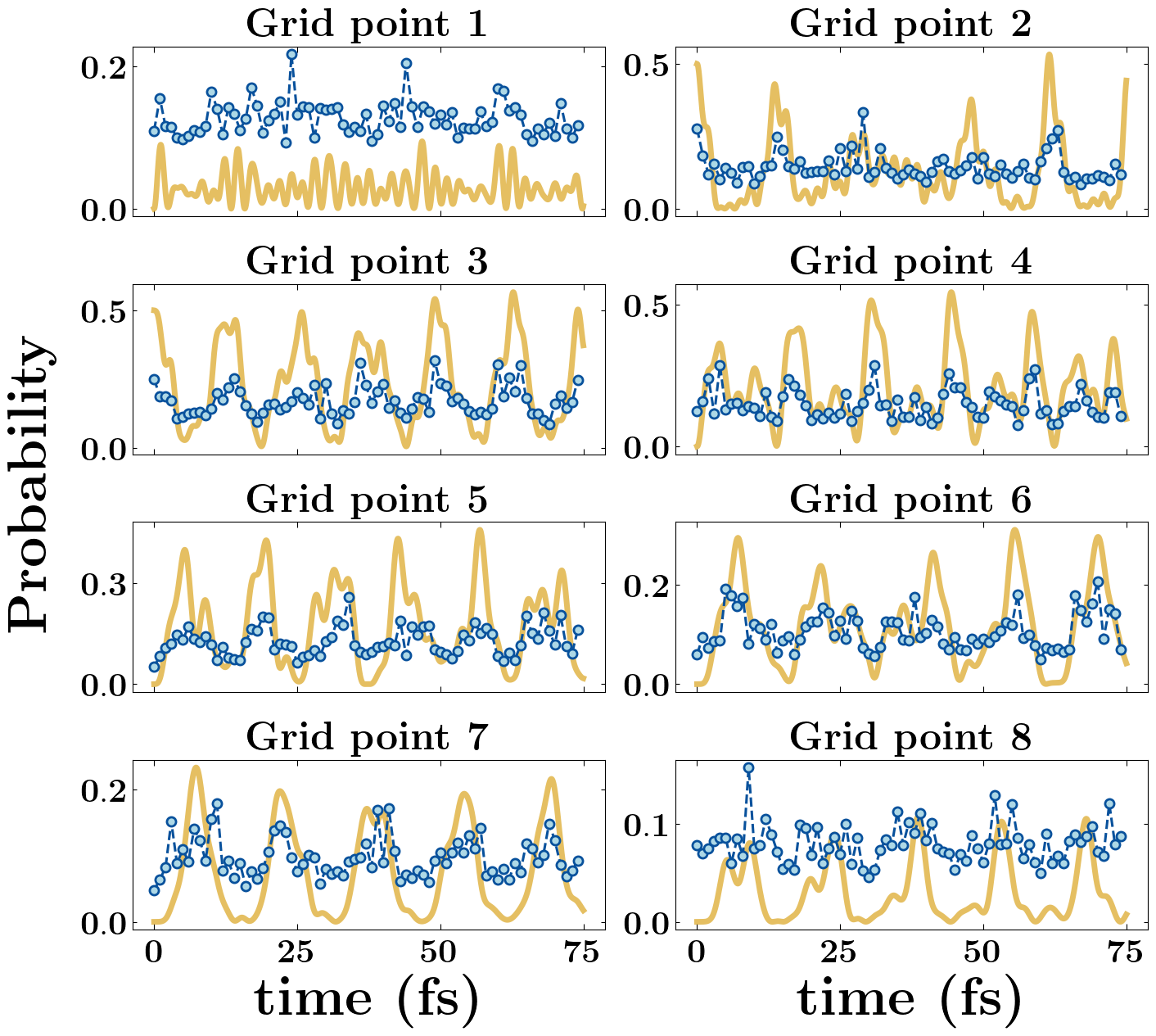}\label{Fig:average_right1_wp}}
    \subfigure[Quantum (blue) and classical (yellow) time-traces of $\ket{x_2}$]{\includegraphics[width=0.45\textwidth]{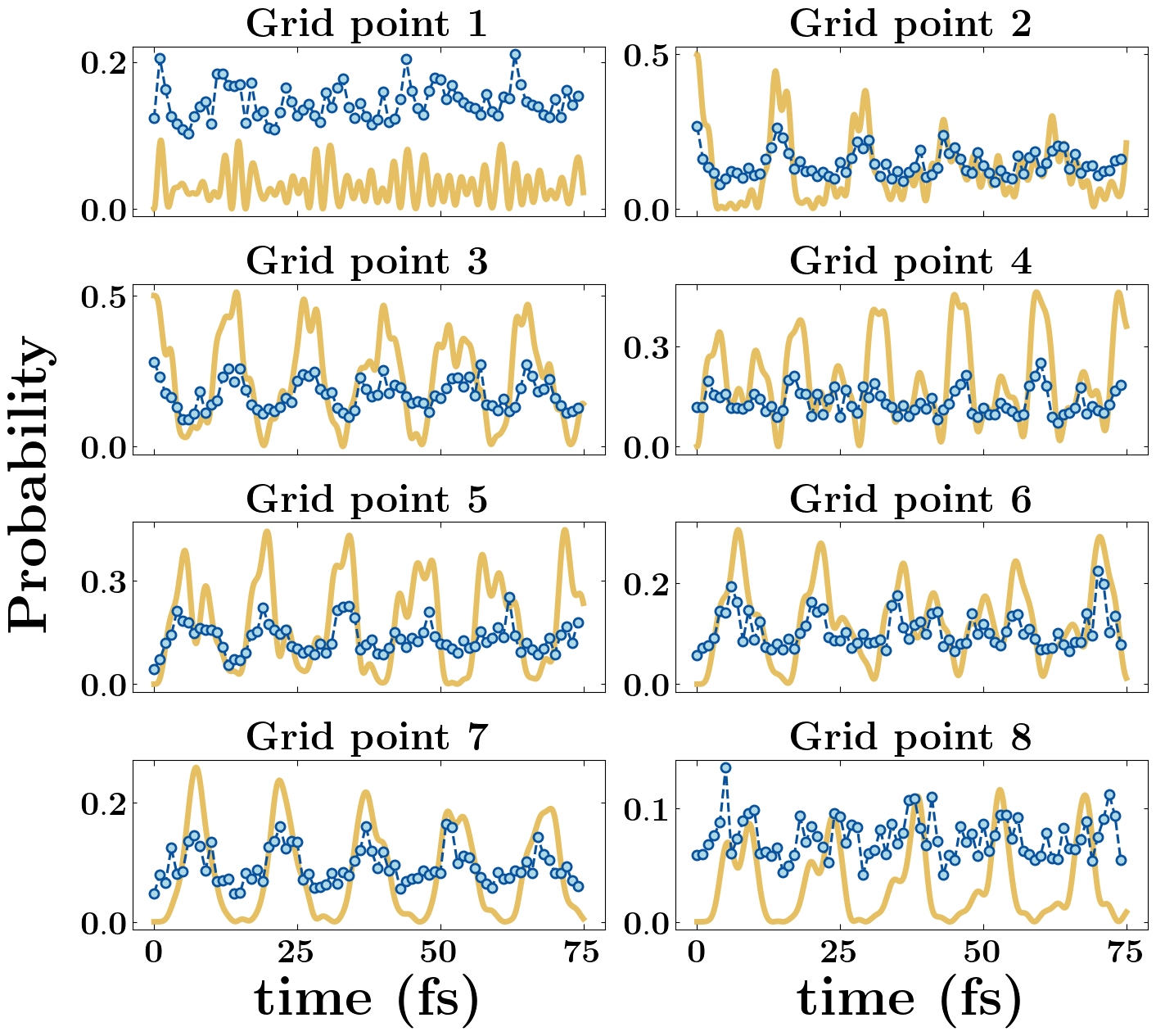}\label{Fig:average_right2_wp}}
    
  	\caption{Time-dependent populations of the propagated wavepacket projected onto individual grid basis states, comparing numerically exact classical simulations (solid lines) with trapped-ion quantum hardware results (blue markers). The dynamics are resolved into effective one-dimensional subsystems generated through the tensor-network decomposition of the multidimensional propagator. Panels (a) and (b) display the dynamics 
    %governed by the Hamiltonians $\left\{H^{[1]}_{1}\right\}$ and $\left\{H^{[1]}_{2}\right\}$, respectively, 
    corresponding to the two entanglement components associated with the vibrational coordinate $x_1$ along the proton-transfer direction connecting the two oxygen atoms. Panels (c) and (d) show the dynamics 
    %associated with the Hamiltonians $\left\{H^{[2]}_{1}\right\}$ and $\left\{H^{[2]}_{2}\right\}$, respectively, 
    corresponding to the two entanglement components associated with the orthogonal vibrational coordinate $x_2$, as illustrated in \cref{Fig:2D_molecule_and_pes}. Although significant deviations in the oscillation amplitudes are observed between the quantum-hardware results and the exact classical propagation, the overall waveform structure remain in strong agreement throughout the evolution.
    %The close agreement between the classical and quantum results demonstrates that the distributed trapped-ion implementation captures the multidimensional quantum dynamics and coherent vibrational motion of the \ce{H7O3+} system.
   }
    \label{Fig:Wavepacket}
\end{figure*}
%From each one-dimensional propagator component,
%$\mathcal{V}^{[j]}_{\beta}$,
%effective one-dimensional potentials are extracted and combined with the corresponding kinetic operators to construct the reduced-dimensional Hamiltonians
%$\left\{H^{[1]}_{\beta}\right\}$ and
%$\left\{H^{[2]}_{\beta}\right\}$. These Hamiltonians define independent one-dimensional quantum subsystems, each of which is mapped directly onto a separate trapped-ion quantum processor. 

%The complete multidimensional propagation is then recovered through tensor-network recombination of the independently evolved components.

For the quantum dynamics presented here, the initial wavepacket, $\chi_{0}(x_1,x_2)$, illustrated in \cref{fig:pes_wavepacket_combined}(a) and summarized in \cref{tab:initial_wavepacket_characteristics}, is chosen to be a product state, $\psi(x_1,x_2,0)=\delta_{x_1,0.25}\,\frac{1}{\sqrt{2}}(\delta_{x_2,-0.15}+\delta_{x_2,-0.05})$. 
%uncorrelated, corresponding to a tensor-network bond dimension of $\eta=1$ in \cref{Eq:MPS_WF}. For clarity, in the present two-dimensional case ($N=2$), we denote this quantity as $\eta_1=\eta$. 
The wavepacket is spatially localized in the proton-transfer coordinate $x_1$ and prepared as an equal superposition of two neighboring grid points along $x_2$. This initialization places the proton predominantly within one channel of the double-well landscape, enabling us to directly monitor population transfer and coherence between the two channels during the subsequent time evolution.

The spectral composition of the initial state is shown in \cref{fig:pes_wavepacket_combined}(b), which presents the overlap probabilities $|\langle \phi_n | \psi(0) \rangle|^2$ with the exact vibrational eigenstates of the system. The localized nature of the wavepacket results in a distribution of amplitudes over multiple eigenstates rather than a single vibrational level, with the five largest contributions explicitly indicated in the \cref{fig:pes_wavepacket_combined}(b). The resulting non-equilibrium state %superposition 
generates the characteristic oscillatory dynamics observed in the proton-transfer process and provides a suitable initial state for a robust assessment of the accuracy of the distributed quantum simulation framework.

In contrast, the potential energy surface exhibits a modest degree of inter-dimensional correlation, characterized by the presence of only two terms in the summation in Eq. (\ref{Eq:UV-TN-1}).
%tensor-network bond dimension of $N_{\beta}=\max(\beta)=2$
%in \cref{Eq:propagator-TN} (see also \cref{Fig:MPS_tEvo_trotter-c}). 
Within this framework, each product potential propagator term in Eq. (\ref{Eq:UV-TN-1}) %value of $\beta$ 
%generates an independent effective one-dimensional Hamiltonian for every nuclear degree of freedom. Consequently, for $N_{\beta}=2$, 
generates two distinct one-dimensional propagators %Hamiltonians are obtained 
per dimension. For the present two-dimensional proton-transfer system, the resulting dynamics can therefore be distributed across four independent quantum processors and propagated simultaneously. More generally, the achievable level of parallelization is determined jointly by the number of nuclear dimensions and the degree of correlation encoded within the tensor-network representation of the multidimensional potential energy surface.

\begin{figure*}[tbp]
 	\centering
    \subfigure[]{\includegraphics[width=0.45\textwidth]{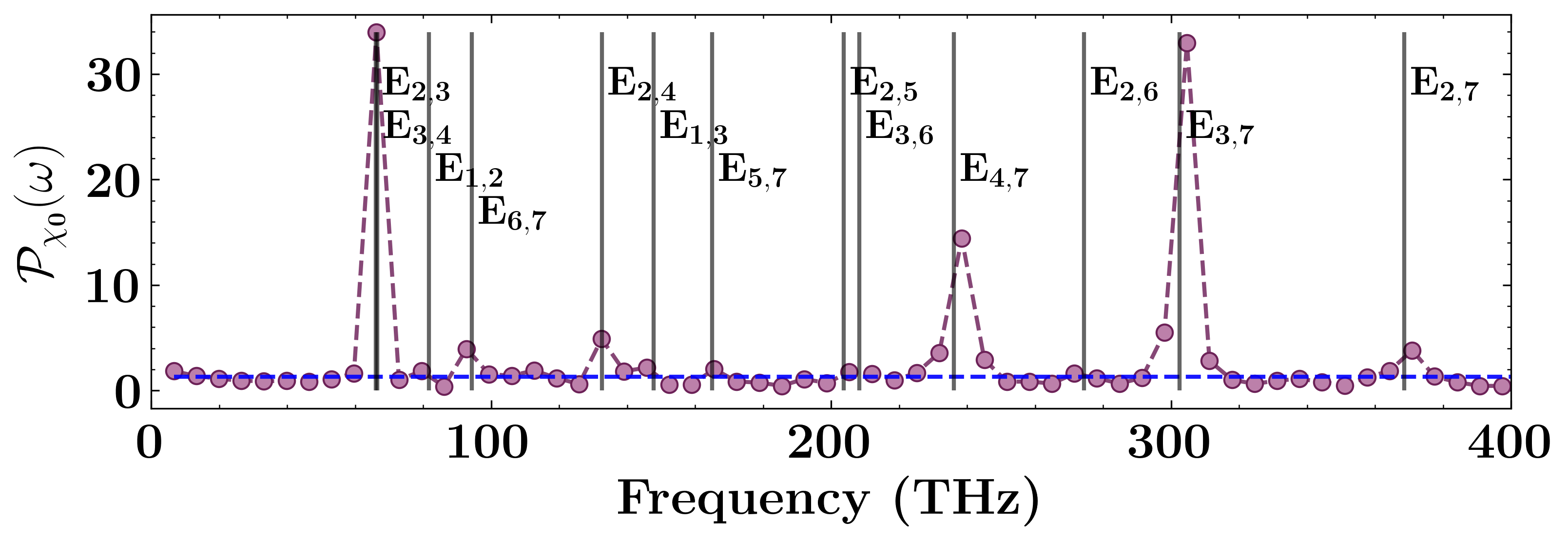}\label{Fig:average_left1_power_vs_freq}}
    \subfigure[]{\includegraphics[width=0.45\textwidth]{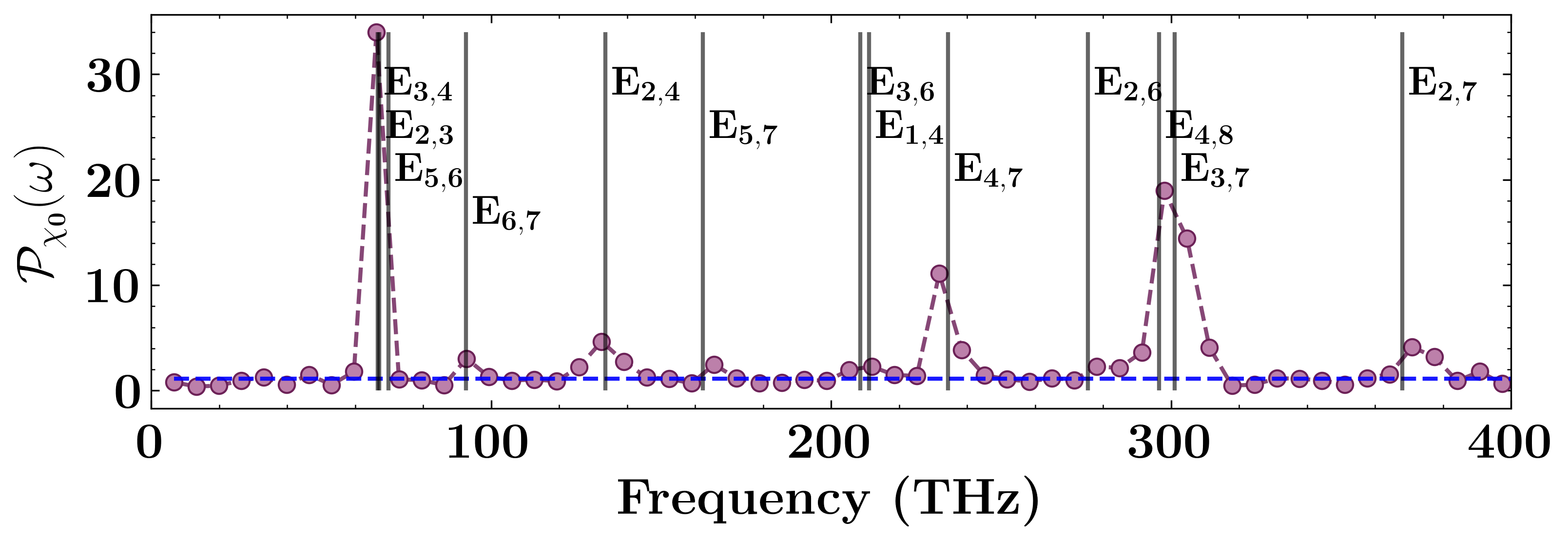}\label{Fig:average_left2_power_vs_freq}}\\
    \subfigure[]{\includegraphics[width=0.45\textwidth]{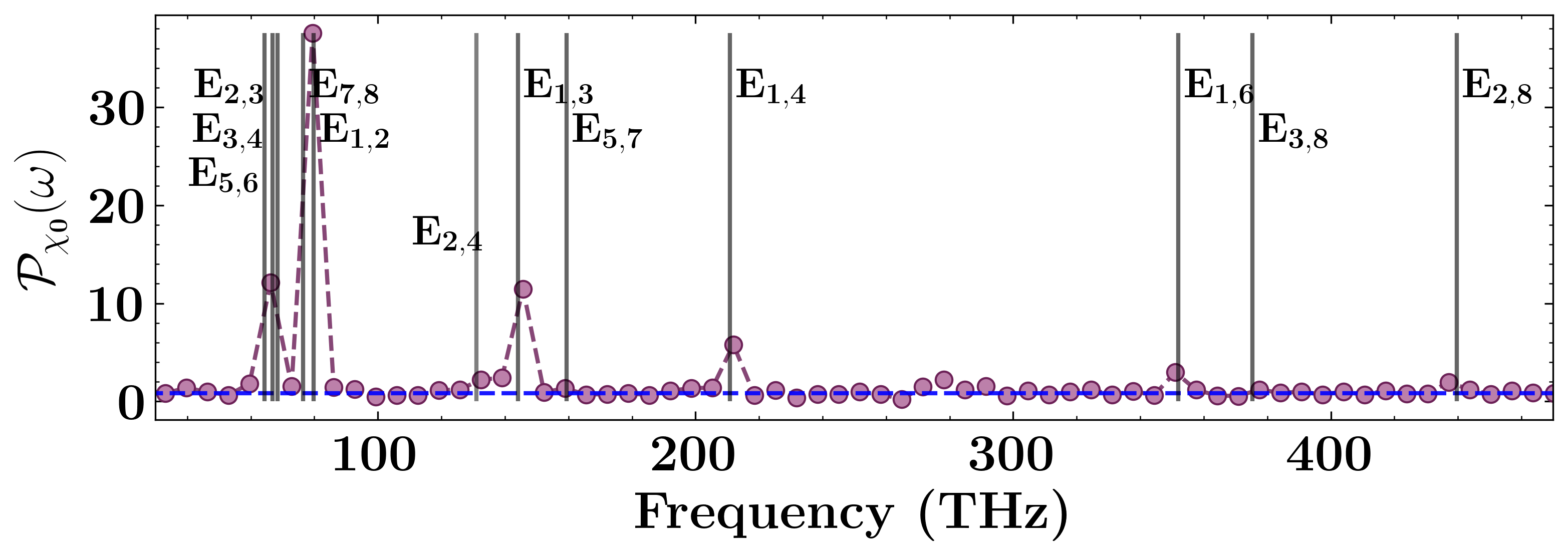}\label{Fig:average_right1_power_vs_freq}}
    \subfigure[]{\includegraphics[width=0.45\textwidth]{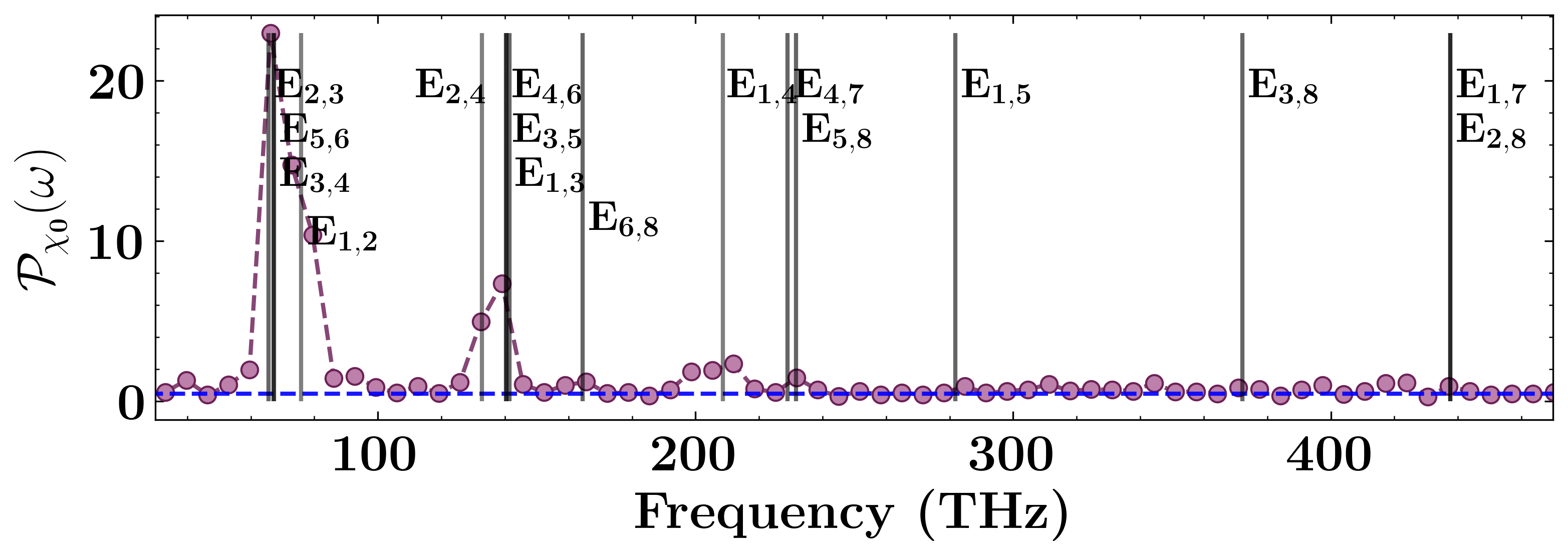}\label{Fig:average_right2_power_vs_freq}}
    
  	 \caption{Frequency spectra obtained from the Fourier transform of the time-dependent wavepacket populations shown in \cref{Fig:Wavepacket}. Panels (a) and (b) correspond to the 
    %effective Hamiltonians $\left\{H^{[1]}_{1}\right\}$ and $\left\{H^{[1]}_{2}\right\}$, 
    two entanglement components associated with the proton-transfer coordinate $x_1$ connecting the two oxygen nuclei, while panels (c) and (d) correspond to the %Hamiltonians $\left\{H^{[2]}_{1}\right\}$ and $\left\{H^{[2]}_{2}\right\}$, 
     entanglement components associated with orthogonal vibrational coordinate $x_2$, shown in \cref{Fig:2D_molecule_and_pes}. The magenta curves represent spectra extracted from trapped-ion quantum simulations using \cref{Eq:Intensity_FT1}, and the solid vertical black lines denote vibrational transition frequencies obtained from exact diagonalization. 
    %Although significant deviations in the oscillation amplitudes are observed between the quantum-hardware results and the exact classical propagation, the overall waveform structure remain in strong agreement throughout the evolution. 
    The close agreement between the quantum-computed spectra and exact results demonstrates that the distributed trapped-ion implementation accurately captures the vibrational structure and multidimensional proton-transfer dynamics of the \ce{H7O3+} system.}
    \label{Fig:Frequency Spectra}
\end{figure*}

To assess the accuracy of the quantum simulations, benchmark dynamics are first computed using numerically exact propagation, where the time-evolution operator is evaluated in the eigenbasis of the full nuclear Hamiltonian. These classical calculations provide reference trajectories against which the experimentally obtained quantum dynamics are compared. The quantum propagators are compiled into quantum circuits as described in \cref{Sec:Experimental} and executed on the QSCOUT trapped-ion quantum computing platform at Sandia National Laboratories.

As shown in \cref{Fig:Wavepacket}, the populations associated with the time-evolved wavepackets obtained from the trapped-ion quantum hardware (blue markers) are compared directly with the numerically exact classical simulations (solid lines) for each grid basis state of the multidimensional system. The time-dependent dynamics are resolved into the effective one-dimensional subsystems generated through the tensor-network decomposition of the multidimensional propagator as described in \cref{Logham}. In particular, \cref{Fig:average_left1_wp,Fig:average_left2_wp} present the time traces associated with the two entanglement components of dimension $x_1$
%the Hamiltonians
%$\left\{H^{[1]}_{1}\right\}$ and
%$\left\{H^{[1]}_{2}\right\}$,
%which describe the vibrational dynamics along the coordinate $x_1$, 
%corresponding to the proton-transfer direction along the internuclear axis connecting the oxygen atoms. 
and similarly, \cref{Fig:average_right1_wp,Fig:average_right2_wp} display the dynamics associated with the two entanglement components of dimension $x_2$. %generated by
%$\left\{H^{[2]}_{1}\right\}$ and
%$\left\{H^{[2]}_{2}\right\}$,
%corresponding to the orthogonal vibrational coordinate $x_2$, as illustrated in 
(See \cref{Fig:2D_molecule_and_pes}.)

Although significant deviations in the oscillation amplitudes are observed between the quantum-hardware results and the exact classical propagation, as we will see in the next section, the overall waveform structure and oscillation frequencies remain in strong agreement throughout the evolution. 
The preservation of the characteristic dynamical frequencies indicates that the tensor-network decomposition together with the trapped-ion implementation captures the essential multidimensional quantum coherence and coupled vibrational motion of the \ce{H7O3+} system. This agreement becomes particularly evident in the Fourier-domain analysis presented later in this work, where the vibrational spectra extracted from the quantum simulations reproduce the dominant spectral features of the exact multidimensional dynamics.

\subsection{Vibrational spectral results from wavepacket dynamics on quantum computers}
\label{Sec:Vibrational}
%{\todo {Anurag: Describe the spectra and agreement.}}

The time evolution of the shared-proton wavepacket provides direct access to the vibrational structure of the multidimensional system. In particular, vibrational frequencies are extracted from the Fourier transform of the density--density time-correlation function defined in \cref{Eq:Density-timecorrelation-FT-Final}. The resulting spectra exhibit distinct peaks whose positions correspond to energy differences between the underlying vibrational eigenstates.

Within the tensor-network framework employed in this work, the spectral analysis is performed independently for each entanglement copy, that is for each effective one-dimensional subsystem generated from the decomposition of the multidimensional propagator. Specifically, the Fourier transforms are computed using the \cref{Eq:Intensity_FT1}, thereby allowing the vibrational response associated with each tensor-network leg to be analyzed separately. A detailed derivation of the corresponding spectral expression is provided in Appendix \ref{Appendix:appendix-FT}.

The frequency spectra obtained from the time-dependent populations shown in \cref{Fig:Wavepacket} are presented in \cref{Fig:Frequency Spectra}. In particular, \cref{Fig:average_left1_power_vs_freq,Fig:average_left2_power_vs_freq} display the Fourier spectra associated with the two entanglement components of dimension $x_1$, while 
%the Hamiltonians
%$\left\{H^{[1]}_{1}\right\}$ and
%$\left\{H^{[1]}_{2}\right\}$,
%which describe the proton-transfer dynamics along the coordinate $x_1$ connecting the two oxygen nuclei. Similarly,
\cref{Fig:average_right1_power_vs_freq,Fig:average_right2_power_vs_freq}
display the Fourier spectra associated with the two entanglement components of dimension $x_2$. 
%present the spectra corresponding to the Hamiltonians
%$\left\{H^{[2]}_{1}\right\}$ and
%$\left\{H^{[2]}_{2}\right\}$,
%associated with the orthogonal vibrational coordinate $x_2$, as illustrated in \cref{Fig:2D_molecule_and_pes}. 
Thus, panels (a)--(d) in \cref{Fig:Frequency Spectra} represent the Fourier transforms of the corresponding time-domain dynamics shown in panels (a)--(d) of \cref{Fig:Wavepacket}.

The spectra extracted from the trapped-ion quantum simulations are shown as magenta curves and are obtained directly from the Fourier transform of the experimentally measured density dynamics. For comparison, the solid vertical black lines denote the vibrational transition frequencies obtained from exact diagonalization of the corresponding Hamiltonians. The observed agreement between the quantum-computed spectra and the exact vibrational energy differences demonstrates that the distributed trapped-ion implementation accurately captures the essential vibrational structure and coherent proton-transfer dynamics of the multidimensional \ce{H7O3+} system.

\begin{figure}[t]
 	\centering
    \subfigure[$\Delta E^{(1D)}$ (Quantum) vs. $\Delta E^{(2D)}$ (Classical)]{\includegraphics[width=0.48\columnwidth]{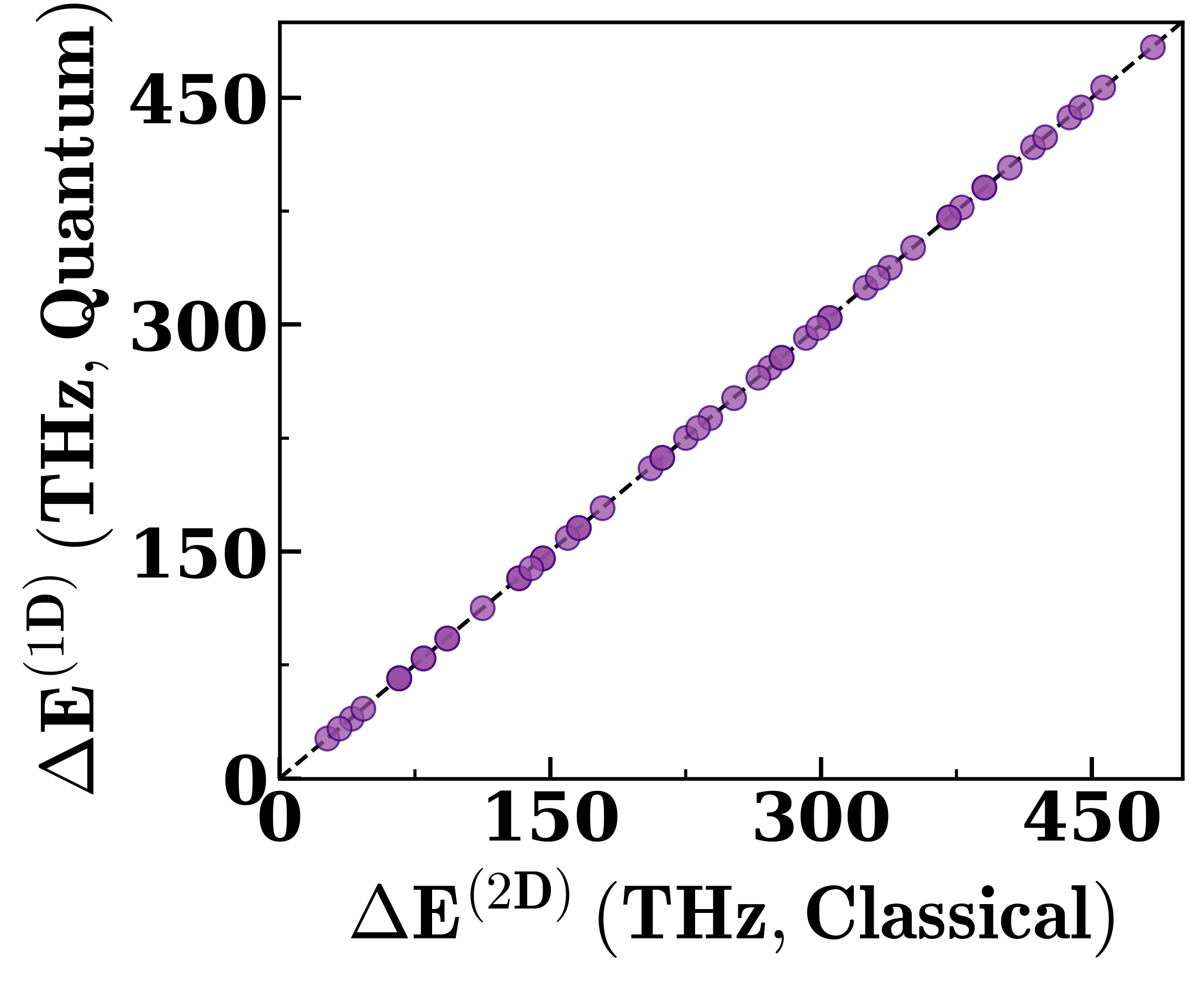}}
    \subfigure[Zoomed low-frequency region]{\includegraphics[width=0.48\columnwidth]{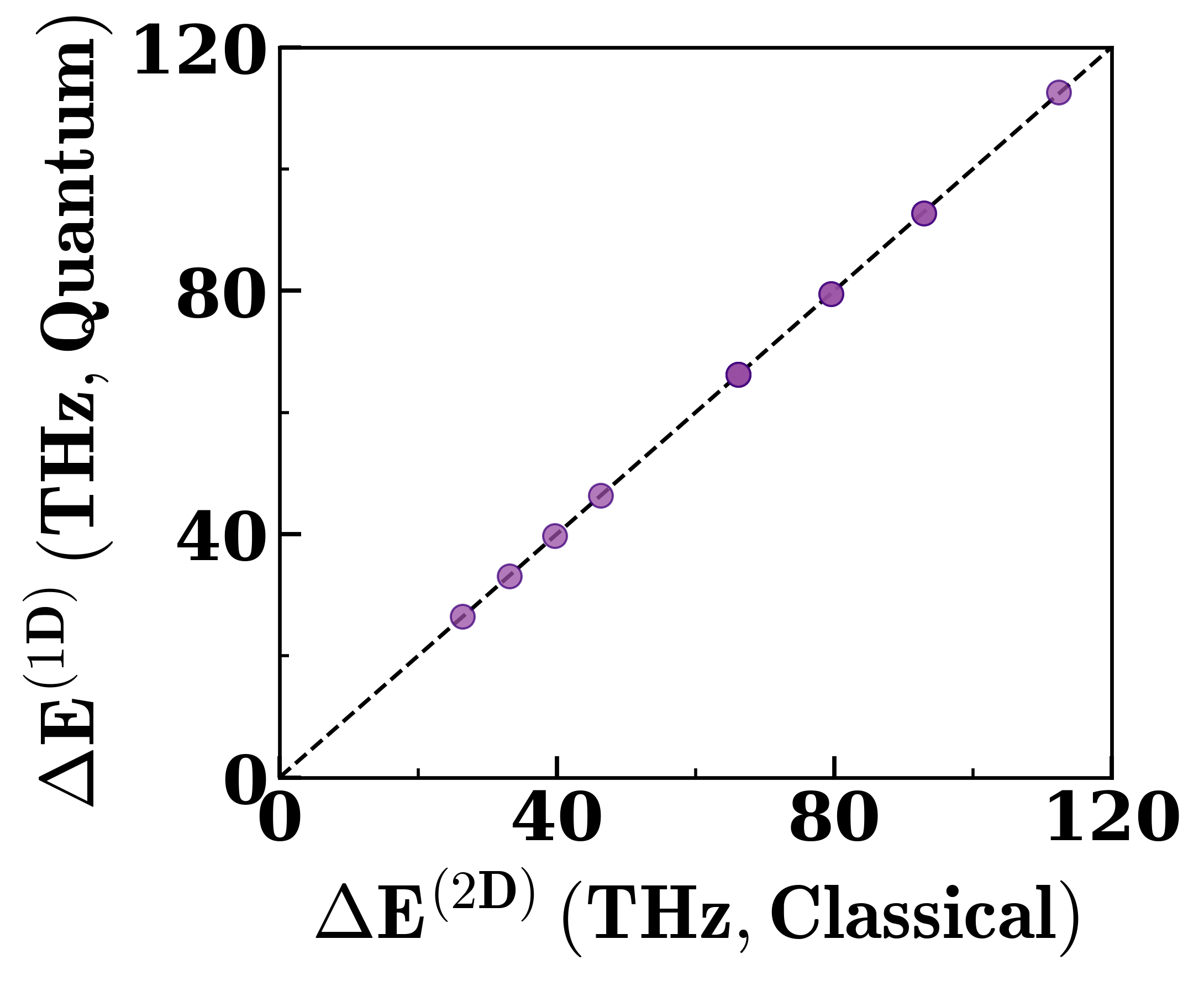}}
  	\caption{Comparison between the vibrational transition frequencies computed using trapped ion quantum hardware and extracted from the effective one-dimensional Hamiltonians and those obtained from exact diagonalization of the full two-dimensional Hamiltonian. Figure (a) Transition energies extracted from the effective one-dimensional Hamiltonians simulated on trapped-ion quantum hardware, $\Delta E^{(1D)}$, plotted against the corresponding transition energies of the full two-dimensional Hamiltonian, $\Delta E^{(2D)}$, obtained through exact diagonalization. Figure (b) presents a magnified view of the low-frequency region, highlighting that the tensor-network-based quantum simulation accurately reproduces the dominant vibrational transition energies of the full \ce{H7O3+} Hamiltonian. 
   }
    \label{Fig:2D_comparison}
\end{figure}

\begin{comment}
To quantify the accuracy of the distributed quantum simulation framework, we compared the vibrational energy differences extracted from trapped-ion quantum simulations of the effective one-dimensional Hamiltonians,
$\left\{
H^{[1]}_{1},
H^{[1]}_{2},
H^{[2]}_{1},
H^{[2]}_{2}
\right\}$
with the corresponding transition energies obtained from exact diagonalization of the full two-dimensional Hamiltonian.
\end{comment}

Additionally, a quantitative comparison of the vibrational transition frequencies is presented in \cref{Fig:2D_comparison}. Panel (a) shows the transition energies obtained from exact diagonalization of the full two-dimensional Hamiltonian along the horizontal axis, %$\Delta E^{(2D)}$, 
plotted against the transition energies extracted from the trapped-ion simulations of the effective entangled one-dimensional propagations along the vertical axis.
%Hamiltonians, $\Delta E^{(1D)}$. 
%Each point corresponds to the closest spectral match identified using the minimum-distance criterion described above, while 
The dashed diagonal line represents the guide for perfect agreement, and very clearly from the plots, the agreement is at an extremely high level. 
%ideal relation,
%$\Delta E^{(1D)}=\Delta E^{(2D)}$
Panel (b) provides a magnified view of the low-frequency region, where deviations between the exact multidimensional spectrum and the reduced-dimensional quantum-computed spectra can be resolved more clearly.

The comparison demonstrates that the dominant vibrational transition energies obtained from the trapped-ion simulations remain in extremely close agreement with the exact multidimensional eigenenergy differences. In particular, the mean absolute deviation between the quantum-computed and exact transition energies,
%\begin{align}
%\mathrm{\epsilon} = \frac{1}{N}
%\sum_{i=1}^{N} \left(\Delta E^{(1D)}_i - \Delta E^{(2D)}_i \right)
%,
%\end{align}
is $4.17~\mathrm{cm}^{-1}$, which is within spectroscopic accuracy. These results indicate that the distributed quantum implementation of the tensor-network decomposition preserves the essential spectral structure of the full proton-transfer Hamiltonian. 
\section{Discussion and Outlook}
\label{Sec:Outlook}
We introduce a tensor-network based distributed quantum algorithm and implement the algorithm on ion-trap quantum computers. The method is demonstrated by computing vibrational spectra, from wavepacket dynamics, constructed on accurate potentials obtained from electronic structure. The approach also utilizes a new resource-optimized modified phase estimation algorithm, where we directly find energy differences, rather than absolute energies and use these to compute vibrational frequencies. At the end, the agreement between the quantum computed frequencies and classically computed frequencies are of the order of 4 cm$^{-1}$, that is within spectroscopoc accuracy. Thus, we believe, this paper constitutes an important step in the area of quantum computation of chemical dynamics. 

The present work also establishes a unified connection between tensor network representations, structured quantum circuit decompositions, and distributed quantum computing architectures and implements the same on real quantum computing devices. This perspective provides a pathway to leverage the entanglement structure as a computational resource for hardware parallelism and to extend the reach of quantum simulations beyond the limits imposed by current device capabilities.

The present work is closely connected to several recent advances in quantum simulations of molecular dynamics and spectroscopy. In particular, related progress has been made in the quantum simulation of vibronic spectra and nonadiabatic molecular processes \cite{sparrow2018simulating,wang2020efficient,huh2015boson,mazziotti2023bosons}, the study of model systems involving conical intersections and geometric phase effects \cite{wang2022observation,nature2023geometricphase-1,nature2023geometricphase-2,mazziotti2024conical}, and the development of quantum algorithms for reduced-dimensional reactive scattering and molecular dynamics simulations \cite{xing2023hybrid,kale2024simulation}. The tensor-network-based distributed quantum dynamics framework introduced here is also expected to benefit substantially from ongoing developments in quantum electronic-structure algorithms, including fermion-to-qubit mappings, variational quantum eigensolvers, adaptive ansatz constructions, correlation-driven methods, and fragmentation-based approaches for molecular systems \cite{Jordan-Wigner,Ortiz-JW,BRAVYI2002210,Aspuru-Guzik-Science-2005,PeterLove-PRL-VQE-SparseMatrices,meta-VQE,o2016scalable,CQE-Mazziotti,kandala2017hardware,xia2018quantum,gorman2018engineering,nam2020ground,Photosynthesis-Quantum-Sim,LHC-quantum-circuit,LHC-superconducting-circuits,peruzzo2014variational,grimsley2019adaptive,Google-12qubit-HF,Martinez-VQE-Solver-Excited-States,tkachenko2020correlation,cervera2020meta,huggins2021efficient,mcclean2020openfermion,motta2020quantum,Artur-gatecount1,Artur-QCC-gatecount,Artur-gatecount-unitary-partitioning,frag-QC-Harry,SSI-Review1-QC-ES-QN}. These developments, together with continuing progress in experimental quantum hardware implementations \cite{lanyon2010towards,lu2011simulation,peruzzo2014variational,kandala2017hardware,hempel2018quantum,nam2020ground,google2020hartree}, provide an important foundation for scalable quantum simulations of multidimensional molecular dynamics and vibrational spectroscopy.

\section{Acknowledgment} 
This research was supported by the National Science Foundation under Grant CHE-2102610 awarded to SSI, including support provided through a Special Creativity Extension. Author AD acknowledges support from the Siedle Materials Fellowship Foundation and Lynne L.
Merritt Fellowship Foundation. These fellowships were awarded to AD by the department of chemistry, Indiana University. Author PR is supported by the Gordon and Betty Moore Foundation, grant DOI 10.37807/GBMF12963. The QSCOUT open-access testbed is funded by the U.S. Department of Energy, Office of Science, Office of Advanced Scientific Computing Research Quantum Testbed Program. Sandia National Laboratories is a multimission laboratory managed and operated by National Technology \& Engineering Solutions of Sandia, LLC, a wholly owned subsidiary of Honeywell International Inc., for the U.S. Department of Energy’s National Nuclear Security Administration under contract DE-NA0003525. This paper describes objective technical results and analysis. Any subjective views or opinions that might be expressed in the paper do not necessarily represent the views of the U.S. Department of Energy or the United States Government. SAND2026-22099O

\appendix
\begin{figure}[tbp]
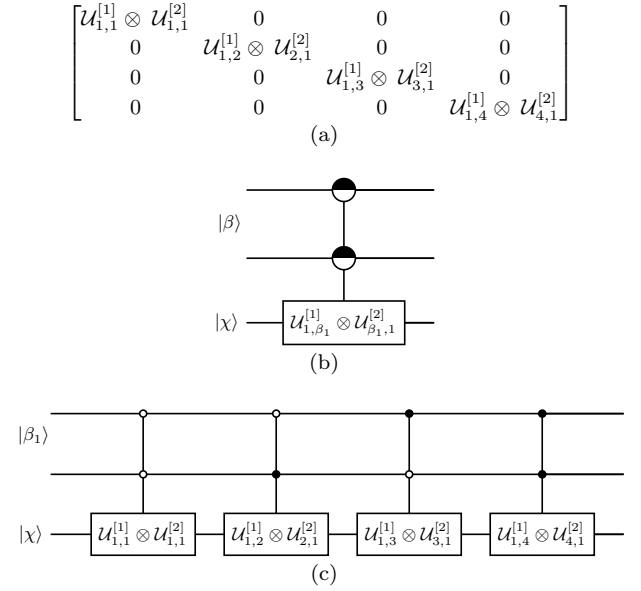

\centering
\subfigure[]{%
    \resizebox{0.38\textwidth}{!}{\input{Figures/matrix_dim2_beta4}}}
\hfill
\subfigure[]{%
    \resizebox{0.18\textwidth}{!}{\input{Figures/figure_ugc_tn1}}}
\hfill
\subfigure[]{%
   \resizebox{0.46\textwidth}{!}{\input{Figures/Figure_TN_UCG}}}
\caption{(a) Block diagonal form of the overall unitary in the elevated Hilbert space including entanglement, (b) associated uniformly controlled gate-like representation of Figure (a) with two control qubits encoding $\beta_1$, and (c) explicit form of the circuit corresponding to Figure (b). As in Figure \ref{fig:2D_2beta_ucg}, the individual blocks $\mathcal{U}_{1,\beta_1}^{[1]} \otimes \;\mathcal{U}_{\beta_1,1}^{[2]}$ are direct product operators that act on product states of dimensions ``$[1]$'' and ``$[2]$''.}
\label{fig:2D_4beta_ucg}
\end{figure}
\begin{figure*}[tbp]
    \centering
    \resizebox{0.92\textwidth}{!}{%
        \input{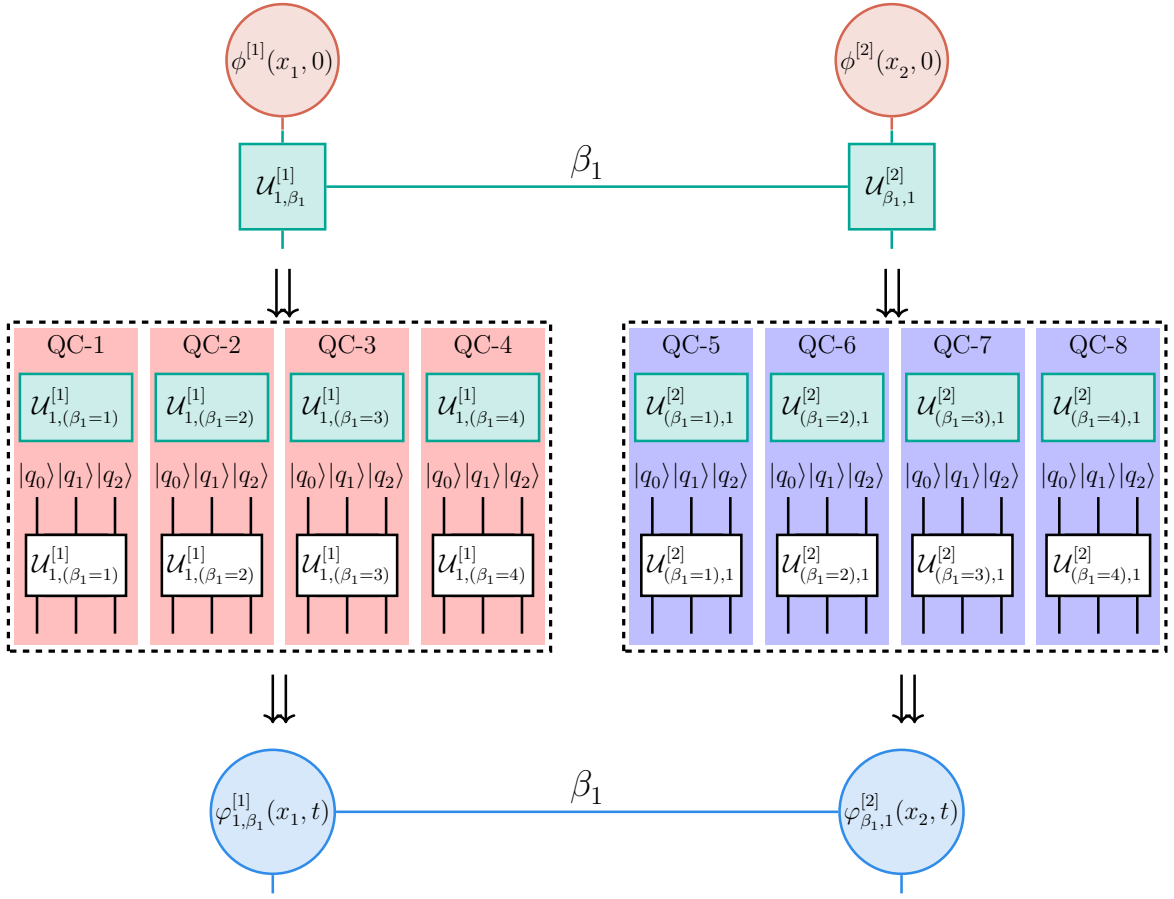}
    }
    \caption{The circuit in \cref{fig:2D_4beta_ucg} can, in principle, be distributed across eight quantum computers, with each constituent sub-circuit being executed in parallel as explained in Figure \ref{Fig:MPS_tEvo_trotter-e}.  But, as noted in Figure \ref{Fig:MPS_tEvo_trotter-e}, for simplicity, in this paper, these operations are performed sequentially on a single quantum device.}
    \label{Fig:MPS_tEvo_trotter-f}
\end{figure*}
\section{Bipartite systems with more entanglement}
\label{Bipartite-beta2}
For $N=2$ and and with maximum value of the entanglement index, $\beta_1 = 4$, the operator consists of four blocks, requiring two control qubits to select the corresponding channel. The associated matrix and circuit representation are shown in \cref{fig:2D_4beta_ucg}, where each computational basis state of the control register encodes a specific value of $\beta_1$. The parallel streams that arise from such a situation are given in Figure \ref{Fig:MPS_tEvo_trotter-f}. In both cases discussed above, the number of ancilla needed for a uniformly controlled gate implementation is $\ln \eta_1$ and the number of parallel streams is  $2\eta_1$. %Again, the quantity $\ln \eta_1$ is the maximum possible entanglement entropy of the tensor network configuration. As we will see below, this idea generalizes to higher dimensions. 
\section{A Green's function description of the Phase estimation algorithm: relation to vibrational spectroscopy from wavepacket dynamics}
\label{PEA-SDO}
In quantum phase estimation, illustrated in Figure \ref{Fig:PEA-ckt1}, a specific unitary is used to construct the time-series that is stored as ``Prop'' on the quantum computer, and can be written as,
\begin{align}
     \ket{\text {Prop}} = \frac{1}{\sqrt{2^a}} \sum_{j=0}^{2^a-1} \ket{j}_a  \left[ U^j \ket{\chi_0}_q \right].
    \label{Eq:Shor-propagated}
\end{align}
Here the kets, $\left\{ \ket{j}_a \right\}$ represent the states of the ancilla, that is 
\begin{align}
    \left\{ \ket{j}_a \vert \; \forall j \in \left\{ 00\cdots00, 00\cdots01, \cdots, 2^a-1 \right\} \right\}
\end{align}
and for each state on the ancilla, the state $\left[ U^j \ket{\chi_0}_q \right]$ is stored in the qubits thus representing an entangled ancilla-qubit state at the $\ket{\text {Prop}}$ stage. 
%Thus, the Fourier transform simply rotates the vector from the time representation to the conjugate energy representation. 
The state $\ket{\chi_0}_q$ is an initial wavepacket state that is represented on the computational basis of $q$ (bottom set of qubits in Figure \ref{Fig:PEA-ckt1}) and here, the computational basis corresponding to $\ket{q}$ are mapped onto a multi-dimensional grid basis representation to describe the quantum nuclear wavepacket. More details on the general map between the continuous representation $\ket{x}$ and discrete qubit computational basis representation for the nuclear dynamics problem treated here are presented in discussed in Section \ref{mapping}. In general, for a discrete grid index representation, $\ket{k}_q \rightarrow x_k$, with 
\begin{align}
    \left\{ \ket{k}_q \vert \; \forall q \in \left\{ 00\cdots00, 00\cdots01, \cdots, 2^q-1 \right\} \right\}
\end{align}
and $x_k$ being a specific multi-dimensional grid point, Eq. (\ref{Eq:Shor-propagated}) may be explicitly written as 
\begin{align}
     \ket{\text {{Prop}}} &= \frac{1}{\sqrt{2^a2^q}} \sum_{j=0}^{2^a-1} \sum_{k=0}^{2^q-1} \ket{j}_a  \ket{k}_q \left[ \bra{k}_q U^j \ket{\chi_0} \right] \nonumber \\ &\equiv
     \frac{1}{\sqrt{2^a2^q}} \sum_{j=0}^{2^a-1} \sum_{k=0}^{2^q-1} \ket{j}_a  \ket{x_k} \left[ \bra{x_k} U^j \ket{\chi_0} \right] 
     %\nonumber \\ &=
     %\frac{1}{\sqrt{2^a2^q}} \sum_{j=0}^{2^a-1} \sum_{k=0}^{2^q-1} \ket{j}_a  \ket{x_k} {\chi}(x_k; t_j) 
    \label{Eq:Shor-propagated-q}
\end{align}
where in the last equation we have explicitly noted the grid representation and time representation at ``Prop''. Thus, if $U = \exp\{-\imath H \Delta t / \hbar\}$, then 
\begin{align}
     \ket{\text {Prop}} &= 
     \frac{1}{\sqrt{2^a2^q}} \sum_{j=0}^{2^a-1} \sum_{k=0}^{2^q-1} \ket{j}_a  \ket{x_k} \left[ \bra{x_k} e^{\frac{-\imath H \left[j\Delta t\right]}{\hbar}} \ket{\chi_0} \right] \nonumber \\ &= \frac{1}{\sqrt{2^a2^q}} \sum_{j=0}^{2^a-1} \sum_{k=0}^{2^q-1} \ket{j}_a  \ket{x_k} {\chi}(x_k;j\Delta t)
     \label{Stage2=chi-t}
\end{align}
Finally, the Fourier transform simply rotates this vector from the time representation to the conjugate energy representation and hence the operation being performed at the QFT stage is given by,
\begin{align}
     \ket{\text {Final}} =& 
     \frac{1}{\sqrt{2^a2^a2^q}} \sum_{e=0}^{2^a-1} \sum_{j=0}^{2^a-1} \sum_{k=0}^{2^q-1} \ket{e}_a \bra{e}_a \ket{j}_a  \ket{x_k} {\chi}(x_k;j\Delta t) \nonumber \\ =& 
     \frac{1}{\sqrt{2^a2^q}} \sum_{e=0}^{2^a-1} \sum_{k=0}^{2^q-1} \ket{e}_a \ket{x_k} \nonumber \\ & 
     \left[ \sum_{j=0}^{2^a-1} \frac{1}{\sqrt{2^a}} e^{\imath (e\Delta E)(j\Delta t) / \hbar} {\chi}(x_k;j\Delta t)  \right]
\end{align}
where the last line above is of course the Fourier transform, and in rotating the ancilla from time to energy, we have also introduced an energy representation:
\begin{align}
    \left\{ \ket{e}_a \vert \; \forall e \in \left\{ 00\cdots00, 00\cdots01, \cdots, 2^a-1 \right\} \right\},
\end{align}
hence the state of the system prior to the measurement can also be written as
\begin{align}
     \ket{\text {Final}} =& 
     \frac{1}{\sqrt{2^a2^q}} \sum_{e=0}^{2^a-1} \sum_{k=0}^{2^q-1} \ket{e}_a \ket{x_k} \xi(x_k,E_e) 
     \label{PEA-Final}
\end{align}
where 
\begin{align}
\xi(x,E) =& \sum_{j=0}^{2^a-1} \frac{1}{\sqrt{2^a}} e^{\imath (e\Delta E)(j\Delta t) / \hbar} {\chi}(x_k;j\Delta t) \nonumber \\ \equiv& \int dt \exp{\imath E t / \hbar} \chi(x,t) 
\end{align}
The measurement of $\ket{\text {Final}}$ gives you the state $\left\{ \ket{k}; \ket{e} \right\}$, with probability, ${\left\vert \xi(x_k,E_e) \right\vert}^2 $
%\begin{align}
%{\left\vert \xi(x_k,E_e) \right\vert}^2 \equiv {\left\vert  \int dt \exp{-\imath E t / \hbar} \chi(x,t) \right\vert}^2
%\label{FT-noabs}
%\end{align}
at a specific value of $x_k$ (system qubit computational basis value) and $E_e$ (ancilla computational basis value). In passing, we also note that $\xi(x,E)$ can also be written as 
\begin{align}
\xi(x,E) =& \int dt \exp{\imath E t / \hbar} \exp{-\imath H t / \hbar} \chi(x,0) \nonumber \\ =& \delta(E-H) \chi(x,0)
\label{deltaEminusH}
\end{align}
where $\delta(E-H)$ is a Green's function, also known as the spectral density operator, and is the difference between the advanced and retarded Green's functions $\delta(E-H) = {\cal G}^+(E) - {\cal G}^-(E)$\cite{RGNewton,TIW}. 
\section{Qubit resource optimized phase estimation using tensor networks}
\label{Appendix:appendix-FT}
Using the tensor-network representation introduced above, the density--density correlation spectrum in \cref{Eq:Density-timecorrelation-FT-Final} can be expressed in terms of the tensor-network wavefunction components as
\begin{widetext}
\begin{align}
{\cal P}(\omega)
&=
\int d\bar{x}\,
\left|
\int_{-\infty}^{+\infty}
dt\,
e^{i\omega t}\,
\rho(\bar{x},\bar{x};t)
\right|^2
\nonumber\\
&=
\int d\bar{x}\,
\left|
\sum_{\bar{\boldsymbol\alpha}}^{}
\sum_{\bar{\boldsymbol\beta}}^{}
\int_{-\infty}^{+\infty}
dt\,
e^{i\omega t}
\prod_{j=1}^{N}
\left(\tensor*{\phi}{^{[j]}_{}^{x_j,t}_{\alpha_{j-1},\alpha_j}}\right)
\left({\tensor*{\phi}{^{[j]}_{}^{x_j,t}_{\beta_{j-1}\beta_j}}}\right)^*
\right|^2 .
\label{Eq:Power_FT_TN_full}
\end{align}
\end{widetext}

The expression above contains both diagonal and off-diagonal tensor-network channels. Separating these contributions yields
\begin{widetext}
\begin{align}
{\cal P}(\omega)
&=
\int d\bar{x}\,
\Bigg|
\sum_{\bar{\boldsymbol\alpha}}^{}
\int_{-\infty}^{+\infty}
dt\,
e^{i\omega t}
\prod_{j=1}^{N}
\left|
\tensor*{\phi}{^{[j]}_{}^{x_j,t}_{\alpha_{j-1},\alpha_j}}
\right|^2
\nonumber\\
&\qquad+
\sum_{\bar{\boldsymbol\alpha}}^{}
\sum_{\bar{\boldsymbol\beta}\neq\bar{\boldsymbol\alpha}}^{}
\int_{-\infty}^{+\infty}
dt\,
e^{i\omega t}
\prod_{j=1}^{N}
\left(\tensor*{\phi}{^{[j]}_{}^{x_j,t}_{\alpha_{j-1},\alpha_j}}\right)
\left({\tensor*{\phi}{^{[j]}_{}^{x_j,t}_{\beta_{j-1}\beta_j}}}\right)^*
\Bigg|^2 .
\label{Eq:Power_FT_TN_split}
\end{align}
\end{widetext}

In the present work, we specifically evaluate the diagonal contribution associated with the individual tensor-network legs. The corresponding reduced spectral function is therefore written as
\begin{widetext}
\begin{align}
{\cal P}'(\omega)
=
\int d\bar{x}\,
\left|
\sum_{\bar{\boldsymbol\alpha}}^{\bar{\boldsymbol\eta}}
\int_{-\infty}^{+\infty}
dt\,
e^{i\omega t}
\prod_{j=1}^{N}
\left|
\tensor*{\phi}{^{[j]}_{}^{x_j,t}_{\alpha_{j-1},\alpha_j}}
\right|^2
\right|^2 .
\label{Eq:Power_FT_TN}
\end{align}
\end{widetext}

As shown in \cref{Eq:Power_FT_TN}, restricting the analysis to the diagonal tensor-network channels modifies the relative spectral intensities while preserving the locations of the vibrational transition frequencies.

To illustrate the resulting spectral structure, we consider the special case of an initially separable wavepacket corresponding to the product-state limit of \cref{Eq:MPS_WF},
\begin{align}
\chi_0(\bar{\vb{x}})
=
\prod_{j=1}^{N}
\tensor*{\phi}{^{[j]}_{}^{x_j,0}}.
\label{Eq:product_WF}
\end{align}

Each tensor-network leg then evolves independently under its corresponding effective one-dimensional propagator,
\begin{align}
\tensor*{\phi}{^{[j]}_{}^{x_j,t}_{\alpha_{j-1},\alpha_j}}
=
\int dx'_j\, \tensor*{\mathcal{U}}{^{[j]}_{}^{x_jx'_j}_{\alpha_{j-1},\alpha_j}}
\tensor*{\phi}{^{[j]}_{}^{x'_j,0}},
\label{eq:phi_Uchi}
\end{align}
%where repeated coordinate indices imply integration over the intermediate coordinate $x'_j$.

The initial wavepacket associated with the $j$th tensor-network leg is expanded in the eigenbasis of the corresponding effective Hamiltonian,
\begin{align}
\tensor*{\phi}{^{[j]}_{}^{x_j,0}}
=
\sum_{l_{j,\alpha_{j-1},\alpha_j}}
c^{[j]}_{l_{j,\alpha_{j-1},\alpha_j}}\,
\varphi^{[j]}_{l_{j,\alpha_{j-1},\alpha_j}}(x_j).
\label{eq:chi0_expansion}
\end{align}

The eigenstates satisfy
\begin{widetext}
\begin{align}
\int dx'_j\,
\tensor*{\mathcal{U}}{^{[j]}_{}^{x_jx'_j}_{\alpha_{j-1},\alpha_j}}
\varphi^{[j]}_{l_{j,\alpha_{j-1},\alpha_j}}(x'_j)
=
e^{-iE^{[j]}_{l_{j,\alpha_{j-1},\alpha_j}}t/\hbar}
\varphi^{[j]}_{l_{j,\alpha_{j-1},\alpha_j}}(x_j),
\label{eq:U_eigen}
\end{align}
\end{widetext}
where, for clarity, we retain only the unitary component of the tensor-network propagator. (As noted in Section \ref{Logham} the non-unitarity of $\left\{ \tensor*{\mathcal{U}}{^{[j]}_{}^{x_jx'_j}_{\alpha_{j-1},\alpha_j}} \right\}$ is a key feature in this algorithm.)
%The nonunitary prefactors discussed previously contribute only to the overall amplitudes and spectral intensities, while leaving the vibrational transition frequencies and eigenspectrum unchanged. Consequently, the spectral structure is determined entirely by the unitary phase evolution generated by the effective Hamiltonians.

From \cref{eq:phi_Uchi,eq:chi0_expansion,eq:U_eigen}, the propagated tensor-network leg therefore becomes
\begin{widetext}
\begin{align}
\tensor*{\phi}{^{[j]}_{}^{x_j,t}_{\alpha_{j-1},\alpha_j}}
=
\sum_{l_{j,\alpha_{j-1},\alpha_j}}
c^{[j]}_{l_{j,\alpha_{j-1},\alpha_j}}\,
\varphi^{[j]}_{l_{j,\alpha_{j-1},\alpha_j}}(x_j)\,
e^{-iE^{[j]}_{l_{j,\alpha_{j-1},\alpha_j}}t/\hbar}.
\label{eq:phi_time}
\end{align}
The corresponding density associated with the $j$th tensor-network leg is therefore
\begin{align}
\left|
\tensor*{\phi}{^{[j]}_{}^{x_j,t}_{\alpha_{j-1},\alpha_j}}
\right|^2
=
\sum_{l_{j,\alpha_{j-1},\alpha_j}}\sum_{m_{j,\alpha_{j-1},\alpha_j}}
c^{[j]}_{l_{j,\alpha_{j-1},\alpha_j}}
c^{[j]*}_{m_{j,\alpha_{j-1},\alpha_j}}\,
\varphi^{[j]}_{l_{j,\alpha_{j-1},\alpha_j}}(x_j)\,
\varphi^{[j]*}_{m_{j,\alpha_{j-1},\alpha_j}}(x_j)
e^{-i\left(
E^{[j]}_{l_{j,\alpha_{j-1},\alpha_j}}
-
E^{[j]}_{m_{j,\alpha_{j-1},\alpha_j}}
\right)t/\hbar}.
\label{eq:phi_density}
\end{align}
\end{widetext}

%which explicitly contains oscillatory contributions arising from energy differences between the eigenstates of the effective one-dimensional Hamiltonian. 
Hence, the time-dependent density associated with an individual tensor-network leg can be expressed as a coherent superposition of oscillatory contributions arising from energy differences between the eigenstates of the corresponding effective entangled lower-dimensional Hamiltonians,
\begin{widetext}
\begin{align}
\left|
\tensor*{\phi}{^{[j]}_{}^{x_j,t}_{\alpha_{j-1},\alpha_j}}
\right|^2
=
\sum_{l_{j,\alpha_{j-1},\alpha_j}}\sum_{m_{j,\alpha_{j-1},\alpha_j}}
C^{[j]}_{l_j,m_{j,\alpha_{j-1},\alpha_j}}(x_j)\,
e^{-i\Delta E^{[j]}_{l_j,m_{j,\alpha_{j-1},\alpha_j}} t/\hbar},
\label{eq:leg_density}
\end{align}
\end{widetext}
where the coefficients
$C^{[j]}_{l_j,m_{j,\alpha_{j-1},\alpha_j}}(x_j)$
contain the spatial overlap and population amplitudes associated with the corresponding eigenstate pair,
\begin{widetext}
\begin{align}
C^{[j]}_{l_j,m_{j,\alpha_{j-1},\alpha_j}}(x_j)
&=
c^{[j]}_{l_{j,\alpha_{j-1},\alpha_j}}
c^{[j]*}_{m_{j,\alpha_{j-1},\alpha_j}}\,
\varphi^{[j]}_{l_{j,\alpha_{j-1},\alpha_j}}(x_j)\,
\varphi^{[j]*}_{m_{j,\alpha_{j-1},\alpha_j}}(x_j),
\\
\Delta E^{[j]}_{l_j,m_{j,\alpha_{j-1},\alpha_j}}
&=
E^{[j]}_{l_{j,\alpha_{j-1},\alpha_j}}
-
E^{[j]}_{m_{j,\alpha_{j-1},\alpha_j}}.
\end{align}
\end{widetext}

The quantity
$\Delta E^{[j]}_{l_j,m_{j,\alpha_{j-1},\alpha_j}}$
therefore represents the transition energy between the eigenstates
$l_j$
and
$m_j$
of the effective Hamiltonian associated with the $j$th tensor-network leg and $\alpha_{j-1},\alpha_j$ entanglement index. 

For the full tensor-network wavefunction, the multidimensional density is obtained by combining the contributions from all tensor-network legs,
\begin{widetext}
\begin{align}
\prod_{j=1}^{N}
\left|
\tensor*{\phi}{^{[j]}_{}^{x_j,t}_{\alpha_{j-1},\alpha_j}}
\right|^2
=
\sum_{\{l_{j,\alpha_{j-1},\alpha_j},m_{j,\alpha_{j-1},\alpha_j}\}}
\left[
\prod_{j=1}^{N}
C^{[j]}_{l_j,m_{j,\alpha_{j-1},\alpha_j}}(x_j)
\right]
\exp\!\left[
-\frac{i t}{\hbar}
\sum_{j=1}^{N}
\Delta E^{[j]}_{l_j,m_{j,\alpha_{j-1},\alpha_j}}
\right],
\label{eq:prod_expanded}
\end{align}
\end{widetext}
where
$\sum_{\{l_{j,\alpha_{j-1},\alpha_j},m_{j,\alpha_{j-1},\alpha_j}\}}$
denotes the summation over all eigenstate pairs for every tensor-network leg. Equation~\eqref{eq:prod_expanded} shows that the multidimensional time evolution contains oscillatory components whose frequencies are determined by sums of transition energies originating from the effective one-dimensional subsystems.

The Fourier transform of the multidimensional density then becomes
\begin{widetext}
\begin{align}
\int_{-\infty}^{+\infty}\dd{t}\,
e^{i\omega t}
\prod_{j=1}^{N}
\left|
\tensor*{\phi}{^{[j]}_{}^{x_j,t}_{\alpha_{j-1},\alpha_j}}
\right|^2
=
\sum_{\{l_{j,\alpha_{j-1},\alpha_j},m_{j,\alpha_{j-1},\alpha_j}\}}
\left[
\prod_{j=1}^{N}
C^{[j]}_{l_j,m_{j,\alpha_{j-1},\alpha_j}}(x_j)
\right]
\delta\!\left[
\omega
-
\frac{1}{\hbar}
\sum_{j=1}^{N}
\Delta E^{[j]}_{l_j,m_{j,\alpha_{j-1},\alpha_j}}
\right].
\label{eq:time_integrated}
\end{align}
\end{widetext}

Equation~\eqref{eq:time_integrated} shows that the multidimensional Fourier spectrum contains peaks corresponding both to sums of transition energies across multiple tensor-network legs and to individual transition energies associated with a particular subsystem. In particular, when
$l_{j,\alpha_{j-1},\alpha_j}=m_{j,\alpha_{j-1},\alpha_j}$
for all dimensions except a specific leg $j_1$, the spectral contribution reduces to
\[
\omega
=
\frac{
\Delta E^{[j_1]}_{l_{j_1}m_{j_1},\alpha_{j_1-1},\alpha_{j_1}}
}{\hbar},
\]
yielding peaks corresponding directly to the eigenenergy differences of the effective one-dimensional Hamiltonian associated with the $j_1$th tensor-network leg. Consequently, the multidimensional Fourier spectrum naturally contains both collective multidimensional excitations and the individual vibrational transition frequencies of the effective one-dimensional subsystems.

Substituting Eq.~\eqref{eq:time_integrated} into Eq.~\eqref{Eq:Power_FT_TN} yields
\begin{widetext}
\begin{align}
{\cal P}'(\omega)
=
\int d\bar{x}\,
\left|
\sum_{\bar{\boldsymbol\alpha}}^{\bar{\boldsymbol\eta}}
\sum_{\{l_{j,\alpha_{j-1},\alpha_j},m_{j,\alpha_{j-1},\alpha_j}\}}
\left[
\prod_{j=1}^{N}
C^{[j]}_{l_j,m_{j,\alpha_{j-1},\alpha_j}}(x_j)
\right]
\delta\!\left[
\omega
-
\frac{1}{\hbar}
\sum_{j=1}^{N}
\Delta E^{[j]}_{l_j,m_{j,\alpha_{j-1},\alpha_j}}
\right]
\right|^2 .
\label{Eq:Power_FT_TN_final}
\end{align}
\end{widetext}

The expression above explicitly connects the spectral peaks observed in the tensor-network quantum dynamics to the transition energies of the effective one-dimensional Hamiltonians. As demonstrated in the Results section, these transition frequencies closely reproduce the vibrational energy differences obtained from exact diagonalization of the full multidimensional Hamiltonian.

\section{Decomposing the family of ${\left\{ \tensor*{\mathcal{U}}{^{[j]}_{}^{x_j x'_j}_{\beta_{j-1},}_{\beta_j}} \right\}}$ into native ion-trap gates}
\label{Sec:QSD}
This appendix provides the full decomposition used to implement the operations $\left\{ \tensor*{\mathcal{U}}{^{[j]}_{}^{x_j x'_j}_{\beta_{j-1},}_{\beta_j}} \tensor*{\phi}{^{[j]}_{}^{x_j}_{\alpha_{j-1},\alpha_j}} \right\}$ discussed in \cref{Sec:TN}.

\begin{figure}[hbtp]
\begin{center}
\begin{quantikz}[row sep=0.2cm,column sep=0.3cm]
& & \gate[3]{MR_z} & & \gate[3]{MR_y} & & \gate[3]{MR_z} & & \\
& \gate[2]{A_1} & & \gate[2]{B_1} & & \gate[2]{A_2} & & \gate[2]{B_2} & \\
& & & & & & & & \\
\end{quantikz}
\end{center}
\caption{
Three-qubit QSD decomposition grouped into alternating two-qubit and multicontrolled three-qubit subroutines. 
}
\label{fig:qsd7sub}
\end{figure}
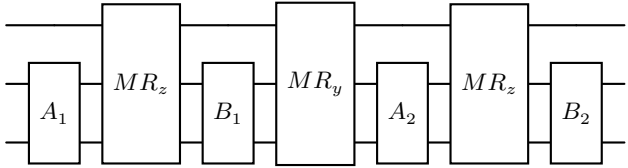
The QSD approach described in Ref. \onlinecite{Debadrita-water} requires 24 fully-entangling two-qubit gates to implement an arbitrary three-qubit unitary. For quantum hardware platforms that admit partial-angle entangling gates, such as trapped ions, the circuit depth and complexity may be further reduced. We begin by writing the three-qubit QSD in terms of 7 subroutines, as shown in Fig.~\ref{fig:qsd7sub}.

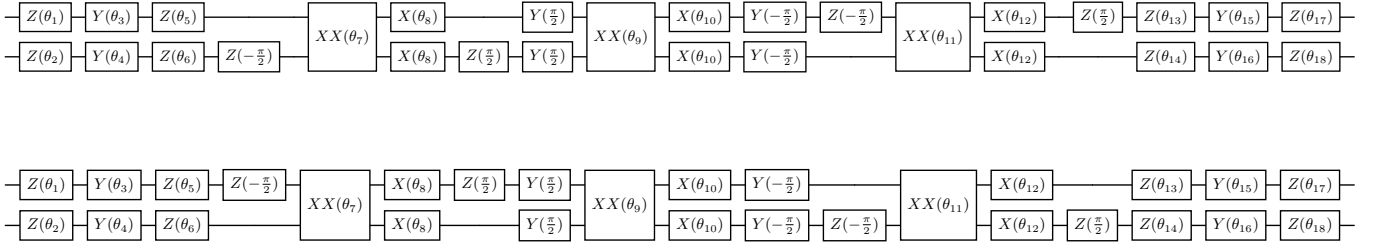
\begin{figure*}[thbp]
\begin{center}
\begin{adjustbox}{width=\textwidth}
\begin{quantikz}[row sep=0.2cm,column sep=0.3cm]
& \gate{Z(\theta_1)} & \gate{Y(\theta_3)} & \gate{Z(\theta_5)} & & & \gate[2]{XX(\theta_7)} & \gate{X(\theta_{8})} & & \gate{Y(\frac{\pi}{2})} & \gate[2]{XX(\theta_9)} & \gate{X(\theta_{10})} & \gate{Y(-\frac{\pi}{2})} & \gate{Z(-\frac{\pi}{2})} & \gate[2]{XX(\theta_{11})} & \gate{X(\theta_{12})} & & \gate{Z(\frac{\pi}{2})} & \gate{Z(\theta_{13})} & \gate{Y(\theta_{15})} & \gate{Z(\theta_{17})} &\\
& \gate{Z(\theta_2)} & \gate{Y(\theta_4)} & \gate{Z(\theta_6)} & \gate{Z(-\frac{\pi}{2})} & & & \gate{X(\theta_{8})} & \gate{Z(\frac{\pi}{2})} & \gate{Y(\frac{\pi}{2})} & & \gate{X(\theta_{10})} & \gate{Y(-\frac{\pi}{2})} & & & \gate{X(\theta_{12})} & & & \gate{Z(\theta_{14})} & \gate{Y(\theta_{16})} & \gate{Z(\theta_{18})} &\\
\end{quantikz}
\end{adjustbox}

\vspace{30pt}

\begin{adjustbox}{width=\textwidth}
\begin{quantikz}[row sep=0.2cm,column sep=0.3cm]
& \gate{Z(\theta_1)} & \gate{Y(\theta_3)} & \gate{Z(\theta_5)} & \gate{Z(-\frac{\pi}{2})} & \gate[2]{XX(\theta_7)} & \gate{X(\theta_{8})} & \gate{Z(\frac{\pi}{2})} & \gate{Y(\frac{\pi}{2})} & \gate[2]{XX(\theta_9)} & \gate{X(\theta_{10})} & \gate{Y(-\frac{\pi}{2})} & & \gate[2]{XX(\theta_{11})} & \gate{X(\theta_{12})} & & \gate{Z(\theta_{13})} & \gate{Y(\theta_{15})} & \gate{Z(\theta_{17})} &\\
& \gate{Z(\theta_2)} & \gate{Y(\theta_4)} & \gate{Z(\theta_6)} & & & \gate{X(\theta_{8})} & & \gate{Y(\frac{\pi}{2})} & & \gate{X(\theta_{10})} & \gate{Y(-\frac{\pi}{2})} & \gate{Z(-\frac{\pi}{2})} & & \gate{X(\theta_{12})} & \gate{Z(\frac{\pi}{2})} & \gate{Z(\theta_{14})} & \gate{Y(\theta_{16})} & \gate{Z(\theta_{18})} &\\
\end{quantikz}
\end{adjustbox}

\end{center}
\caption{
Full gate sequence for subroutines $A_{1,2}$ (top) and $B_{1,2}$ (bottom) used in the partial-angle QSD decomposition. Each circuit has 18 adjustable angles $\theta_i$ which are used to specify the desired two-qubit unitary.
}
\label{fig:qsd2qubit}
\end{figure*}

In Fig.~\ref{fig:qsd7sub}, the four subroutines $A_{1,2}$ and $B_{1,2}$ operate only on two qubits and may be fully decomposed using the KAK method\cite{tucci2005introduction}. This yields a minimal quantum circuit with 15 single-qubit gates and 3 fully entangling two-qubit operations. In our implementation, the full two-qubit entanglement gates are replaced by partial-angle $XX(\theta)$ gates, with the surrounding single-qubit gates adjusted accordingly. The resulting sequences for these two-qubit unitary subroutines is shown in Fig.~\ref{fig:qsd2qubit}.

\begin{figure*}[thbp]
\begin{center}
\begin{adjustbox}{width=\textwidth}
\begin{quantikz}[row sep=0.2cm,column sep=0.3cm]
& \gate{Z(\theta_1)} & \gate{Y(\frac{\pi}{2})} & \gate[2]{XX_{12}(\theta_2)} & \gate{X(\theta_{3})} & \gate{Y(-\frac{\pi}{2})} & \gate[3]{XX_{13}(\frac{\pi}{2})} & \gate{X(-\frac{\pi}{2})} & \gate{Y(\frac{\pi}{2})} & \gate[2]{XX_{12}(\theta_4)} & \gate{X(\theta_{5})} & \gate{Y(-\frac{\pi}{2})} & \gate{Z(\theta_6)} & \gate[3]{XX_{13}(\frac{\pi}{2})} & \gate{X(-\frac{\pi}{2})} &\\
& & \gate{Y(\frac{\pi}{2})} & & \gate{X(\theta_{3})} & & & & & & \gate{X(\theta_{5})} & \gate{Y(-\frac{\pi}{2})} & & & &\\
& & \gate{Y(\frac{\pi}{2})} & & & & & \gate{X(\pi)} & & & & & & & \gate{Y(-\frac{\pi}{2})} &\\
\end{quantikz}
\end{adjustbox}

\vspace{30pt}

\begin{adjustbox}{width=\textwidth}
\begin{quantikz}[row sep=0.2cm,column sep=0.3cm]
& \gate{Y(\theta_1)} & \gate{Z(-\frac{\pi}{2})} & \gate[2]{XX_{12}(\theta_2)} & \gate{X(\theta_{3})} & \gate{Z(\frac{\pi}{2})} & \gate[3]{XX_{13}(\frac{\pi}{2})} & \gate{X(-\frac{\pi}{2})} & \gate{Z(-\frac{\pi}{2})} & \gate[2]{XX_{12}(\theta_4)} & \gate{X(\theta_{5})} & \gate{Z(\frac{\pi}{2})} & \gate{Y(\theta_6)} & \gate[3]{XX_{13}(\frac{\pi}{2})} & \gate{X(-\frac{\pi}{2})} &\\
& & \gate{Y(\frac{\pi}{2})} & & \gate{X(\theta_{3})} & & & & & & \gate{X(\theta_{5})} & \gate{Y(-\frac{\pi}{2})} & & & &\\
& & \gate{Y(\frac{\pi}{2})} & & & & & \gate{X(\pi)} & & & & & & & \gate{Y(-\frac{\pi}{2})} &\\
\end{quantikz}
\end{adjustbox}

\end{center}
\caption{
Gate decomposition for multicontrolled three-qubit subroutines $MR_z$ (top) and $MR_y$ (bottom). Each circuit contains 6 adjustable angles $\theta_i$ used to specify the target multicontrolled unitary.
}
\label{fig:qsd3qubit}
\end{figure*}
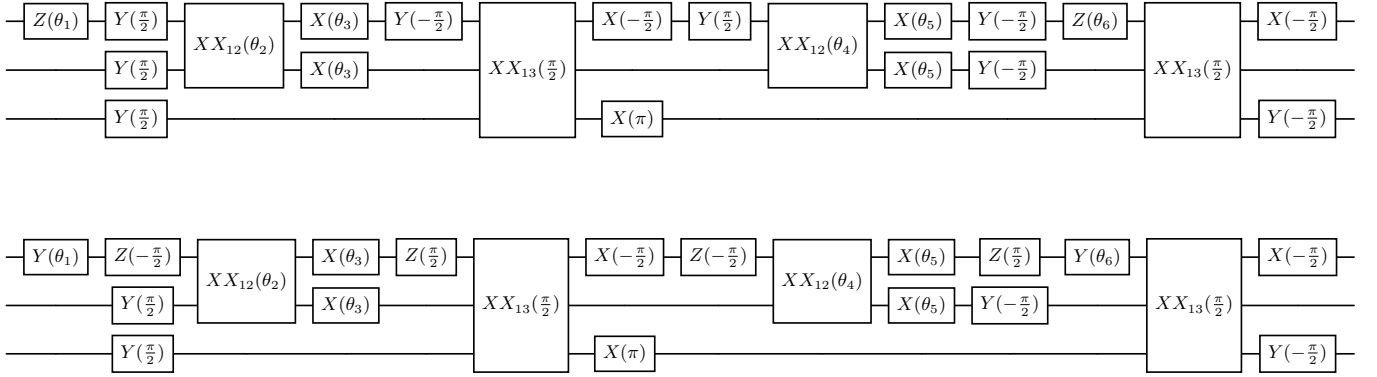

The full QSD additionally requires three multi-controlled operations acting on all three qubits. In Fig.~\ref{fig:qsd7sub}, subroutines $MR_z$ and $MR_y$ indicate $R_z$ and $R_y$ rotations (respectively), with rotation angles controlled by the four possible states of the top two qubits. In Fig.~\ref{fig:qsd3qubit}, we provide the full decomposition of these subroutines using partial-angle entangling gates. Compared to a typical multicontrolled gate on three qubits, which requires four fully-entangling operations, our decomposition executes the same unitary using two fully-entangling and two partially-entangling operations. In total, six adjustable angles are used to specify the desired $MR_z$ and $MR_y$ subroutines.

%\onecolumngrid


%%%%%%%%%%%%%%%%%%%%%%%%%%%%%%%%%%%%%%%%%%%%%%%%%%%
%\bibliography{References/electronic-structure,References/21mer-cc,References/Quantum-Computing,References/Quantum-sim-chemistry,References/TN-paper,References/bio,References/water-clusters,References/SLO1,References/srini-water,References/Bohm,References/dafsmooth,References/wavepacket,References/htransfer,References/aimdrefs,References/PBC,References/jeremy,References/qwaimdrefs,References/math,References/PRX-refs,References/SSI-molecular-fragmentation,References/Xiaohu-5th-refs,more-refs,References/waterrefs,References/isoprene-related,References/admprefs,References/PES,References/exp-vib-action,References/bipy,References/qc_refs}
\end{document}